\documentclass[epsfig,12pt,onecolumn]{article}
\usepackage{amsfonts}
\usepackage{amssymb}
\usepackage{multicol}
\usepackage{graphicx}
\usepackage{float}
\usepackage{caption}
\usepackage{xcolor}
\usepackage[utf8]{inputenc}
\usepackage{amsmath}

\makeatletter \@addtoreset{equation}{section}
\renewcommand{\theequation}{\thesection.\arabic{equation}}
\def \be{\begin{equation}}
\def \ee{\end{equation}}
\def \bea{\begin{eqnarray}}
\def \eea{\end{eqnarray}}

\newcommand{\nc}{\newcommand}
\nc{\al}{\alpha} \nc{\bib}{\bibitem} \nc{\la}{\lambda}
\nc{\C}{\mbox{\hspace{1.24mm}\rule{0.2mm}{2.5mm}\hspace{-2.7mm} C}}
\nc{\R}{\mbox{\hspace{.04mm}\rule{0.2mm}{2.8mm}\hspace{-1.5mm} R}}
\input{tcilatex}
\begin{document}

\title{\rightline{\mbox{\small {LPHE-MS-March-23}} \vspace
{1 cm}}\textbf{'t Hooft lines of ADE-type and Topological Quivers}}
\author{Y. Boujakhrout, E.H Saidi, R. Ahl Laamara, L.B Drissi \\
{\small 1. LPHE-MS, Science Faculty}, {\small Mohammed V University in
Rabat, Morocco}\\
{\small 2. Centre of Physics and Mathematics, CPM- Morocco, Rabat}}
\maketitle

\begin{abstract}
We investigate 4D Chern-Simons theory with ADE gauge symmetries in the
presence of interacting Wilson and 't Hooft line defects. We analyse the
intrinsic properties of these lines' coupling and explicate the building of
oscillator-type Lax matrices verifying the RLL integrability equation. We
propose gauge quiver diagrams Q$_{G}^{\mu }$ encoding the topological data
carried by the Lax operators and give several examples where Darboux
coordinates are interpreted in terms of topological bi-fundamental matter.
We exploit this graphical description $\left( i\right) $ to give new results
regarding solutions in representations beyond the fundamentals of $sl_{N}$, $%
so_{2N}$ and $e_{6,7}$, and $\left( ii\right) $ to classify the Lax
operators for simply laced symmetries in a unified E$_{7}$ CS theory. For
quick access, a summary list of the leading topological quivers Q$%
_{ADE}^{\mu }$ is given in the conclusion section [Figures \textbf{\ref{TA}}%
.(a-e)\textbf{, \ref{TD}}.(a-d)\textbf{\ }and\textbf{\ \ref{TE}}.(a-d)].

\textbf{Keywords}: 4D CS theory, Wilson /'t Hooft lines, Lax
operators, Oscillator realisation, Gauge quiver diagrams, Topological
matter, E$_{7}$ Unified Theory.
\end{abstract}

\section{ Introduction}

Integrable two-dimensional field theories and spin models represent a
significant area in classical and quantum physics that still bear\ several
open questions intending to explicitly describe the interactions between
fundamental particles \cite{1A}-\cite{7A}. The investigation of\textrm{\
special features of these low dimensional theories has aroused much interest
since the integrable spin chains advent \cite{11} and the factorisation of
many body scattering amplitudes of relativistic QFT \cite{11A,12}. In these
regards,} tremendous efforts have been deployed to deal with the basic
equations underlying these systems by following various approaches such as
the Bethe Ansatz \cite{5}-\cite{13A}, quantum groups \cite{3} and algebraic
methods involving Yangian and graded-Yangian representations \cite{14}-\cite%
{18}. \

Recently, these efforts gained a big impulse after the setup by Costello,
Witten and Yamazaki of a Chern-Simons -like theory living on
four-dimensional $M_{4}$ with the typical (rational) fibration $\mathbb{R}%
^{2}\times C$, and having a complexified gauge symmetry $G$ \cite{19}-\cite%
{21}. This topological gauge theory \textrm{represents a higher dimensional
field framework to approach quantum integrability and offers a new form of
the gauge/Integrability correspondence \cite{Y1}-\cite{23}. On another side,}
it bridges to $N=(1,1)$ supersymmetric Yang Mills theory in six and lower
dimensions \cite{20C}-\cite{20H} and to supersymmetric quiver gauge theories
\cite{20F}-\cite{20N}. \textrm{It also} allows for an \textrm{interesting}
realisation of solvable systems in terms of intersecting M-branes of the 11d
M-theory and, via dualities, in terms of intersecting branes in type II
strings with NS5- and D-branes as the main background \cite{20B}-\cite{27}.\
\

The main\textrm{\ ingredients of} the 4D Chern-Simons theory are line and
surface defects \cite{29}-\cite{34}; these topological quantities play a
fundamental role \textrm{in the study of} this theory and the realisation of
lower dimensional solvable systems. \textrm{In particular,} we distinguish
two basic line operators: $\left( i\right) $ Electrically charged Wilson
lines which, roughly speaking, are \textrm{assimilated} to worldlines of
particles in 2D space-time with a spectral parameter $z$ related to
rapidity; they are characterised by highest weights $\lambda $ of
representations $R$ of the gauge symmetry $G$. $\left( ii\right) $
Magnetically charged 't Hooft lines characterised by coweights $\mu $\ of $G$
\textrm{and acting like Dirac monopoles.} The coupling of these lines
\textrm{in the 4D gauge theory} is behind \textrm{important results} of
quantum integrability. In these regards, recall that crossing Wilson lines
yield a nice realisation of the famous R-matrix and the Yang-Baxter equation
of integrable two-dimensional QFTs \cite{19}.\

Regarding the magnetically charged line defects to be further explored in
this paper, they have recently been subject to particular interest \textrm{%
where they were interpreted} in terms of the monodromy matrix for non
compact spin chains, the transfer matrix for compact spin chains \cite{31,32}
and more specifically as the Baxter Q-operator \cite{34A}. They have also
been implemented in various contexts as boundaries of surface defects \cite%
{34C}, or as type II branes intersecting along distinguished directions \cite%
{34D}. Moreover, these brane realisations open windows to links between
integrable spin and superspin chains and supersymmetric gauge quiver
theories via correspondences like the so-called Bethe/gauge correspondence
\cite{37}-\cite{40}. \

In what concerns us here, a quantum integrable XXX spin chain with $N$ nodes
can be generated in the framework of the 4D CS by taking $N$ parallel Wilson
lines perpendicularly crossed by a 't Hooft line standing for the magnetic
field created by the system of the spin chain particles \cite{34A}. In this
spirit, one can calculate the Lax operator for each node of the chain as a
coupling of Wilson and 't Hooft lines in the gauge theory. The power of
\textrm{this construction} with interacting lines in 4D comes from: $\left(
i\right) $ the topological invariance on the real plane $\mathbb{R}^{2}$
\textrm{that translates into the RLL integrability equation, }$\left(
ii\right) $ the Dirac -like singularity of the topological gauge
configuration in the presence of 't Hooft line \textrm{yielding the
oscillator realisation of the Lax operator, }$\left( iii\right) $ the
holomorphy of observables on the \textrm{Riemann surface }$C$\textrm{\ where
the complex parameter }$z$\textrm{\ allows for realisations in the Yangian
representation.} These features constitute the common thread of the
fascinating results derived from this Gauge/Integrability correspondence. In
particular, it was shown in \cite{34A} that for the special case where the
magnetic charge of the 't Hooft line is given by a minuscule coweight $\mu $%
\ of the gauge group $G$, the oscillator realisation of the Lax operator for
a spin chain with internal symmetry given by $g$, the lie algebra of $G$, is
easily constructed in 4D CS as the parallel transport of the topological
field connexion through the singular 't Hooft line. \textrm{This yields a
general formula permitting to explicitly realise the Lax or the L-operator
in the fundamental representation} of any lie algebra $g$\textrm{\ having at
least one minuscule representation, in terms of harmonic oscillators.}\

\textrm{The main goal of this paper is to deeply analyse the data carried by
the Lax operator and encode it into a simple gauge quiver description
unveiling interesting common features of this quantity. These properties are
relevant for both the study of integrable spin chains and of the gauge
fields behavior in the presence of disorder operators. To this end, we
investigate} 4D Chern-Simons theories on $\mathbb{R}^{2}\times \mathbb{CP}%
^{1}$ with complex gauge symmetries $G=$A$_{n}$, D$_{n}$, E$_{6,7}$ by
implementing Wilson and 't Hooft line defects and studying intrinsic
topological features of their coupling. In these regards, notice that the
oscillator realisation of Lax matrices for minuscule nodes of $sl_{N}$ and $%
so_{2N}$ was firstly recovered from 4D CS in \cite{34A}; the exceptional E$%
_{6}$ and E$_{7}$ minuscule Lax operators were constructed in details in
\cite{54}, while a full list of ABCDE minuscule Lax matrices is collected in
\cite{54A} where the absence of a Lax matrix for the E$_{8}$ symmetry is
because this group has no minuscule coweight. Here, in order to graphically
visualise the effect of the Dirac-like singularity induced by a 't Hooft
line on a deep level of the gauge configuration, we treat each case
separately by demystifying the Lie algebra components appearing in the
construction of the L-operator and derive its action on the internal quantum
states by using a projector basis in the electric representation.
Eventually, we can build the corresponding topological quivers Q$_{G}^{\mu }$%
\textrm{\ where we translate the topological data of the lines' coupling
into quiver-like diagrams with nodes and edges as inspired from
supersymmetric quiver gauge theories (see subsection 3.1 for motivation).
This graphical representation allows to }$\left( i\right) $\textrm{\
interpret sub-blocks of the L-matrices in terms of topological adjoint and
bi-fundamental matter, }$\left( ii\right) $\textrm{\ forecast the form of
cumbersome Lax matrices without explicit calculation, }$\left( iii\right) $
link Levi decompositions of ADE Lie algebras to exceptional symmetry
breaking chains of a unified E$_{7}$ Chern-Simons theory. These results are
summarized in the conclusion section, see Figures \ref{TA}.(a-e), \ref{TD}%
.(a-d),\ and\ \ref{TE}.(a-d).\

The presentation is as follows: In section $2$, we begin by considering the
4D CS theory with $SL_{N}$ gauge symmetry as a reference model where we
describe in details the \textrm{implementation of} the electrically and
magnetically charged line defects and the calculation of their coupling in
the topological theory. We revisit the oscillator realisation of the A-type
minuscule Lax operators in the fundamental representation and then extend
the construction by discussing other cases where electric charges of the
Wilson lines correspond to representations of $sl_{N}$ beyond the
fundamental. In section $3$, we derive the topological gauge quiver diagrams
corresponding to the A-type L-operators calculated in section $2$, and give
an interpretation of their nodes and links in terms of topological matter.
Moreover, we yield quiver diagrams describing the form of L-operators for
the symmetric $\boldsymbol{N}\vee \boldsymbol{N}\vee \boldsymbol{N}$ and
adjoint representations of $sl_{N}$. In section $4$, we study the minuscule
D-type line defects in 4D CS theory with $SO_{2N}$ gauge invariance. Here,
we distinguish two sub-families given by the vector-like minuscule coweight,
and the two spinorial ones. Focussing on the vector-like family, we
calculate the corresponding L-operator and construct the associated
topological gauge quiver. In section $5$, we move on to the minuscule
spinor-like D-type L-operators where we also build the associated
topological quiver. Other aspects concerning fermionic lines and the link
with the $sl_{N}$ family are also discussed. In section $6$ and $7$, we
similarly treat the 4D CS theories with exceptional E$_{6}$ and E$_{7}$
gauge symmetries in order, we focus on the minuscule topological lines and
their associated topological quivers. The conclusion is devoted to a summary
of the results. The appendix section regards the derivation of a Lax matrix
from the corresponding topological gauge quiver in 4D CS.

\section{Wilson and 't Hooft lines of A- type}

In this section, we begin by focusing on the 4D Chern-Simons theory of
\textrm{\cite{21}} with $sl_{N}$ gauge symmetry where we introduce the
basics of this theory and the implementation of topological line defects. We
consider the various types of minuscule 't Hooft lines for the $sl_{N}$-
family with $N\geq 2$ and investigate their interaction with electric Wilson
lines. We show how the symplectic oscillators of the phase space of 't Hooft
lines allow for an explicit realisation of the Lax operators. We moreover
extend the results by considering Wilson lines for different representations
of $sl_{N}$ and investigating their properties according to the nature of
their electric charges.

\subsection{Electric/Magnetic lines in $sl_{N}$ Chern-Simons theory}

\begin{figure}[h]
\begin{center}
\includegraphics[width=14cm]{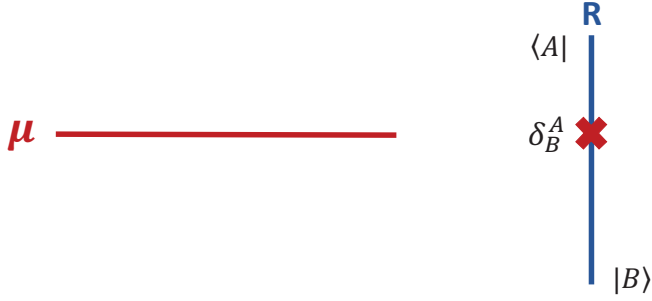}
\end{center}
\par
\vspace{-0.5cm}
\caption{Line defects in the real plane $\mathbb{R}^{2}$. On the left, a
horizontal 't Hooft line with magnetic charge $\protect\mu $ expanding along
the x-axis ($y=0$) at $z=0$. On the right, a vertical Wilson line expanding
along the y-axis ($x=0$) at $z\neq 0$\textrm{\ with electric charge in some
representation }$\boldsymbol{R}$. \textrm{Notice that the 't Hooft line is
in fact paired to a similar one located at }$z=\infty $\textrm{\ with
magnetic charge }$\mathrm{-}\protect\mu $\textrm{\ \protect\cite{34A}.}}
\label{twf}
\end{figure}

\begin{figure}[h]
\begin{center}
\includegraphics[width=12cm]{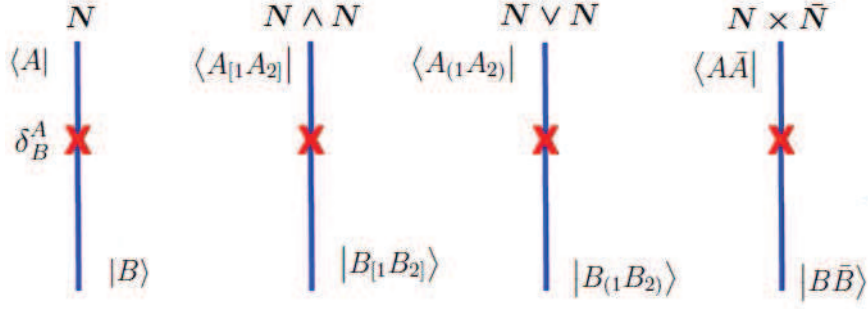}
\end{center}
\caption{Four examples of Wilson lines in different representations of $%
sl_{N}$\textrm{\ occupying vertical lines in }$\mathbb{R}^{2}$; they carry
different electric charges. The representation $\boldsymbol{R}$ and the type
of incoming quantum states \textrm{are} indicated above each line, while the
outgoing states are given at the bottom of the line. The red cross indicates
a local interaction point.}
\label{RR}
\end{figure}

\begin{figure}[h]
\begin{center}
\includegraphics[width=10cm]{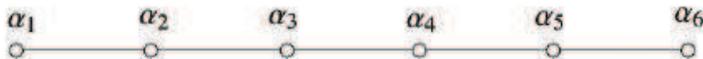}
\end{center}
\caption{The Dynkin diagram for the $sl_{N}$ family, it has $N-1$ simple
roots, all corresponding to minuscule coweights}
\label{A6}
\end{figure}

\begin{figure}[h]
\begin{center}
\includegraphics[width=14cm]{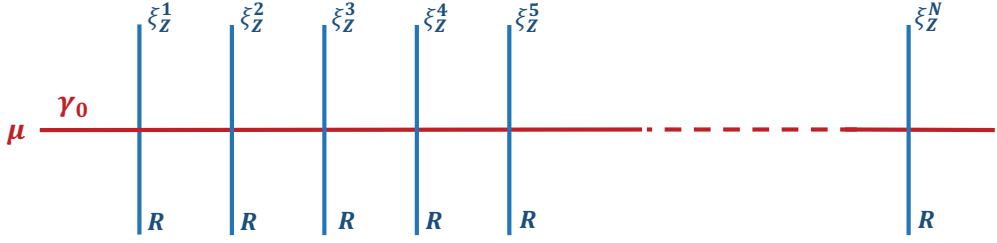}
\end{center}
\caption{The spin chain configuration in the Chern-Simons theory: N Wilson
lines represented by the blue vertical lines crossed by a 'tH$_{\protect%
\gamma }^{\protect\mu }$ represented by the red horizontal line.}
\label{Itw}
\end{figure}

In order to study the A- type electric Wilson lines and magnetic 't Hooft
line defects as well as their interpretation in quantum integrable systems,
we begin by briefly recalling some useful aspects of the 4D Chern-Simons
theory with $SL_{N}$ gauge symmetry over complex numbers. This is an
unconventional topological field theory living on a 4D space $\boldsymbol{M}%
_{4}$ that we take as $\mathbb{R}^{2}\times \mathbb{CP}^{1}$ parameterised
by $\left( x,y;z\right) $ with real $\left( x,y\right) $ for $\mathbb{R}^{2}$
and local complex $z=Z_{1}/Z_{2}$ \textrm{for }$C=\mathbb{CP}^{1}$. The
field action of the topological theory was first constructed in \textrm{\cite%
{Y1}} and reads as follows
\begin{equation}
S_{4dCS}=\int_{\mathbb{R}^{2}\times \mathbb{CP}^{1}}dz\wedge tr\Omega _{3},
\label{ac}
\end{equation}%
where $\Omega _{3}$\ is the CS 3-form%
\begin{equation}
\Omega _{3}=\mathcal{A}\wedge d\mathcal{A}+\frac{2}{3}\mathcal{A}\wedge
\mathcal{A}\wedge \mathcal{A},
\end{equation}%
with 1-form gauge potential $\mathcal{A}=t_{a}\mathcal{A}^{a}$ where $t_{a}$
stand for the generators of $sl_{N}$ and $\mathcal{A}^{a}$ is a partial
gauge connection as follows \textrm{\cite{22}}
\begin{equation}
\mathcal{A}^{a}=dx\mathcal{A}_{x}^{a}+dy\mathcal{A}_{y}^{a}+d\bar{z}\mathcal{%
A}_{\bar{z}}^{a}
\end{equation}%
The equation of motion of the potential field $\mathcal{A}$ is given by the
vanishing gauge curvature
\begin{equation}
\mathcal{F}_{2}=dz\wedge \left( d\mathcal{A}+\mathcal{A}\wedge \mathcal{A}%
\right) =0
\end{equation}%
This flat curvature indicates that the system is in the ground state with
zero energy. To deform this state, we consider observables given by line or
surface defects such as the Wilson $W_{\xi _{z}}^{\boldsymbol{R}}$ and 't
Hooft $tH_{\mathrm{\gamma }_{0}}^{\mu }$ lines that we are interested in
here. These are represented by curves in the topological plane $\mathbb{R}%
^{2}$ and located at positions z in $\mathbb{CP}^{1}$; they can be
represented as in the Figure \ref{twf}. \newline
\newline
Regarding the Wilson lines expanding along $\xi _{z}\subset \mathbb{R}^{2}$
with $z\in \mathbb{CP}^{1}$\textrm{, they} are semi-classical line defects,
electrically charged, defined as
\begin{equation}
W_{\xi _{z}}^{\boldsymbol{R}}=Tr_{\boldsymbol{R}}\left[ \text{P}\exp \left(
\int_{\xi _{z}}\mathcal{A}\right) \right]  \label{wr}
\end{equation}%
This shows that they are functions of $\xi _{z}$ and $\boldsymbol{R}$ which
is here a representation of $sl_{N}$ characterised by a highest weight state
$\left\vert \omega _{\boldsymbol{R}}\right\rangle $ with $\omega _{%
\boldsymbol{R}}=\sum_{i=1}^{N-1}n_{i}^{\boldsymbol{R}}\omega _{i}$. Notice
that at the quantum level, $\boldsymbol{R}$ is lifted to a representation of
the Yangian $Y(sl_{N})$ \ \cite{20,24AB,54A}. Here, to perform explicit
calculations, $\boldsymbol{R}$ is often taken as the (anti-) fundamental $%
\boldsymbol{N}$ representation of $sl_{N}$ with fundamental weight $\omega
_{1}$; however this construction can be extended at the classical level to
other $sl_{N}$ representations $n_{i}^{\boldsymbol{R}}\omega _{i}$ such as
the family of completely antisymmetric representations $\boldsymbol{N}%
^{\wedge k}\sim $ $\omega _{k}$, the family of completely symmetric $%
\boldsymbol{N}^{\vee n}\sim $ $n\omega _{1}$ and the adjoint representation $%
\boldsymbol{N}^{2}\boldsymbol{-1}$. As examples, Wilson lines with electric
weight charges in the representations $\boldsymbol{N}\wedge \boldsymbol{N}$
and $\boldsymbol{N}\vee \boldsymbol{N}$ as well as in the adjoint are
depicted in the Figure \ref{RR}. The interest into Wilson lines $W_{\xi
_{z}}^{\boldsymbol{R}}$ with generic $\boldsymbol{R}$ can be motivated by
the two following: \newline
$\left( \mathbf{1}\right) $ The special $sl_{N}$ representation theory where
from fundamental objects like $\boldsymbol{R}=\boldsymbol{N}$ and/or $%
\boldsymbol{\bar{N}}$ with weight $\omega _{N-1}$, one can construct many
composites carrying higher weight charges and describing higher conserved
quantities. For example, the particles' current running along $W_{\xi _{z}}^{%
\boldsymbol{R}}$ is given by quadratic composites transforming like $%
\boldsymbol{N}\otimes \boldsymbol{\bar{N}=1+adj}$. In this \textrm{regard},
notice that for the fundamental $W_{\xi _{z}}^{\boldsymbol{R}=\boldsymbol{N}%
},$ we have N quantum states $\left\vert A\right\rangle $ traveling along
the vertical blue line of the Figure \textbf{\ref{twf}}. They couple to the
CS gauge field like\textbf{\ }$\mathcal{J}_{a}\mathcal{A}^{a}$ with $%
\mathcal{J}_{a}\sim $\textbf{\ }$\left\langle A|t_{a}|B\right\rangle $.
\newline
$\left( \mathbf{2}\right) $ \textrm{Knowing the action of the minuscule
coweight }$\mu $\textrm{\ on the fundamental representation of }$sl_{N}$%
\textrm{, we can deduce its action on higher dimensional representations by
help of the tensor product properties. To fix ideas, see eq.(\ref{n2}).}%
\newline
\newline
Concerning the 't Hooft lines that we denote like tH$_{\mathrm{\gamma }%
_{0}}^{\mu },$ they are magnetically charged semi-classical line defects
with magnetic charge given by a minuscule coweight $\mu $ of the \textrm{%
complex Lie algebra }$sl_{N}$\textrm{. The curve} $\gamma _{0}$\textrm{\
belongs to }$\mathbb{R}^{2}$\textrm{\ and sits at a point }$z_{0}$\textrm{\
in the holomorphic plane\ that we take at the origin; it is imagined in the
4D CS theory }as the intersection $\mathcal{U}_{1}\cap \mathcal{U}_{2}$ of
two patches $\mathcal{U}_{1}$ and $\mathcal{U}_{2}$ of the topological plane
$\mathbb{R}^{2}$. Following \cite{34A},\ the topological field $\mathcal{A}^{%
\left[ \mu \right] }$ sourced by \textrm{the} magnetic 't Hooft line defect $%
\mathrm{\gamma }_{0}$ is generated by a \emph{singular} gauge transformation
$g=g\left( z\right) $ from the patch $\mathcal{U}_{1}$ to the patch $%
\mathcal{U}_{2}.$ By thinking of $\mathrm{\gamma }_{0}$ as coinciding with
the x-axis in the topological plane, meaning that
\begin{equation}
\mathrm{\gamma }_{0}=\mathbb{R}_{y\leq 0}^{2}\cap \mathbb{R}_{y\geq
0}^{2}\qquad ,\qquad \mathbb{R}^{2}=\mathbb{R}_{y\leq 0}^{2}\cup \mathbb{R}%
_{y\geq 0}^{2},
\end{equation}%
the gauge configuration $\mathcal{A}^{\left[ \mathbf{\mu }\right] }$ in the
presence of singularity $\mu $ is generated by a parallel transport of the
gauge field bundles from $\mathbb{R}_{y\leq 0}^{2}$ towards $\mathbb{R}%
_{y\geq 0}^{2}$. In this case, the transport path is then given by the
y-axis and the topological gauge configuration is given by
\begin{equation}
\mathcal{A}_{y}^{\left[ \mu \right] }=g_{_{1}}z^{\mathbf{\mu }}g_{_{2}}
\label{12}
\end{equation}%
with gauge \textrm{transformations} $g_{_{1}}\left( z\right) $ and $%
g_{_{2}}\left( z\right) $ singular near $z=0$ but regular in \textrm{the}
neighbourhood of $z=\infty $ with the limit $g_{_{1}}\left( \infty \right)
=g_{_{2}}\left( \infty \right) =I_{id}$. \textrm{Notice that }$z^{\mathbf{%
\mu }}$\textrm{\ is the operator }$exp(log(z)\mu )$\textrm{\ with }$\mu $%
\textrm{\ referring to the adjoint action of the coweight operating as in
eqs(\ref{act},\ref{215}).} Using this configuration, one can associate to
the $tH_{\mathrm{\gamma }_{0}}^{\mu }$ the following gauge invariant
observable measuring the parallel transport from $y\leq 0$ to $y\geq 0$ as
follows
\begin{equation}
L^{\left[ \mu \right] }\left( z\right) =\text{P}\exp \left( \int_{\mathrm{y}%
}dy\mathcal{A}_{y}^{\left[ \mu \right] }\left( z\right) \right)  \label{ml}
\end{equation}%
This $L^{\left[ \mu \right] }$ is a holomorphic function of z valued in the $%
SL_{N}$ gauge group; it may have poles and zeros at z= 0 and z=$\infty $
arising from the $tH_{\mathrm{\gamma }_{0}}^{\mu }$ at $z=0$ and \textrm{the}
mirror $tH_{\mathrm{\gamma }_{\infty }}^{-\mu }$ line at $z=\infty $ \cite%
{34A}. The gauge singularity is implemented in \textrm{this} construction by
thinking of $\mathcal{A}_{y}^{\left[ \mu \right] }$ as valued in the Levi
decomposition of $sl_{N}$ with respect to the minuscule coweight $\mu ,$
namely \cite{51}
\begin{equation}
\begin{tabular}{lll}
$sl_{N}$ & $\rightarrow $ & $\mathbf{n}_{-}\oplus \boldsymbol{l}_{\mu
}\oplus \mathbf{n}_{+}$ \\
$\mathcal{A}^{\left[ \mu \right] }$ & $\sim $ & $\mathcal{A}_{\mathbf{n}%
_{-}}+\mathcal{A}_{\boldsymbol{l}_{\mu }}+\mathcal{A}_{\mathbf{n}_{+}}$%
\end{tabular}
\label{acta}
\end{equation}%
\textrm{Notice that this decomposition is due to the fact that the minuscule
coweight }$\mu $\textrm{\ acts on the Lie algebra elements with only three
eigenvalues }$0;\pm 1$.\textrm{\ Therefore, a Lie algebra is decomposed to
three subspaces;} the $\boldsymbol{l}_{\mu }$ is a Levi subalgebra, and $%
\mathbf{n}_{\pm }$ are nilpotent subalgebras constrained as follows, with
Levi charge $q=\pm 1$:%
\begin{equation}
\left[ \mu ,\boldsymbol{l}_{\mu }\right] =0\qquad ,\qquad \left[ \mu ,%
\mathbf{n}_{q}\right] =q\mathbf{n}_{q}\qquad ,\qquad \left[ \mathbf{n}_{q},%
\mathbf{n}_{q}\right] =0  \label{act}
\end{equation}%
\textrm{In these regards, notice that for the case\ of} the topological $%
sl_{N}$ gauge theory,\textrm{\ we can define } $N-1$ minuscule 't Hooft
lines carrying different magnetic charges :
\begin{equation}
\begin{tabular}{lllll}
tH$_{\mathrm{\gamma }_{z_{1}}}^{\mu _{1}}$ & $,$ & $...$ & $,$ & tH$_{%
\mathrm{\gamma }_{z_{N-1}}}^{\mu _{N-1}}$%
\end{tabular}%
\end{equation}%
They are in 1:1 correspondence with the $N-1$ minuscule coweights $\mu
_{1},...,\mu _{N-1}$ of the $sl_{N}$ Lie algebra of the gauge symmetry
\textrm{(as listed in (\ref{fr}))}; and eventually with the $N-1$ simple
roots $\alpha _{1},...,\alpha _{N-1}$ of the Dynkin diagram of $sl_{N}$ as
depicted in Figure \textbf{\ref{A6}}. \newline
\newline
In what follows, we focus our attention on the XXX spin chain construction
in the framework of the 4D CS theory. As described in the figure \ref{Itw},
we need to take N vertical (parallel) Wilson lines $W_{\xi _{z}^{i}}^{%
\boldsymbol{R}}$\ in the topological plane $\mathbb{R}^{2}$ traversed by a
horizontal 't Hooft\ line tH$_{\mathrm{\gamma }_{0}}^{\mu }$\ (in red
color). The $W_{\xi _{z}^{i}}^{\boldsymbol{R}}$s\ sit\ at the position $%
z\neq 0$\ in the holomorphic plane while the tH$_{\mathrm{\gamma }_{0}}^{\mu
}$\ is in $z=0.$ From the integrable spin chain point of view, \textrm{every
Wilson line presents a node of the chain and the 't Hooft line is
interpreted as the Baxter Q-operator \cite{34A}. This way, we have a 't
Hooft-Wilson coupling in the topological plane at every node as depicted by
the Figure.}
\ The interaction by line-crossing is interesting as it allows to define the
Lax operator in every node of the spin chain which plays an important role
in the study of integrable systems\textbf{.} Because $W_{\xi _{z}}^{%
\boldsymbol{R}}$ is characterised by $\left( \xi _{z};\boldsymbol{R}\right) $
and tH$_{\mathrm{\gamma }_{0}}^{\mu }$ by $\left( \mathrm{\gamma }_{0};\mu
\right) $, the coupling between them should carry all this data and can be
defined as follows
\begin{equation}
L_{\boldsymbol{R}}^{\mu }\left( \mathrm{\gamma }_{0},\mathrm{\xi }%
_{z}\right) =\left\langle tH_{\mathrm{\gamma }_{0}}^{\mu },W_{\mathrm{\xi }%
_{z}}^{\boldsymbol{R}}\right\rangle  \label{lrm}
\end{equation}%
Following \textrm{\cite{34A}}, this L-operator, denoted from now on like $%
\mathcal{L}_{\boldsymbol{R}}^{\mu }$, is precisely given by (\ref{ml})%
\textrm{\ such that the transport path is identified with the Wilson line.}
Moreover, it can be \textrm{put} into a simpler form using the Levi-like
factorisation%
\begin{equation}
\mathcal{L}_{\boldsymbol{R}}^{\mu }\left( z\right) =e^{X_{\boldsymbol{R}}}z^{%
\mathbf{\mu }_{R}}e^{Y_{\boldsymbol{R}}}
\end{equation}%
where $X_{\boldsymbol{R}}$ is a nilpotent matrix valued in the nilpotent
algebra $\mathbf{n}_{+},$ and $Y_{\boldsymbol{R}}$ is also a nilpotent
matrix but valued in the nilpotent algebra $\mathbf{n}_{-}.$ These matrices
are constrained by the Levi decomposition requiring
\begin{equation}
\left[ \mathbf{\mu }_{R},X_{\boldsymbol{R}}\right] =+X_{\boldsymbol{R}%
}\qquad ,\qquad \left[ \mathbf{\mu }_{R},Y_{\boldsymbol{R}}\right] =-Y_{%
\boldsymbol{R}}  \label{215}
\end{equation}

\subsection{Interacting tH$_{\mathrm{\protect\gamma }_{0}}^{\protect\mu %
_{i}} $-$W_{\protect\xi _{z}}^{\boldsymbol{R}}$ lines in CS theory}

For the next step in the study of minuscule tH$_{\mathrm{\gamma }_{0}}^{\mu
_{i}}$ lines interacting with $W_{\xi _{z}}^{\boldsymbol{R}}$ in 4D CS
theory, it is interesting to explore the algebraic structure of the magnetic
charges $\mu _{i}$ of the tH$_{\mathrm{\gamma }_{0}}^{\mu _{i}}$'s. As these
charges are given by the minuscule coweights of the $sl_{N}$ Lie algebra, we
give below some useful tools regarding their properties and then turn to
study their coupling with $W_{\xi _{z}}^{\boldsymbol{R}}$.

\subsubsection{Minuscule coweights of $sl_{N}$}

First, we recall that there are $N-1$ fundamental coweights $\omega _{i}$
\textrm{for the} $sl_{N}$\textrm{\ Lie algebra, they are }defined as the
algebraic dual of the $N-1$ simple roots $\alpha _{i}$; \textrm{which means
that} $\omega _{i}.\alpha _{j}=\delta _{ij}.$ These simple roots of $sl_{N}$
are realised in terms of a weight basis vectors $\left \langle
e_{i}\right
\rangle $ like $\alpha _{i}=e_{i}-e_{i+1}.$ So, the fundamental
coweights solving $\omega _{i}.\alpha _{j}=\delta _{ij}$ read in terms of
the $e_{i}$'s as follows%
\begin{equation}
\omega _{i}=\frac{N-i}{N}\left( e_{1}+...+e_{i}\right) -\frac{i}{N}\left(
e_{i+1}+...+e_{N}\right)
\end{equation}%
It turns out that in the case of the $sl_{N}$ Lie algebra, the fundamental
coweights are all minuscule \textrm{\cite{51}}. So, the magnetic charges of
the $\left( N-1\right) $ lines tH$_{\mathrm{\gamma }_{0}}^{\mu _{i}}$ of the
A$_{N-1}$- CS theory are given by
\begin{equation}
\begin{tabular}{lll}
$\mu _{1}$ & $=$ & $\frac{N-1}{N}e_{1}-\frac{1}{N}\left(
e_{2}+...+e_{N}\right) $ \\
$\mu _{l}$ & $=$ & $\frac{N-l}{N}\left( e_{1}+...+e_{l}\right) -\frac{l}{N}%
\left( e_{l+1}+...+e_{N}\right) $ \\
$\mu _{N-1}$ & $=$ & $\frac{1}{N}\left( e_{1}+...+e_{N-1}\right) -\frac{N-1}{%
N}e_{N}$%
\end{tabular}
\label{fr}
\end{equation}%
with $2\leq l\leq N-2$. Obviously one can treat all these coweights
collectively; but it is interesting to cast them as we have done.\newline
As illustrating examples, we have for the $sl_{2}$ model, one minuscule
charge $\mu =\frac{1}{2}\left( e_{1}-e_{2}\right) .$ For the $sl_{3}$
theory, we have two minuscule coweights $\mu _{1}=\frac{2}{3}e_{1}-\frac{1}{3%
}\left( e_{2}+e_{3}\right) $ and $\mu _{2}=\frac{1}{3}\left(
e_{1}+e_{2}\right) -\frac{2}{3}e_{3}$; and for the $sl_{4}$ CS theory, we
have three minuscule charges given by%
\begin{equation}
\begin{tabular}{lll}
$\mu _{1}$ & $=$ & $\frac{3}{4}e_{1}-\frac{1}{4}\left(
e_{2}+e_{3}+e_{4}\right) $ \\
$\mu _{2}$ & $=$ & $\frac{1}{2}\left( e_{1}+e_{2}\right) -\frac{1}{2}\left(
e_{3}+e_{4}\right) $ \\
$\mu _{3}$ & $=$ & $\frac{1}{4}\left( e_{1}+e_{2}+e_{3}\right) -\frac{3}{4}%
e_{4}$%
\end{tabular}
\label{cw}
\end{equation}%
As far as the $sl_{4}$ example is concerned, notice that using the
isomorphism $sl_{4}\sim so_{6},$ these fundamental weights can be also
viewed as the fundamental of $so_{6}.$ Here, the $\mu _{2}$ corresponds to
the vector of $so_{6}$ while the $\mu _{1}$ and $\mu _{3}$ correspond to the
two Weyl spinors of orthogonal groups in even dimensions, they will be
encountered later when we study the L-operators of D-type.\newline
Notice also that given a minuscule coweight $\mu $ of $sl_{N}$, one defines
its adjoint form by help of the $e_{j}^{\ast }$'s obeying $e_{j}^{\ast
}\left( e_{i}\right) =\delta _{i}^{j}$. We denote the adjoint form of the
coweight $\mu _{l}$ by the bold symbol $\mathbf{\mu }_{l}$ and express it as
follows%
\begin{equation}
\mathbf{\mu }_{l}=\frac{N-l}{N}\Pi _{l}-\frac{l}{N}\bar{\Pi}_{l}  \label{pi}
\end{equation}%
with projector $\Pi _{l}$ and co-projector $\bar{\Pi}_{l}=I_{id}-\Pi _{l}$
as follows
\begin{equation}
\Pi _{l}=\sum_{i=1}^{l}e_{i}e_{i}^{\ast }\qquad ,\qquad \bar{\Pi}%
_{l}=\sum_{i=l+1}^{N}e_{i}e_{i}^{\ast }  \label{ip}
\end{equation}%
The use of this projector in the above decomposition is crucial in our
modeling; it is at the basis of our way to approach the coupling between the
$tH_{\mathrm{\gamma }_{0}}^{\mu }$ and $W_{\mathrm{\xi }_{z}}^{\boldsymbol{R}%
}$ as well as in the construction of the topological gauge quivers Q$_{%
\boldsymbol{R}}^{\mu }$ describing the A-type L-operators.

\subsubsection{the tH$_{\mathrm{\protect\gamma }_{0}}^{\protect\mu _{i}}$ - W%
$_{\protect\xi _{z}}^{\boldsymbol{R}}$ coupling}

To properly define the coupling between $W_{\xi _{z}}^{\boldsymbol{R}}$ and
a given minuscule 't Hooft line tH$_{\mathrm{\gamma }_{0}}^{\mu _{k}}$ with
a magnetic charge $\mu _{k}$ in the 4D Chern-Simons theory living in $%
\mathbb{R}^{2}\times \mathbb{CP}^{1}$, we follow \textrm{\cite{34A}} and
proceed as summarised below:

$\left( \mathbf{i}\right) $ tH$_{\mathrm{\gamma }_{0}}^{\mu _{k}}$\emph{\ as
a horizontal magnetic defect in }$\mathbb{R}^{2}$\newline
We think of the 't Hooft tH$_{\mathrm{\gamma }_{0}}^{\mu _{k}}$ as the curve
$\mathrm{\gamma }_{0}$ extending in the topological plane $\mathbb{R}^{2}$
of the 4D space. The defect $\mathrm{\gamma }_{0}$ is located at a given
point $z$ in $\mathbb{CP}^{1}$ that we take as $z=0$; say the south pole of $%
\mathbb{S}^{2}\sim \mathbb{CP}^{1}.$ For convenience, we think of $\mathrm{%
\gamma }_{0}$ as the horizontal line given by the x-axis of the plane $%
\mathbb{R}^{2}$ with $\left( x,y\right) $ coordinates; see the red line in
the Figure \textbf{\ref{Itw}}. Topologically speaking, this $\mathrm{\gamma }%
_{0}$ can be also imagined as the intersection of two patches like $\mathrm{%
\gamma }_{0}=\mathbb{R}_{y\leq 0}^{2}\cap \mathbb{R}_{y\geq 0}^{2}.$ Along
with this tH$_{\mathrm{\gamma }_{0}}^{\mu _{k}},$ we also have \textrm{a} tH$%
_{\mathrm{\gamma }_{\infty }}^{-\mu _{k}}$ sitting at $z=\infty $
corresponding to the north pole of $\mathbb{S}^{2}$.

$\left( \mathbf{ii}\right) $ \emph{crosses a vertical Wilson line}\newline
The horizontal tH$_{\mathrm{\gamma }_{0}}^{\mu _{k}}$ crosses a vertical
Wilson line $W_{\mathrm{\xi }_{z}}^{\boldsymbol{R}}$ with $\mathrm{\xi }_{z}$
located at a generic point $z$ of $\mathbb{CP}^{1}$. We imagine $\mathrm{\xi
}_{z}$ as coinciding with the y-axis in $\mathbb{R}^{2}$, i.e. $\mathrm{\xi }%
_{z}=\left \{ \left( x,y\right) |=x=0,y\in \mathbb{R}\right \} $. Recall
that the quantum states $\left \vert A\right \rangle $ propagating in the
electrically charged line $W_{\mathrm{\xi }_{z}}^{\boldsymbol{R}}$ are in
the representation $\boldsymbol{R}$ \textrm{which is taken for instance as
the fundamental} $\mathbf{N}$ of $sl_{N}$. The incoming particle states are
denoted by the bra $\left \langle A\right \vert $ and the outgoing states by
the ket $\left \vert B\right \rangle $ with
\begin{equation}
\left \langle A|B\right \rangle =\delta _{A}^{B}
\end{equation}%
in the case of free propagation. In the presence of interaction, the above $%
\delta _{A}^{B}$ is replaced by a multi-label vertex object.

\begin{figure}[h]
\begin{center}
\includegraphics[width=12cm]{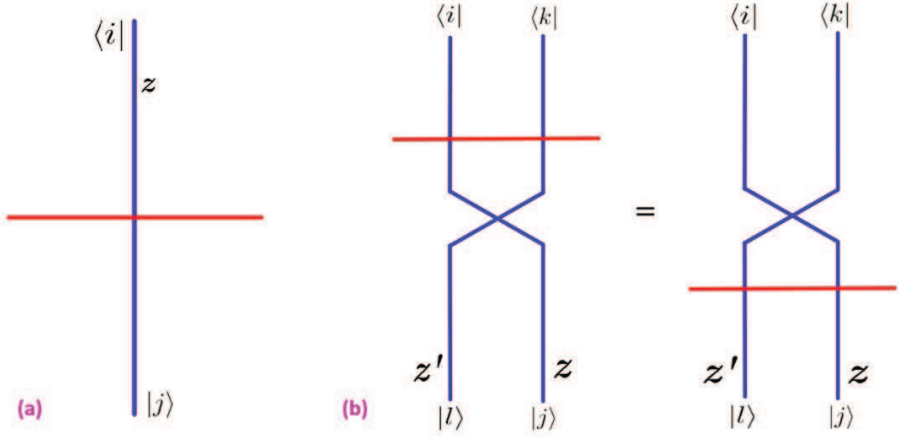}
\end{center}
\par
\vspace{-0.5cm}
\caption{$\left( \mathbf{a}\right) $ The operator $\mathcal{L}\left(
z\right) $ encoding the coupling between a 't Hooft line at z=0 (in red) and
a Wilson line at z (in blue) with incoming $\left \langle i\right \vert $
and out going $\left \vert j\right \rangle $ states. $\left( \mathbf{b}%
\right) $ RLL relations encoding the commutation relations between two
L-operators at z and z'.}
\label{2L}
\end{figure}

$\left( \mathbf{iii}\right) $ $\mathcal{L}$-\emph{operator and phase space}%
\newline
The crossing of the horizontal tH$_{\mathrm{\gamma }_{0}}^{\mu _{k}}$ and
the vertical $W_{\mathrm{\xi }_{z}}^{\boldsymbol{R}}$ lines\ is thought
\textrm{of} in terms of lines' coupling described by the $\mathcal{L}$%
-operator (\ref{lrm}) represented by the typical matrix operator
\begin{equation}
\left \langle A|\mathcal{L}_{\boldsymbol{R}}^{\left( \mu \right) }|B\right
\rangle =\mathcal{L}_{AB}^{\left( \mu \right) }
\end{equation}%
This operator is equivalent to the usual Lax operator of integrable spin
chain systems \textrm{\cite{15,FrA}}. It is a holomorphic function of $z$
and its representative matrix $\mathcal{L}_{AB}^{\left( \mu \right) }$ is
valued in the algebra $\mathfrak{A}$ of functions on the phase space of tH$_{%
\mathrm{\gamma }_{{\small z}}}^{\mu }$. Formally, we have
\begin{equation}
\mathcal{L}_{\boldsymbol{R}}^{\mathbf{\mu }}\in \mathfrak{A}\otimes
End\left( \boldsymbol{R}\right)
\end{equation}%
with $\mathfrak{A}$ generated by Darboux coordinates $\left( b,c\right) $ to
be commented later on; see eq.(\ref{234}). The phase space of the $%
L_{j}^{i}\left( z\right) $ operator is obtained by considering two coupled
vertical Wilson lines $W_{\mathrm{\xi }_{z}}^{\boldsymbol{R}}$ and $W_{%
\mathrm{\xi }_{z^{\prime }}}^{\boldsymbol{R}}$ crossed by a horizontal tH$_{%
\mathrm{\gamma }_{0}}^{\mu _{k}}$ as depicted by the Figure \textbf{\ref{2L}.%
} \newline
\newline
This topological invariant crossing describes integrability as encoded in
the following RLL relations
\begin{equation}
L_{j}^{r}\left( z\right) R_{rs}^{ik}\left( z-z^{\prime }\right)
L_{l}^{s}\left( z^{\prime }\right) =L_{r}^{i}\left( z\right)
R_{jl}^{rs}\left( z-z^{\prime }\right) L_{s}^{k}\left( z^{\prime }\right)
\label{rll}
\end{equation}%
In this equation, $R_{rs}^{ik}\left( z-z^{\prime }\right) $ is the well
known R-operator appearing in the Yang- Baxter equation, it is proportional
to the second Casimir $C_{rs}^{ik}$ of $sl_{N}$ having the value $\delta
_{r}^{i}\delta _{s}^{k}.$ For the trigonometric case corresponding to the
holomorphic line $\mathbb{CP}^{1}$, the structure of this R-matrix as a
series of $\hbar $ has leading terms like $R_{rs}^{ik}\left( z\right)
=\delta _{r}^{i}\delta _{s}^{k}+\frac{\hbar }{z}C_{rs}^{ik}+O\left( \hbar
^{2}\right) .$

\subsubsection{ Levi decomposition of $sl_{N}$}

The RLL relations of eq.(\ref{rll}) can be shown to be equivalent to the
usual Poisson bracket $\left \{ b^{\alpha },c_{\beta }\right \} _{PB}=\delta
_{\beta }^{\alpha }$ of symplectic geometry with $b^{\alpha }$ and $c_{\beta
}$ as phase space coordinates (Darboux coordinates). This equivalence
between the $L_{j}^{i}$ bracket (eq.(\ref{rll})) and the $\left \{ b^{\alpha
},c_{\beta }\right \} _{PB}$ follows from the Levi decompositions of $sl_{N}
$ that we describe here for different coweights of eq(\ref{cw}).

\begin{figure}[h]
\begin{center}
\includegraphics[width=12cm]{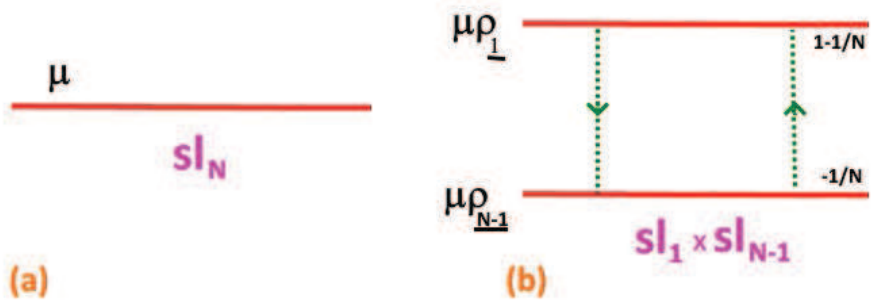}
\end{center}
\par
\vspace{-0.5cm}
\caption{(\textbf{a}) Magnetic 't Hooft with charge $\protect\mu _{1}$ from
the point of view of global$\ sl_{N}$ symmetry. (\textbf{b}) The same line
from the point of view of internal $sl_{1}\oplus sl_{N-1}$. here, the line $%
\protect\mu $ splits into two sub-lines $\mathbf{\protect\mu \protect\varrho
}_{1}$ and $\mathbf{\protect\mu \protect\varrho }_{N-1}$ as described in eq.(%
\protect\ref{ps}).}
\label{m}
\end{figure}

\textbf{1)} \emph{Minuscule coweight} $\mu _{1}$\newline
The Levi decomposition of $sl_{N}$ and its fundamental representation $%
\boldsymbol{N}$ with respect to the minuscule coweight $\mu _{1}$\ reads as
follows\textbf{\ }%
\begin{equation}
\begin{tabular}{lllll}
$\mu _{1}$ & $:$ & $sl_{N}$ & $\rightarrow $ & $sl_{1}\oplus sl_{N-1}\oplus
\mathbf{n}_{+}\oplus \mathbf{n}_{-}$ \\
&  & $\boldsymbol{N}$ & $\rightarrow $ & $\mathbf{1}_{\frac{N-1}{N}}\oplus
\left( \boldsymbol{N}\mathbf{-1}\right) _{-\frac{1}{N}}$%
\end{tabular}
\label{mu1}
\end{equation}%
with $\mathbf{n}_{\pm }=\left( N-1\right) _{\pm }$ and $sl_{1}$ refers to $%
\mathbb{C}.$ Because of this decomposition of $sl_{N},$ one can imagine the
Levi subalgebra as the manifest invariance in dealing with the study of the
tH$_{\mathrm{\gamma }_{0}}^{\mu _{1}}$ lines in the CS gauge theory with $%
sl_{N}$ gauge symmetry. In this view, we use the projectors $\varrho _{%
\mathbf{1}}$ and $\varrho _{\boldsymbol{N}\mathbf{-1}}$ of the irreducible
parts of the decomposition $\boldsymbol{N}=\mathbf{1}_{\frac{N-1}{N}}\oplus
\left( \boldsymbol{N}\mathbf{-1}\right) _{-\frac{1}{N}}$ as well as the
identity $\varrho _{\mathbf{1}}+\varrho _{\boldsymbol{N}\mathbf{-1}}=I_{id}$
to think of the adjoint form $\mathbf{\mu }_{1}$ of the minuscule coweight
as the sum of two contributions, one coming from $\mathbf{1}_{1-1/N}$ and
the other from $\left( \boldsymbol{N}\mathbf{-1}\right) _{-1/N}$ like%
\begin{equation}
\mathbf{\mu }_{1}=\mathbf{\mu }_{1}\varrho _{\underline{\mathbf{1}}}+\mathbf{%
\mu }_{1}\varrho _{\underline{\boldsymbol{N}\mathbf{-1}}}  \label{ps}
\end{equation}%
The projectors $\varrho _{\boldsymbol{R}}$ appearing in the above relation
are as in eqs.(\ref{pi}-\ref{ip}). In this picture the 't Hooft line of the $%
sl_{N}$ gauge symmetry gets splitted into two parallel "sub-lines" as
represented in the Figure \ref{m}. This is our first result regarding the
using the projector basis to understand the intrinsic properties of the
L-operator in the A-series. Clearly, the two 't Hooft "sub-lines" in the
Figure \textbf{\ref{m}}-(b) are coincident in the external space $\mathbb{R}%
^{2}\times \mathbb{CP}^{1}$ of the CS theory, but are lifted in the $sl_{N}$
internal space where the transitions between the two sub-lines are generated
by operators belonging to the \textrm{nilpotent} subalgebras $\mathbf{n}%
_{\pm }$. \newline
\newline
Moreover, the decomposition $\boldsymbol{N}\rightarrow \boldsymbol{1}%
_{1-1/N}\oplus \left( \boldsymbol{N-1}\right) _{-1/N}$ can be extended to
higher dimensional representations $\boldsymbol{R}$ of the $sl_{N}$ gauge
symmetry. This follows with the previous discussion concerning $W_{\mathrm{%
\xi }_{z}}^{\boldsymbol{R}}$ beyond the fundamental weight $\boldsymbol{N}$.
\textrm{For example}, the antisymmetric $\boldsymbol{N}\wedge \boldsymbol{N}$%
, the symmetric $\boldsymbol{N}\vee \boldsymbol{N}$ and the $\boldsymbol{adj}
$ representations of $sl_{N}$ decompose with respect to the minuscule
coweight $\mu _{1}$ as follows%
\begin{equation}
\begin{tabular}{||l||l||}
\hline\hline
\ $\ sl_{N}$ & \ \ \ \ \ \ \ \ \ $sl_{1}\oplus sl_{N-1}$ \\ \hline\hline
$\boldsymbol{N}\wedge \boldsymbol{N}$ & $\boldsymbol{F}_{1-\frac{2}{N}%
}\oplus \left( \boldsymbol{F}\wedge \boldsymbol{F}\right) _{-\frac{2}{N}}$
\\ \hline
$\boldsymbol{N}\vee \boldsymbol{N}$ & $\boldsymbol{1}_{2-\frac{2}{N}}\oplus
\boldsymbol{F}_{1-\frac{2}{N}}\oplus \left( \boldsymbol{F}\vee \boldsymbol{F}%
\right) _{-\frac{2}{N}}$ \\ \hline
$\boldsymbol{N}\times \boldsymbol{\bar{N}}$ & $\boldsymbol{1}_{1-\frac{1}{N}%
}\otimes \mathbf{1}_{\frac{1}{N}-1}+\boldsymbol{1}_{1-\frac{1}{N}}\otimes
\boldsymbol{F}_{\frac{1}{N}}+\boldsymbol{F}_{-\frac{1}{N}}\otimes \mathbf{1}%
_{\frac{1}{N}-1}+\boldsymbol{F}_{-\frac{1}{N}}\otimes \boldsymbol{F}_{\frac{1%
}{N}}$ \\ \hline
$\boldsymbol{adj}\left( sl_{N}\right) $ & $\boldsymbol{F}_{-1}\oplus \left[
\boldsymbol{1}_{0}\oplus \boldsymbol{adj}\left( sl_{N-1}\right) _{0}\right]
\oplus \boldsymbol{F}_{+1}$ \\ \hline\hline
\end{tabular}
\label{n2}
\end{equation}%
where $\boldsymbol{F=N-1}$. Notice also that compared to $\boldsymbol{N}%
\rightarrow \boldsymbol{1}_{1-1/N}\oplus \left( \boldsymbol{N-1}\right)
_{-1/N}$, the symmetric $\boldsymbol{N}\vee \boldsymbol{N}$ reduces to three
$sl_{1}\oplus sl_{N-1}$ representations namely $\boldsymbol{1}_{2-\frac{2}{N}%
}$ and $\boldsymbol{F}_{1-\frac{2}{N}}$ as well as $\left( \boldsymbol{F}%
\vee \boldsymbol{F}\right) _{-\frac{2}{N}}$; the same holds for $\boldsymbol{%
adj}\left( sl_{N}\right) $. This feature is interesting as it indicates that
the \textrm{corresponding }Lax operators $\mathcal{L}_{\boldsymbol{N}\vee
\boldsymbol{N}}$ and $\mathcal{L}_{adj\left( sl_{N}\right) }$ have a richer
intrinsic structure compared to $\mathcal{L}_{\boldsymbol{N}}$, see
subsection 2.3.

\textbf{2)} \emph{Minuscule coweights} $\mu _{k}$ for $2\leq k\leq N-2.$%
\newline
Levi decomposition of $sl_{N}$ and its fundamental representation with
respect to $\mu _{k}$ leads to%
\begin{equation}
\begin{tabular}{lllll}
$\mu _{k}$ & $:$ & $sl_{N}$ & $\rightarrow $ & $sl_{k}\oplus sl_{N-k}\oplus
sl_{1}\oplus k\left( N-k\right) _{+}\oplus k\left( N-k\right) _{-}$ \\
&  & $\boldsymbol{N}$ & $\rightarrow $ & $\boldsymbol{k}_{\frac{N-k}{N}%
}\oplus \left( \boldsymbol{N-k}\right) _{-\frac{k}{N}}$%
\end{tabular}
\label{app}
\end{equation}%
where the Levi subalgebra is $sl_{k}\oplus sl_{N-k}\oplus sl_{1}$ and the
\textrm{nilpotent}s are $k\left( N-k\right) _{\pm }$. For the example of $%
sl_{4}$ with $k=2,$ we have%
\begin{equation}
\begin{tabular}{lllll}
$\mu _{2}$ & $:$ & $sl_{4}$ & $\rightarrow $ & $sl_{2}\oplus sl_{2}\oplus
sl_{1}\oplus 4_{+}\oplus 4_{-}$ \\
&  & $4$ & $\rightarrow $ & $2_{+\frac{1}{2}}\oplus 2_{-\frac{1}{2}}$%
\end{tabular}%
\end{equation}%
Notice that for this case as well, we have a splitting picture as in the
Figure \textbf{\ref{m}}-(b) where eq.(\ref{ps}) should be replaced by%
\begin{equation}
\mathbf{\mu }_{k}=\mathbf{\mu }_{k}\varrho _{\underline{\mathbf{k}}}+\mathbf{%
\mu }_{k}\varrho _{\underline{\mathbf{N-k}}}
\end{equation}

\textbf{3}) \emph{Minuscule coweight} $\mu _{N-1}$ \newline
The Levi- decomposition of $sl_{N}$ with respect to $\mu _{N-1}$ reads as
follows%
\begin{equation}
\begin{tabular}{lllll}
$\mu _{N-1}$ & $:$ & $sl_{N}$ & $\rightarrow $ & $sl_{N-1}\oplus
sl_{1}\oplus F_{+}\oplus F_{-}$ \\
&  & $N$ & $\rightarrow $ & $1_{\frac{N-1}{N}}\oplus \left( N-1\right) _{-%
\frac{1}{N}}$%
\end{tabular}%
\end{equation}%
It has a similar structure to eq.(\ref{mu1}), so we can omit the details
regarding this $\mu _{N-1}$ case; it can also be recovered from the generic $%
\mu _{k}$ with $k=N-1$.

\subsection{the $\mathcal{L}$-operators in $sl_{N}$ theory}

\begin{figure}[h]
\begin{center}
\includegraphics[width=10cm]{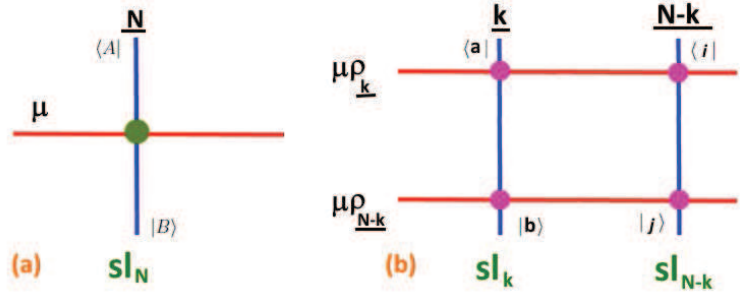}
\end{center}
\par
\vspace{-0.5cm}
\caption{(a) A horizontal minuscule 't Hooft line with magnetic charge $%
\protect\mu _{k}$ crossing a vertical Wilson line with electric charge $%
\boldsymbol{R}=\mathbf{N}$. The green dot describes the coupling given by
the Lax operator $\left \langle A|L_{\boldsymbol{R}}^{\mathbf{\protect\mu }%
_{k}}|B\right \rangle $. (b) Intrinsic structure of the Lax operator taking
into account the Levi decomposition of $sl_{N}$ with respect to $\protect\mu %
_{k}.$ }
\label{1N}
\end{figure}

The expression of the $\mathcal{L}^{\mathbf{\mu }_{k}}$-operator in terms of
the adjoint form of the minuscule coweight $\mathbf{\mu }_{k}$ and the
Darboux coordinates $b^{a}$ and $c_{a}$ is given by
\begin{equation}
\mathcal{L}^{\mathbf{\mu }_{k}}\left( z\right) =e^{X}z^{\mathbf{\mu }%
_{k}}e^{Y}  \label{xy}
\end{equation}%
with
\begin{equation}
X=\sum_{a=1}^{k\left( N-k\right) }b^{a}X_{a}\qquad ,\qquad
Y=\sum_{a=1}^{k\left( N-k\right) }c_{a}Y^{a}  \label{234}
\end{equation}%
In eq(\ref{xy}), the minuscule coweight acts \textrm{like}
\begin{equation}
\left[ \mathbf{\mu }_{k},X_{a}\right] =+X_{a}\qquad ,\qquad \left[ \mathbf{%
\mu }_{k},Y^{a}\right] =-Y^{a}  \label{mxy}
\end{equation}%
with the adjoint action $\mathbf{\mu }_{k}=\mu _{k}^{i}\varrho _{i}$ where $%
\varrho _{i}=\left \vert i\right \rangle \left \langle i\right \vert $ and
where the $\mu _{k}^{i}$'s are fractions of the unity given by (\ref{fr}).
See also the Figure \textbf{\ref{1N}}-(a,b) representing our vision
regarding the topology of the L-operators of A-type series. For the
expressions of the generators $X_{a}$ and $Y^{a}$ solving the constraints of
eq(\ref{mxy}), they are constructed below depending on the value of the
level $k$.

\subsubsection{'t Hooft line with magnetic charge $\mathbf{\protect\mu }_{1}$%
}

In the case of a 't Hooft line with a magnetic charge $\mu _{1}$ crossing a
Wilson line $W_{\mathrm{\xi }_{z}}^{\boldsymbol{R}=N}$ of $sl_{N}$, we have $%
N-1$ generators $X_{a}$ and $N-1$ generators $Y^{a}$ in the fundamental
representation. These are $N\times N$ triangular matrices solving eq.(\ref%
{mxy}) and\textrm{\ given} by
\begin{equation}
\begin{tabular}{lll}
$X_{a}$ & $=$ & $\left \vert 1\right \rangle \left \langle a+1\right \vert $
\\
$Y^{a}$ & $=$ & $\left \vert a+1\right \rangle \left \langle 1\right \vert $
\\
$\mathbf{\mu }_{1}$ & $=$ & $\frac{N-1}{N}\varrho _{1}-\frac{1}{N}\left(
\varrho _{2}+...+\varrho _{N}\right) $%
\end{tabular}
\label{rn}
\end{equation}%
where we have set $\varrho _{i}=\left \vert i\right \rangle \left \langle
i\right \vert $ with $\sum_{i=1}^{N}\varrho _{i}=I_{N\times N}$.\ Moreover,
by taking $\varrho _{\bar{1}}=\varrho _{2}+...+\varrho _{N}$ with $\varrho
_{1}+\varrho _{\bar{1}}=I,$ the adjoint form $\mathbf{\mu }_{1}$ can be
written in the following form
\begin{equation}
\mathbf{\mu }_{1}=\frac{N-1}{N}\varrho _{1}-\frac{1}{N}\varrho _{\bar{1}}
\label{mm1}
\end{equation}%
These projectors play an important role in the study of the L-operator of
the 4D CS theory with $SL_{N}$ gauge invariance: $\left( \mathbf{1}\right) $
They single out the Levi charges of the two internal subspaces in the Levi
decomposition $\boldsymbol{N}=\boldsymbol{1}_{1-1/N}\oplus \left(
\boldsymbol{N-1}\right) _{-1/N}$. For example, by multiplying eq(\ref{mm1})
first by $\varrho _{1}$ and then by $\varrho _{\bar{1}}$, we obtain
\begin{equation}
\mathbf{\mu }_{1}\varrho _{1}=\frac{N-1}{N}\varrho _{1}\qquad ,\qquad
\mathbf{\mu }_{1}\varrho _{\bar{1}}=-\frac{1}{N}\varrho _{\bar{1}}
\end{equation}%
which describe the two horizontal sub-lines in the Figure \textbf{\ref{m}}%
-(b). $\left( \mathbf{2}\right) $ They allow to write interesting properties
verified by the realisation (\ref{rn}) such as
\begin{equation}
\begin{tabular}{lllllll}
$X_{a}\varrho _{1}$ & $=$ & $0$ & $\qquad ,\qquad $ & $\varrho _{1}Y^{a}$ & $%
=$ & $0$ \\
$\varrho _{\bar{1}}X_{a}$ & $=$ & $0$ & $\qquad ,\qquad $ & $Y^{a}\varrho _{%
\bar{1}}$ & $=$ & $0$%
\end{tabular}%
\end{equation}%
indicating that $\mathcal{L}_{\boldsymbol{R=N}}^{\mathbf{\mu }_{1}}$ can be
presented as a matrix with sub-blocks given in terms of the projectors $%
\varrho _{1}$ and $\varrho _{\bar{1}}$. \newline
We can check the relations (\ref{mxy}) by computing the quantities $\mathbf{%
\mu }_{1}X_{a}$ and $X_{a}\mathbf{\mu }_{1}$ using the above realisation, we
have%
\begin{equation}
\mathbf{\mu }_{1}X_{a}=\frac{N-1}{N}X_{a}\qquad ,\qquad X_{a}\mathbf{\mu }%
_{1}=-\frac{1}{N}X_{a}
\end{equation}%
thus giving $\left[ \mathbf{\mu }_{1},X_{a}\right] =X_{a}$; the same can be
done for the generators $Y^{a}$. \newline
Now, in order to explicitly calculate the L-operator, we need to evaluate
the exponentials $e^{X}$ and $e^{Y}$ such that $X$ and $Y$ are given by%
\begin{equation}
X=\sum_{a=1}^{N-1}b^{a}\left \vert 1\right \rangle \left \langle a+1\right
\vert \qquad ,\qquad Y=\sum_{a=1}^{N-1}c_{a}\left \vert a+1\right \rangle
\left \langle 1\right \vert
\end{equation}%
These matrices obey the property $X^{2}=Y^{2}=0$, so we have $e^{X}=I+X$ and
$e^{Y}=I+Y$, consequently%
\begin{equation}
\begin{tabular}{lll}
$\mathcal{L}_{\boldsymbol{N}}^{\mathbf{\mu }_{1}}\left( z\right) $ & $=$ & $%
\left( I+X\right) z^{\mathbf{\mu }_{1}}\left( I+Y\right) $ \\
& $=$ & $z^{\mathbf{\mu }_{1}}I+Xz^{\mathbf{\mu }_{1}}+z^{\mathbf{\mu }%
_{1}}Y+Xz^{\mathbf{\mu }_{1}}Y$%
\end{tabular}%
\end{equation}%
Using $\mathbf{\mu }_{1}=\frac{N-1}{N}\varrho _{1}-\frac{1}{N}\varrho _{\bar{%
1}}$ with $\varrho _{\bar{1}}=I-\varrho _{1}$ and $z^{\mathbf{\mu }%
_{1}}=z^{\mu _{k}^{i}}\varrho _{i},$ we can express the L-operator in terms
of projectors as follows%
\begin{equation}
\begin{tabular}{lll}
$\mathcal{L}_{\boldsymbol{N}}^{\mathbf{\mu }_{1}}\left( z\right) $ & $=$ & $%
z^{\frac{N-1}{N}}\varrho _{1}+z^{-\frac{1}{N}}X\varrho _{\bar{1}}Y+$ \\
&  & $z^{-\frac{1}{N}}\left( X\varrho _{\bar{1}}+\varrho _{\bar{1}}Y\right)
+z^{-\frac{1}{N}}\varrho _{\bar{1}}$%
\end{tabular}
\label{m1}
\end{equation}%
\textrm{This form of the Lax operator is a result of the projectors basis
that we choose above}, this unique writing is particularly significant for
the quiver description of the L-operator as well as for the straightforward
extension to other electric charges of the 4D CS gauge theory with $sl_{N}$
gauge symmetry.

\subsubsection{Magnetic charge $\mathbf{\protect\mu }_{k}$ with $2\leq k\leq
N-2$}

In this generic case, we have $k\left( N-k\right) $ generators $X_{\alpha }$
and $k\left( N-k\right) $ generators $Y^{\alpha }$ generating the nilpotent $%
\boldsymbol{k}\left( \boldsymbol{N-k}\right) _{+}$ and $\boldsymbol{k}\left(
\boldsymbol{N-k}\right) _{-}$ of the Levi decomposition of $sl_{N}$ with
respect to the minuscule coweight $\mathbf{\mu }_{k}.$ In fact, the $%
X_{i\alpha }$ and the $Y^{i\alpha }$ of $\mathbf{n}_{\pm }$ are $N\times N$
triangular matrices realised as follows%
\begin{equation}
\begin{tabular}{lllll}
$X_{i\alpha }$ & $=$ & $\left \vert i\right \rangle \left \langle k+\alpha
\right \vert $ & $,$ & $1\leq i\leq k$ \\
$Y^{i\alpha }$ & $=$ & $\left \vert k+\alpha \right \rangle \left \langle
i\right \vert $ & $,$ & $1\leq \alpha \leq N-k$%
\end{tabular}
\label{xia}
\end{equation}%
and the $\mathbf{\mu }_{k}$ is given by%
\begin{equation}
\mathbf{\mu }_{k}=\frac{N-k}{N}\Pi _{k}-\frac{k}{N}\Pi _{\bar{k}}
\end{equation}%
with%
\begin{equation}
\Pi _{k}=\sum \limits_{l=1}^{k}\varrho _{l}\qquad ,\qquad \Pi _{\bar{k}%
}=\sum \limits_{l=k+1}^{N}\varrho _{l}
\end{equation}%
The generators (\ref{xia}) satisfy the Levi decomposition conditions that
read as%
\begin{equation}
\begin{tabular}{lllll}
$\left[ \mathbf{\mu }_{k},X_{ia}\right] $ & $=$ & $\left( \frac{N-l}{N}+%
\frac{l}{N}\right) X_{ia}$ & $=$ & $X_{ia}$ \\
$\left[ \mathbf{\mu }_{k},Y^{ia}\right] $ & $=$ & $\left( -\frac{l}{N}-\frac{%
N-l}{N}\right) Y^{ia}$ & $=$ & $-Y^{ia}$%
\end{tabular}%
\end{equation}%
\newline
This interesting realisation also obeys%
\begin{equation}
X_{a}\Pi _{k}=0\qquad ,\qquad \Pi _{k}Y^{a}=0
\end{equation}%
which indicates the sub-blocks of the matrix $\mathcal{L}_{\boldsymbol{N}}^{%
\mathbf{\mu }_{k}}$. The commutators $\left[ X_{ia},Y^{ia}\right] $ give the
Cartan generators reading as $H_{ia}=\varrho _{i}-\varrho _{a}$ while the
nilpotency $X_{i\alpha }X_{j\beta }=Y^{i\alpha }Y^{j\beta }=0$ leads to $%
e^{X}=I+X$ and $e^{Y}=1+Y.$ Using these features, we obtain%
\begin{equation}
\begin{tabular}{lll}
$\mathcal{L}_{\boldsymbol{N}}^{\mathbf{\mu }_{k}}$ & $=$ & $\left(
I+X\right) z^{\mathbf{\mu }_{k}}\left( I+Y\right) $ \\
& $=$ & $z^{\mathbf{\mu }_{k}}I+Xz^{\mathbf{\mu }_{k}}+z^{\mathbf{\mu }%
_{k}}Y+Xz^{\mathbf{\mu }_{k}}Y$%
\end{tabular}%
\end{equation}%
Moreover, using%
\begin{equation}
\mathbf{\mu }_{k}=\frac{N-k}{N}\Pi _{k}-\frac{k}{N}\Pi _{\bar{k}}
\end{equation}%
with $\Pi _{\bar{k}}=I-\Pi _{k}$ and $z^{\mathbf{\mu }_{1}}=z^{\mu
_{k}^{k}}\Pi _{k}+z^{\mu _{k}^{\bar{k}}}\Pi _{\bar{k}},$ we can express the
operator in terms of the projectors as follows$.$ $\ $%
\begin{equation}
\begin{tabular}{lll}
$\mathcal{L}_{\boldsymbol{N}}^{\mathbf{\mu }_{k}}\left( z\right) $ & $=$ & $%
z^{\frac{N-k}{N}}\Pi _{k}+z^{-\frac{k}{N}}\Pi _{\bar{k}}+$ \\
&  & $z^{-\frac{k}{N}}X\Pi _{\bar{k}}+z^{-\frac{k}{N}}\Pi _{\bar{k}}Y+z^{-%
\frac{k}{N}}X\Pi _{\bar{k}}Y$%
\end{tabular}
\label{mk}
\end{equation}%
This is the generic form of the L-operator in \textrm{the} 4D Chern-Simons
gauge theory with $sl_{N}$\ gauge invariance.

\subsubsection{Magnetic charge $\mathbf{\protect\mu }_{N-1}$}

In this case, the $N-1$ generators $X_{i}$ and $N-1$ generators $Y^{a}$ are
given by
\begin{equation}
X_{a}=\left \vert 1+a\right \rangle \left \langle N\right \vert \qquad
,\qquad Y^{a}=\left \vert N\right \rangle \left \langle 1+a\right \vert
\end{equation}%
with $1\leq a\leq N-1$ and%
\begin{equation}
\mathbf{\mu }_{N-1}=\frac{1}{N}\varrho _{\overline{N}}-\frac{N-1}{N}\varrho
_{N}
\end{equation}%
The Lax operator reads as%
\begin{equation}
\mathcal{L}_{\boldsymbol{N}}^{\mathbf{\mu }_{N-1}}\left( z\right) =z^{\frac{1%
}{N}}\varrho _{\overline{N}}+z^{\frac{1-N}{N}}\varrho _{N}+z^{\frac{1-N}{N}%
}X\varrho _{N}+z^{\frac{1-N}{N}}\varrho _{N}Y+z^{\frac{1-N}{N}}X\varrho _{N}Y
\end{equation}%
which corresponds to setting $k=N-1$ in eq(\ref{mk}).

\section{Topological gauge quivers: A- family}

In this section, we want to construct quiver gauge diagrams corresponding to
the topological L-operators in 4D Chern Simons theory with A-type gauge
symmetry. This graphical description was first proposed in \textrm{\cite{54}}
\textrm{for the} case of exceptional gauge symmetries E$_{6,7}$, and it will
be extended here for the ADE series. First, we begin by defining these
quivers and explaining the procedure of their construction; then we
illustrate this for the topological quivers Q$_{\boldsymbol{N}}^{\mathbf{\mu
}_{k}}$ corresponding to L-operators $\mathcal{L}_{\boldsymbol{N}}^{\mathbf{%
\mu }_{k}}$ of $sl_{N}$ -type with $\mathbf{\mu }_{k}$, $1\leq k\leq N$ and $%
\boldsymbol{R}=\boldsymbol{N}$. This leading model is exploited to build
other quiver diagrams Q$_{\boldsymbol{R}}^{\mathbf{\mu }_{k}}$ in 4D CS with
A-type gauge symmetry; these correspond\ to L-operators in representations $%
\boldsymbol{R}$ beyond the fundamental $\boldsymbol{N}$ of $sl_{N}$ and are
collectively given in the Figure \ref{TA} at the conclusion section.

\subsection{Motivating the topological quivers Q$_{\boldsymbol{R}}^{\mathbf{%
\protect\mu }}$}

The quiver diagrams Q$_{\boldsymbol{R}}^{\mathbf{\mu }}$ that we introduce
here in the framework of 4D Chern Simons theory give a unified graphical
representation of the data carried by the L-operators $\mathcal{L}_{%
\boldsymbol{R}}^{\mathbf{\mu }}$. We refer to these graphs as topological
gauge quivers, first because they have a formal similarity with quiver
diagrams Q$_{gauge}^{susy}$ in supersymmetric quiver gauge theories that we
briefly recall here below; and second because the L-operators they
illustrate match topological 't Hooft line defects in the 4D CS \cite{34A}.

$\bullet $ \emph{Gauge quivers in supersymmetric theory}\newline
For a supersymmetric quiver gauge theory with unitary gauge symmetry $G$
factorised as
\begin{equation}
G=\prod \limits_{i=1}^{n_{0}}U\left( M_{i}\right)  \label{GG}
\end{equation}%
and Lie algebra $\boldsymbol{g}=\oplus _{i=1}^{n_{0}}u\left( M_{i}\right) $,
and where the gauge symmetry factors are imagined in type II strings as
stacks of $M_{i}$ coincident D branes wrapping cycles in Calabi-Yau
compactifications \textrm{\cite{55}}, we have a gauge quiver denoted as Q$%
_{gauge}^{susy}$. This diagram has: $\left( \mathbf{i}\right) $ n$_{0}$
nodes $\mathcal{N}_{1},...,\mathcal{N}_{n_{0}}$ corresponding to the gauge
group factors $G_{1},...,G_{n_{0}}$ describing \textquotedblright adjoint
matter\textquotedblright \ in the gauge theory transforming in the adjoint
representations%
\begin{equation}
adjU\left( M_{i}\right) =\boldsymbol{M}_{i}\times \boldsymbol{\bar{M}}_{i}
\label{adj}
\end{equation}%
$\left( \mathbf{ii}\right) $ a number $n_{link}$ of links $L_{ij}$ between
the nodes $\left( \mathcal{N}_{i},\mathcal{N}_{j}\right) $ describing
bi-fundamental matter transforming in the representations%
\begin{equation}
\boldsymbol{M}_{i}\times \boldsymbol{\bar{M}}_{j}\in U\left( M_{i}\right)
\times U\left( M_{j}\right)
\end{equation}

$\bullet $ \emph{Topological gauge quivers in 4D CS}\newline
Based on general aspects of supersymmetric quivers, we introduce our
topological gauge quiver diagrams Q$_{\boldsymbol{R}}^{\mathbf{\mu }}$
describing the L-operators in 4D Chern Simons theory by focusing in this
section on the A-type symmetry. These have similar features with Q$%
_{gauge}^{susy}$ that allow to interpret the phase space coordinates $%
b^{\alpha }$ and $c_{\alpha }$ in terms of topological variables and
bi-fundamental matter. As for the L-operator, a topological quiver Q$_{%
\boldsymbol{R}}^{\mu }$ is defined for a general gauge group $G,$ by the
choice of a minuscule coweight $\mu $ and a representation $\boldsymbol{R}$
of $g.$\ \textrm{Notice here that only representations that lift to the
Yangian lead to quantum L-operators in the 4D CS, otherwise the obtained
L-operators are to be interpreted semi-classically}.\newline
However, the construction presented here is valid for any representation $%
\boldsymbol{R}$, where the minuscule $\mu $ that decomposes the Lie algebra $%
g$ into a Levi subalgebra $l_{\mu },$ and two nilpotents $n_{\pm }$ as in $%
\left( \ref{acta},\ref{act}\right) ,$ splits the $\boldsymbol{R}$ into $p$
irrepresentations $\boldsymbol{R}_{i}$\ having charges $m_{i}$\ with respect
to the $SO(2)$ of $\mu .$ We write%
\begin{equation}
\boldsymbol{R}=\sum\limits_{i=1}^{p}\boldsymbol{R}_{i},\qquad \mu
=\sum\limits_{i=1}^{p}m_{i}\Pi _{i},\qquad \sum\limits_{i}\Pi _{i}=\mathbf{1}%
_{\boldsymbol{R}}
\end{equation}%
with $\Pi _{i}$ being the projector on the subspace $\boldsymbol{R}_{i}.$ To
such data, we associate a topological gauge quiver Q$_{\boldsymbol{R}}^{\mu
} $ having $p$ nodes, each one given by the couple $\left( \boldsymbol{R}%
_{i},m_{i}\right) $ such that the charge is noted as a subscript of the
irrepresentation. These nodes are ordered such that $m_{i}-m_{i+1}=\pm 1$,
and we have for two nodes $\mathcal{N}_{i}$ and $\mathcal{N}_{i}$, $%
m_{i}-m_{j}=\pm k$\ where $k=1,...,p-1$ is an integer. This property comes
from the branching rules \cite{61},\ and is to be observed from the
different cases studied in the present paper. In the $(dim\boldsymbol{R}%
\times dim\boldsymbol{R})$ matrix representation of the corresponding $%
\mathcal{L}_{\boldsymbol{R}}^{\mathbf{\mu }}$, we have $p$ diagonal
sub-blocks in one to one with the nodes $\mathcal{N}_{i}=\Pi _{i}\mathcal{L}%
\Pi _{i}$ of Q$_{\boldsymbol{R}}^{\mu }.$\newline
The links connecting different nodes of the quiver are therefore given by
off-diagonal blocks $L_{ij}=\Pi _{i}\mathcal{L}\Pi _{j}$ that indeed allow
to transit between the $m_{i}$'s. These carry charges $\pm k$\ because they
contain polynomials of the form $\mathbf{c}^{k+l}\mathbf{b}^{l}$ and $%
\mathbf{b}^{k+l}\mathbf{c}^{l}$ with $l=0,...,p-2.$ Here $\mathbf{b}%
=b^{\alpha }$ and $\mathbf{c}=c_{\alpha }$ are the oscillator vector and
co-vector of dimensions $n_{+}=n_{-};$ they carry charges $\mp 1$ as noticed
from%
\begin{equation}
e^{X}=e^{b_{\left( -1\right) }^{\alpha }X_{\alpha \left( +1\right) }},\qquad
e^{Y}=e^{c_{\alpha \left( +1\right) }Y_{\left( -1\right) }^{\alpha }}
\end{equation}%
For simplicity, the link $L_{i\rightarrow j}$ from $N_{i}$ to $N_{j}$ with $%
\left\vert m_{j}-m_{i}\right\vert =k$ is indexed in the quiver by $\mathbf{c}%
^{k};$\ and similarly $L_{j\rightarrow i}$ is indexed by $\mathbf{b}^{k}$.
Eventually, we should obtain $p-k$ couple of links $\left( \mathbf{b}^{k},%
\mathbf{c}^{k}\right) $ that guarantee the conservation of charges following
the circulation of arrows in the quiver.\newline
The topological aspect of such quivers can be visualised from the key
ingredients $\mathbf{b}$ and $\mathbf{c}$ appearing in the quiver diagram.
In fact, the $b^{\alpha }$ can be expressed in terms of the topological line
defect using eqs.(\ref{xy},\ref{234}) as well as\textrm{\ }$b^{\alpha
}=tr\left( XY^{\alpha }\right) $ and $X=\log \left( \mathcal{L}^{\mathbf{\mu
}_{k}}e^{-Y}z^{-\mathbf{\mu }_{k}}\right) ;$ we have\textrm{\ }%
\begin{equation}
b^{\alpha }=tr\left( \log \left( \mathcal{L}^{\mathbf{\mu }_{k}}e^{-Y}z^{-%
\mathbf{\mu }_{k}}\right) Y^{\alpha }\right)
\end{equation}%
Similar calculations for $c_{\alpha }$ yield%
\begin{equation}
c_{\alpha }=tr\left( \log \left( z^{-\mathbf{\mu }_{k}}e^{-X}\mathcal{L}^{%
\mathbf{\mu }_{k}}\right) X_{\alpha }\right)
\end{equation}%
Concerning the interpretation of the phase space coordinates $b^{\alpha }$\
and $c_{\alpha }$\ as bi-fundamental matter, it follows from the
decomposition of the gauge potential $\mathcal{A}^{\left[ \mu \right] }$\ in
the Lie algebra. For example, for $\mathcal{A}^{\left[ \mu \right] }\sim
adj_{sl_{N}},$ we have the following decompositions (eq.(\ref{muk}))%
\begin{equation}
\begin{tabular}{ccccccccccc}
$sl_{N}$ & $\rightarrow $ & $sl_{k}$ & $\oplus $ & $sl_{N-k}$ & $\oplus $ & $%
sl_{1}$ & $\oplus $ & $n_{+}$ & $\oplus $ & $n_{-}$ \\
$adj_{sl_{N}}$ & $\rightarrow $ & $adj_{sl_{k}}$ & $\oplus $ & $%
adj_{sl_{N-k}}$ & $\oplus $ & $adj_{sl_{1}}$ & $\oplus $ & $\left( k,%
\overline{N-k}\right) $ & $\oplus $ & $\left( \bar{k},N-k\right) $ \\
$\mathcal{A}^{\left[ \mu \right] }$ & $\rightarrow $ & $\mathcal{A}_{sl_{k}}$
& $\oplus $ & $\mathcal{A}_{sl_{N-k}}$ & $\oplus $ & $\mathcal{A}_{sl_{1}}$
& $\oplus $ & $\left\{ b^{a}\right\} $ & $\oplus $ & $\left\{ c_{a}\right\} $%
\end{tabular}%
\end{equation}%
where we see that\ $b^{a}$\textrm{\ }and $c_{a}$\ sit in the bi-fundamental
of the gauge symmetry $SL_{k}\times SL_{N-k}.$\ Therefore, we have a quiver
diagram with\ two nodes corresponding to the adjoints $\left( \mathbf{%
k\times \bar{k}}\right) -1$\ and $\left( \mathbf{N-k}\right) \left(
\overline{\mathbf{N-k}}\right) -1,$\ and two links corresponding to the
bi-fundamentals $\left( \mathbf{k},\overline{\mathbf{N-k}}\right) $%
\thinspace\ and\ $\left( \mathbf{\bar{k}},\mathbf{N-k}\right) $. This quiver
is constructed below by analysis of the elements of the associated L- matrix
(see Figure \ref{fig2}).

\subsection{Topological quiver of $\mathcal{L}_{\boldsymbol{N}}^{\mathbf{%
\protect\mu }_{1}}$}

In this subsection, we want to associate a topological quiver to the $%
\mathcal{L}_{\boldsymbol{N}}^{\mathbf{\mu }_{1}}$\ calculated before in the
framework of 4D CS theory with $SL_{N}$ gauge symmetry for the first
coweight $\mathbf{\mu }_{1}$ and the fundamental representation $\boldsymbol{%
R=N}$. To this end, we exploit the projector basis in the matrix form of the
L-operator to cast its elements corresponding to different representations
of the subalgebras in the Levi $\boldsymbol{l}_{\mu _{1}}$.

\subsubsection{The L-operator in the projector basis}

The expression of $\mathcal{L}_{\boldsymbol{N}}^{\mathbf{\mu }_{1}}$
involves two projectors $\varrho _{1}$ and $\bar{\varrho}_{1}$ corresponding
the representations of the Levi subalgebra $sl_{1}\oplus sl_{N-1}.$ The
presence of these projectors in the explicit expansion of $\mathcal{L}_{%
\boldsymbol{N}}^{\mathbf{\mu }_{1}}$ is interesting in the sense that it
allows to represent it as a four sub-block matrix. We have
\begin{equation}
\mathcal{L}_{\boldsymbol{N}}^{\mathbf{\mu }_{1}}=\left(
\begin{array}{cc}
z^{\frac{N-1}{N}}I_{N-1}+XY & z^{-\frac{1}{N}}X \\
z^{-\frac{1}{N}}Y & z^{-\frac{1}{N}}%
\end{array}%
\right)
\end{equation}%
which is obtained from \ref{m1} using special projectors features of the
realisation of $X_{\alpha }$ and $Y^{\alpha }$\ (\ref{rn}) like $X\varrho
_{1}=0$, $\varrho _{1}Y=0,$ $X\bar{\varrho}_{1}=X$, $\bar{\varrho}_{1}Y=Y$
and $X\bar{\varrho}_{1}Y=XY$. Moreover, we can write%
\begin{equation}
\mathcal{L}_{\boldsymbol{N}}^{\mathbf{\mu }_{1}}=\left(
\begin{array}{cc}
\varrho _{1}[z^{\frac{N-1}{N}}+z^{-\frac{1}{N}}XY]\varrho _{1} & z^{-\frac{1%
}{N}}\varrho _{1}X\varrho _{\bar{1}} \\
z^{-\frac{1}{N}}\varrho _{\bar{1}}Y\varrho _{1} & z^{-\frac{1}{N}}\varrho _{%
\bar{1}}\varrho _{\bar{1}}%
\end{array}%
\right)  \label{lpp}
\end{equation}%
to visualize the correspondence with irrepresentations and bi-modules of $%
sl_{N};$ thus opening a window on a formal similarity with the structure of
supersymmetric quiver graphs. This allows to think of the topological quiver
Q$_{\boldsymbol{N}}^{\mathbf{\mu }_{1}}$ for the $A$- type symmetry as
having two nodes $\mathcal{N}_{i}$ and two links $L_{ij}$ associated to
sub-blocks of $\mathcal{L}_{\boldsymbol{N}}^{\mathbf{\mu }_{1}}$\ as follows
\begin{equation}
\begin{tabular}{lllll}
$\mathcal{N}_{1}$ & $=\left \langle \varrho _{1}\mathcal{L}\varrho
_{1}\right \rangle $ & $\qquad ,\qquad $ & $L_{1\bar{1}}$ & $=\left \langle
\varrho _{1}\mathcal{L}\varrho _{\bar{1}}\right \rangle $ \\
$\mathcal{N}_{\bar{1}}$ & $=\left \langle \varrho _{\bar{1}}\mathcal{L}%
\varrho _{\bar{1}}\right \rangle $ & $\qquad ,\qquad $ & $L_{\bar{1}1}$ & $%
=\left \langle \varrho _{\bar{1}}\mathcal{L}\varrho _{1}\right \rangle $%
\end{tabular}%
\end{equation}%
Moreover, by replacing with $X=b^{\alpha }X_{a}$ and $Y=c_{a}Y^{\alpha }$ as
well as $XY=(b^{\alpha }c_{a})\varrho _{1}$ in the Lax operator$,$ we end up
with the known form of $\mathcal{L}_{\boldsymbol{N}}^{\mathbf{\mu }_{1}}$ in
the literature \textrm{\cite{14}.}
\begin{equation}
\mathcal{L}_{\boldsymbol{N}}^{\mathbf{\mu }_{1}}=\left(
\begin{array}{cc}
z^{\frac{N-1}{N}}+z^{-\frac{1}{N}}\mathbf{b}^{T}\mathbf{c} & z^{-\frac{1}{N}}%
\mathbf{b}^{T} \\
z^{-\frac{1}{N}}\mathbf{c} & z^{-\frac{1}{N}}%
\end{array}%
\right)  \label{lppp}
\end{equation}%
In this oscillator realisation, the $b^{\alpha }$ and $c_{\alpha }$ appear
indeed as fundamental quantities.

\subsubsection{Formal expression of $\mathcal{L}_{\boldsymbol{N}}^{\mathbf{%
\protect\mu }_{1}}$ and the quiver Q$_{\boldsymbol{N}}^{\mathbf{\protect\mu }%
_{1}}$}

\begin{figure}[h]
\begin{center}
\includegraphics[width=12cm]{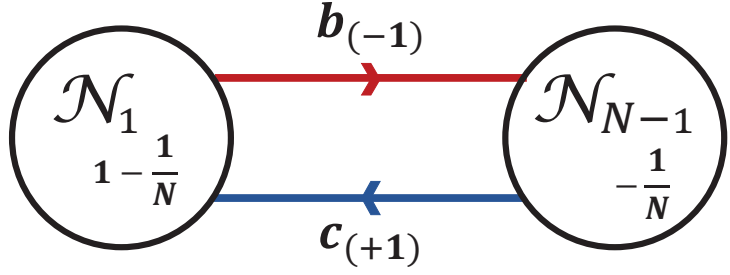}
\end{center}
\par
\vspace{-0.5cm}
\caption{The topological quiver Q$_{\boldsymbol{N}}^{\mathbf{\protect\mu }%
_{1}}$ representing $\mathcal{L}_{\boldsymbol{N}}^{\mathbf{\protect\mu }%
_{1}} $ of $sl_{N}$. It has 2 nodes and 2 links. The nodes describe
self-dual topological matter and the links describe topological bi-matter$.$}
\label{fig1}
\end{figure}

To explicitly match the L-matrix in terms of oscillators \ref{lppp} with
ingredients of the Q$_{\boldsymbol{N}}^{\mathbf{\mu }_{1}}$, we can use the
property $\varrho _{1}+\varrho _{\bar{1}}=I_{\boldsymbol{N}}$ to cast $%
\mathcal{L}_{\boldsymbol{N}}^{\mathbf{\mu }_{1}}$ in different but
equivalent ways: First as $I_{\boldsymbol{N}}\mathcal{L}^{\mathbf{\mu }_{1}}$
and $\mathcal{L}^{\mathbf{\mu }_{1}}I_{\boldsymbol{N}}$ giving%
\begin{equation}
\begin{tabular}{lll}
$\mathcal{L}_{\boldsymbol{N}}^{\mathbf{\mu }_{1}}$ & $=$ & $\varrho _{1}%
\mathcal{L}^{\mathbf{\mu }_{1}}+\varrho _{\bar{1}}\mathcal{L}^{\mathbf{\mu }%
_{1}}$ \\
& $=$ & $\mathcal{L}^{\mathbf{\mu }_{1}}\varrho _{1}+\mathcal{L}^{\mathbf{%
\mu }_{1}}\varrho _{\bar{1}}$%
\end{tabular}%
\end{equation}%
And second, using the form $I_{\boldsymbol{N}}\mathcal{L}^{\mathbf{\mu }%
_{1}}I_{\boldsymbol{N}}$ to express the Lax operator as $\left( \varrho
_{1}+\varrho _{\bar{1}}\right) \mathcal{L}^{\mathbf{\mu }_{1}}\left( \varrho
_{1}+\varrho _{\bar{1}}\right) ,$ namely%
\begin{equation}
\mathcal{L}_{\boldsymbol{N}}^{\mathbf{\mu }_{1}}=\varrho _{1}\mathcal{L}^{%
\mathbf{\mu }_{1}}\varrho _{1}+\varrho _{1}\mathcal{L}^{\mathbf{\mu }%
_{1}}\varrho _{\bar{1}}+\varrho _{\bar{1}}\mathcal{L}^{\mathbf{\mu }%
_{1}}\varrho _{1}+\varrho _{\bar{1}}\mathcal{L}^{\mathbf{\mu }_{1}}\varrho _{%
\bar{1}}
\end{equation}%
Moreover, by help of $\varrho _{1}^{2}=\varrho _{1}$ and $\varrho _{\bar{1}%
}^{2}=\varrho _{\bar{1}}$ as well as $\varrho _{1}\varrho _{\bar{1}}=0$, we
can present $\mathcal{L}_{\boldsymbol{N}}^{\mathbf{\mu }_{1}}$ in the
operator basis $\left( \varrho _{1},\varrho _{\bar{1}}\right) $ like a 2$%
\times $2 blocks matrix as follows
\begin{equation}
\mathcal{L}_{\boldsymbol{N}}^{\mathbf{\mu }_{1}}=\left(
\begin{array}{cc}
\varrho _{1}\mathcal{L}^{\mathbf{\mu }_{1}}\varrho _{1} & \varrho _{1}%
\mathcal{L}^{\mathbf{\mu }_{1}}\varrho _{\bar{1}} \\
\varrho _{\bar{1}}\mathcal{L}^{\mathbf{\mu }_{1}}\varrho _{1} & \bar{\varrho}%
_{1}\mathcal{L}^{\mathbf{\mu }_{1}}\varrho _{\bar{1}}%
\end{array}%
\right)
\end{equation}%
This formulation of the Lax operator was behind the construction of the
topological quivers concerning exceptional 't Hooft lines in 4DCS theories
with E$_{6}$ and E$_{7}$\ gauge symmetries.\ Here, for the minuscule
coweight $\mu _{1}$ and representation $\boldsymbol{R=N}$ of $sl_{N}$, the
topological gauge quiver Q$_{\boldsymbol{N}}^{\mathbf{\mu }_{1}}$ is
depicted in the Figure \textbf{\ref{fig1}.}
Its two nodes $\mathcal{N}_{1}=\varrho _{1}\mathcal{L}^{\mathbf{\mu }%
_{1}}\varrho _{1}$ and $\mathcal{N}_{\bar{1}}=\varrho _{\bar{1}}\mathcal{L}^{%
\mathbf{\mu }_{1}}\varrho _{\bar{1}}$ are interpreted as topological adjoint
matter of the Levi sub-symmetry group $SL_{1}\times SL_{N-1}$ $SL_{1}.$ This
can also be referred to as self dual matter since it is uncharged under the
minuscule coweight operator, like the quantity\ $\mathbf{b}^{T}\mathbf{c}$.
The two links $L_{1\bar{1}}=\varrho _{1}\mathcal{L}^{\mathbf{\mu }%
_{1}}\varrho _{\bar{1}}$ and $L_{\bar{1}1}=\varrho _{\bar{1}}\mathcal{L}^{%
\mathbf{\mu }_{1}}\varrho _{1}$ are given in terms of oscillators $b^{\alpha
}$ and $c_{\alpha }.$ They carry charges $\mp 1$ under $\mathbf{\mu }_{1}$,
and are interpreted in terms of topological bi-fundamental matter of $%
SL_{1}\times SL_{N-1}$. This QFT interpretation matches the supersymmetric
gauge quiver description. Finally, notice that using the Killing form, the $%
b^{\alpha }$ and $c_{\alpha }$ can be related to the links $L_{1\bar{1}}$
and $L_{\bar{1}1}$ as
\begin{equation}
b^{\alpha }=z^{\frac{1}{N}}Tr\left( L_{1\bar{1}}Y^{\alpha }\right) \qquad
,\qquad c_{\alpha }=z^{\frac{1}{N}}Tr\left( L_{\bar{1}1}X_{\alpha }\right)
\end{equation}

\subsection{Topological quivers: case $2\leq k\leq N-2$}

Here, we generalise the construction of subsection 3.2 regarding the
minuscule coweight $\mu _{1}$ to the generic minuscule coweight $\mu _{k}$
with $2\leq k\leq N-2$.

\subsubsection{Generic projectors $\Pi _{k}$ and $\Pi _{\bar{k}}$}

In the generic case, the expression (\ref{mk}) involves the projectors $\Pi
_{k}$ and $\Pi _{\bar{k}}$ on the representations of the Levi subalgebra $%
sl_{k}\oplus sl_{N-k}\oplus sl_{1}.$ Using the properties
\begin{equation}
X\Pi _{k}=0\qquad ,\qquad \Pi _{k}Y=0
\end{equation}%
and the identities
\begin{equation}
X\Pi _{\bar{k}}=X\qquad ,\qquad \Pi _{\bar{k}}Y=Y
\end{equation}%
leading to $X\Pi _{\bar{k}}Y=XY,$ the L- operator $\mathcal{L}_{\boldsymbol{N%
}}^{\mathbf{\mu }_{k}}$ can be presented in block matrices like%
\begin{equation}
\mathcal{L}_{\boldsymbol{N}}^{\mathbf{\mu }_{k}}=\left(
\begin{array}{cc}
z^{\frac{N-k}{N}}+z^{-\frac{k}{N}}XY & z^{-\frac{k}{N}}X \\
z^{-\frac{k}{N}}Y & z^{-\frac{k}{N}}%
\end{array}%
\right)
\end{equation}%
By exhibiting the dependence into the Darboux coordinates while substituting
$X=b^{i\alpha }X_{ia}$ and $Y=c_{j\beta }Y^{j\beta }$ as well as $%
XY=b^{i\alpha }c_{i\alpha },$ we obtain%
\begin{equation}
\mathcal{L}_{\boldsymbol{N}}^{\mathbf{\mu }_{k}}=\left(
\begin{array}{cc}
z^{-\frac{k}{N}}(z+b^{i\alpha }c_{i\alpha }) & z^{-\frac{k}{N}}b^{i\alpha
}X_{ia} \\
z^{-\frac{k}{N}}c_{j\beta }Y^{j\beta } & z^{-\frac{k}{N}}%
\end{array}%
\right)
\end{equation}%
which is also in agreement with \textrm{\cite{14}.}

\subsubsection{Constructing the topological quivers Q$_{\boldsymbol{N}}^{%
\mathbf{\protect\mu }_{k}}$}

\begin{figure}[h]
\begin{center}
\includegraphics[width=12cm]{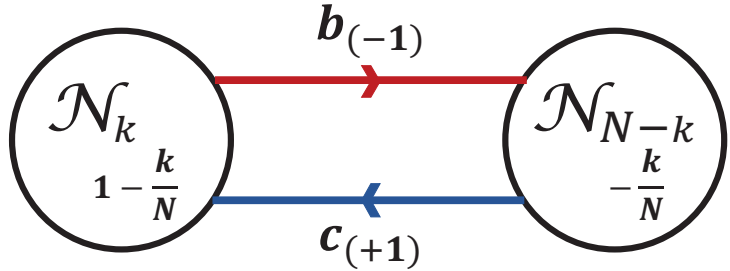}
\end{center}
\par
\vspace{-0.5cm}
\caption{The topological quiver representing $\mathcal{L}_{\boldsymbol{N}}^{%
\protect\mu _{k}}$ of $sl_{N}$. It has 2 nodes and 2 links. The nodes
describe self-dual topological matter and the links describe bi-matter$.$}
\label{fig2}
\end{figure}

By using the property $\Pi _{k}+\Pi _{\bar{k}}=I,$ we can cast $\mathcal{L}_{%
\boldsymbol{N}}^{\mathbf{\mu }_{k}}$ as follows%
\begin{equation}
\mathcal{L}_{\boldsymbol{N}}^{\mathbf{\mu }_{k}}=\left( \Pi _{k}+\Pi _{\bar{k%
}}\right) \mathcal{L}^{\mathbf{\mu }_{k}}\left( \Pi _{k}+\Pi _{\bar{k}%
}\right)
\end{equation}%
Using $\Pi _{k}\Pi _{\bar{k}}=0,$ we can put this $\mathcal{L}^{\left( \mu
_{k}\right) }$ into the following matrix form%
\begin{equation}
\mathcal{L}_{\boldsymbol{N}}^{\mathbf{\mu }_{k}}=\left(
\begin{array}{cc}
\Pi _{k}\mathcal{L}^{\mathbf{\mu }_{k}}\Pi _{k} & \Pi _{k}\mathcal{L}^{%
\mathbf{\mu }_{k}}\Pi _{\bar{k}} \\
\Pi _{\bar{k}}\mathcal{L}^{\mathbf{\mu }_{k}}\Pi _{k} & \Pi _{\bar{k}}%
\mathcal{L}^{\mathbf{\mu }_{k}}\Pi _{\bar{k}}%
\end{array}%
\right)
\end{equation}%
The topological gauge quiver Q$_{\boldsymbol{N}}^{\mathbf{\mu }_{k}}$
associated with this L-operator has two nodes $\mathcal{N}_{k},\mathcal{N}_{%
\bar{k}}$ and two links $L_{k\bar{k}},L_{\bar{k}k}$. It is depicted by the
Figure \textbf{\ref{fig2}}. \newline
\newline
The two nodes
\begin{equation}
\mathcal{N}_{k}=\Pi _{\bar{k}}\mathcal{L}^{\mathbf{\mu }_{k}}\Pi _{k}\qquad
,\qquad \mathcal{N}_{\bar{k}}=\Pi _{\bar{k}}\mathcal{L}^{\mathbf{\mu }%
_{k}}\Pi _{\bar{k}}
\end{equation}%
describe topological adjoint matter of $SL_{k}$ and $SL_{N-k};$ and are
interpreted as topological self dual matter. The two links relating the two
nodes are given by,
\begin{equation}
\Pi _{k}\mathcal{L}^{\mathbf{\mu }_{k}}\Pi _{\bar{k}}\qquad ,\qquad \Pi _{%
\bar{k}}\mathcal{L}^{\mathbf{\mu }_{k}}\Pi _{k}
\end{equation}%
they describe bi-fundamental matter of $SL_{k}\times SL_{N-k}$. These
bi-matters are precisely given by the Darboux variables $b^{i\alpha }$ and $%
c_{ia}$ of the phase space of 't Hooft line tH$_{\mathrm{\gamma }_{0}}^{\mu
_{k}}.$ To end this section, notice the following :

\begin{description}
\item[$\left( \mathbf{1}\right) $] the topological quiver Q$_{\boldsymbol{N}%
}^{\mathbf{\mu }_{1}}$ of the operator $\mathcal{L}_{\boldsymbol{N}}^{%
\mathbf{\mu }_{1}}$ appears just as the leading quiver of the k-family Q$_{%
\boldsymbol{N}}^{\mathbf{\mu }_{k}}$ associated with the family $\mathcal{L}%
_{\boldsymbol{N}}^{\mathbf{\mu }_{k}}$. So, the topological quiver Q$_{%
\boldsymbol{N}}^{\mathbf{\mu }_{N-1}}$ of the $\mathcal{L}_{\boldsymbol{N}}^{%
\mathbf{\mu }_{N-1}}$ turns out be just the last member of the k-family. We
omit its description.

\item[$\left( \mathbf{2}\right) $] the quiver Q$_{\boldsymbol{N}}^{\mathbf{%
\mu }_{k}}$ given in this section concerns Wilson lines with quantum states
in the fundamental $\boldsymbol{R}=\boldsymbol{N}$. For Wilson lines in
other representations of $sl_{N}$ like the completely antisymmetric $%
\boldsymbol{N}^{\wedge k}$\ and the completely symmetric $\boldsymbol{N}%
^{\vee n}$, we can construct the associated the L-operators and the
corresponding quivers Q$_{\boldsymbol{R}}^{\mathbf{\mu }_{k}}.$ Examples of
the topological quivers Q$_{\boldsymbol{N}^{\wedge k}}^{\mathbf{\mu }_{1}}$
and Q$_{\boldsymbol{N}^{\vee n}}^{\mathbf{\mu }_{1}}$\ are given in Figure
\textbf{\ref{TA}}. Their Levi charges reported on the nodes can be read from
the decomposition (\ref{n2}). As an illustration, the quiver Q$_{\boldsymbol{%
N}^{\vee 3}}^{\mathbf{\mu }_{1}}$ corresponding to the representation the
symmetric $\boldsymbol{N\vee N\vee N}$ is depicted by the Figure \textbf{\ref%
{sym3}}.
\begin{figure}[h]
\begin{center}
\includegraphics[width=10
cm]{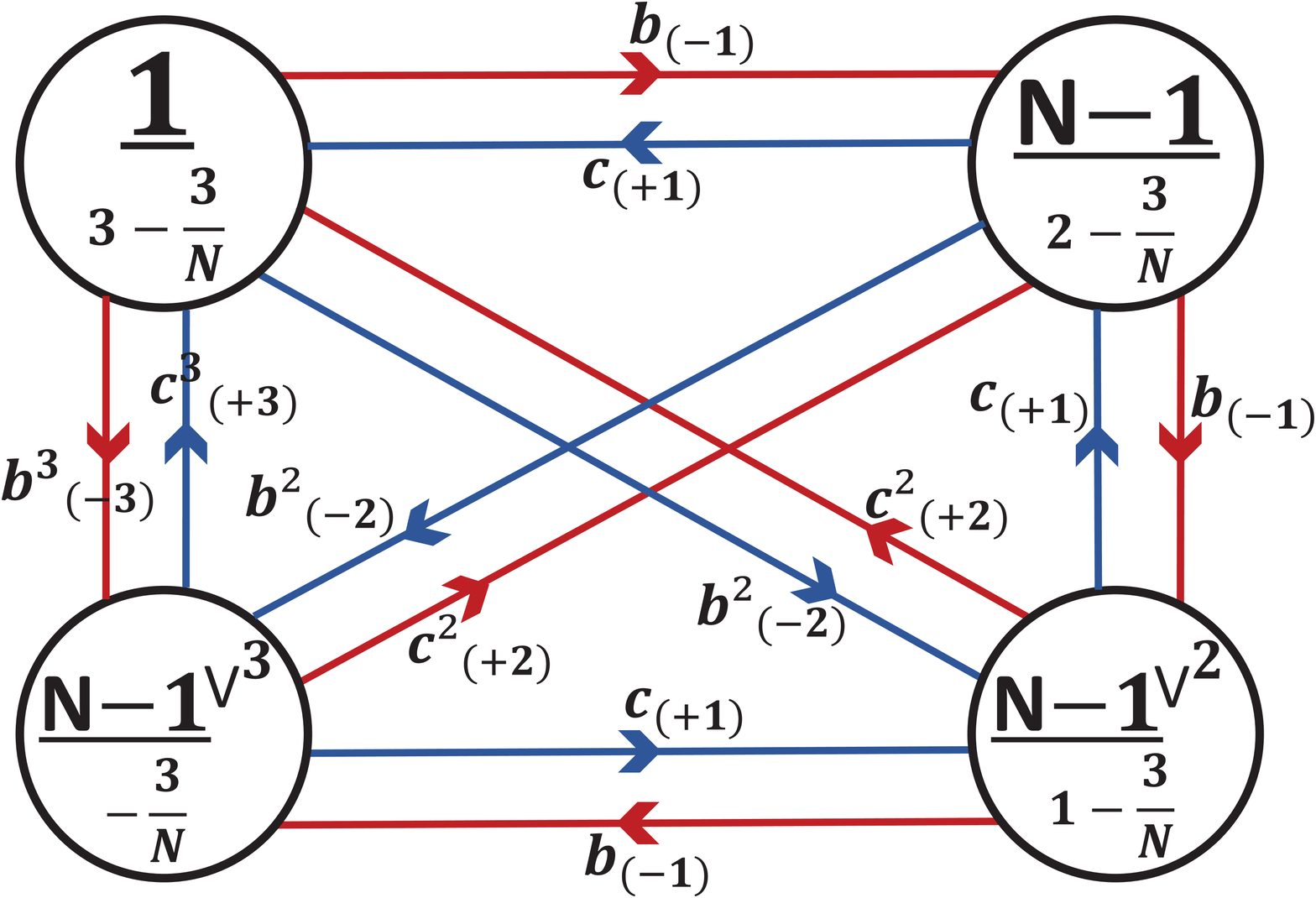}
\end{center}
\par
\vspace{-0.5cm}
\caption{The topological quiver Q$_{\boldsymbol{R}}^{\protect\mu _{1}}$ for
the representation $\boldsymbol{R=N\vee N\vee N}.$ This quiver has four
nodes and 12 links.}
\label{sym3}
\end{figure}

\item[$\left( \mathbf{3}\right) $] An interesting topological quiver diagram
Q$_{\boldsymbol{adj}\left( sl_{N}\right) }^{\mathbf{\mu }_{k}}$ given by the
Figure \textbf{\ref{adjs}}. It\textbf{\ }is the one associated with the
adjoint representation; that is $\boldsymbol{R}=\boldsymbol{adj}\left(
sl_{N}\right) $. From the decomposition given by eq.(\ref{n2}), we see that $%
\boldsymbol{adj}\left( sl_{N}\right) $ splits as $\boldsymbol{n}_{-}\oplus
\boldsymbol{l}_{\mu _{k}}\oplus \boldsymbol{n}_{+}$ with $\boldsymbol{l}%
_{\mu _{k}}=\boldsymbol{adj}\left( sl_{k}\right) _{0}+\boldsymbol{adj}\left(
sl_{N-k}\right) _{0}+sl_{1}$ and $\boldsymbol{n}_{\pm }=\boldsymbol{k}\left(
\boldsymbol{N-k}\right) _{\pm }$. The second concerns $sl_{1}$ with the
representation $\boldsymbol{1}_{0}=\boldsymbol{1}_{1-1/N}\times \mathbf{1}%
_{-1+1/N}$.
\begin{figure}[h]
\begin{center}
\includegraphics[width=12cm]{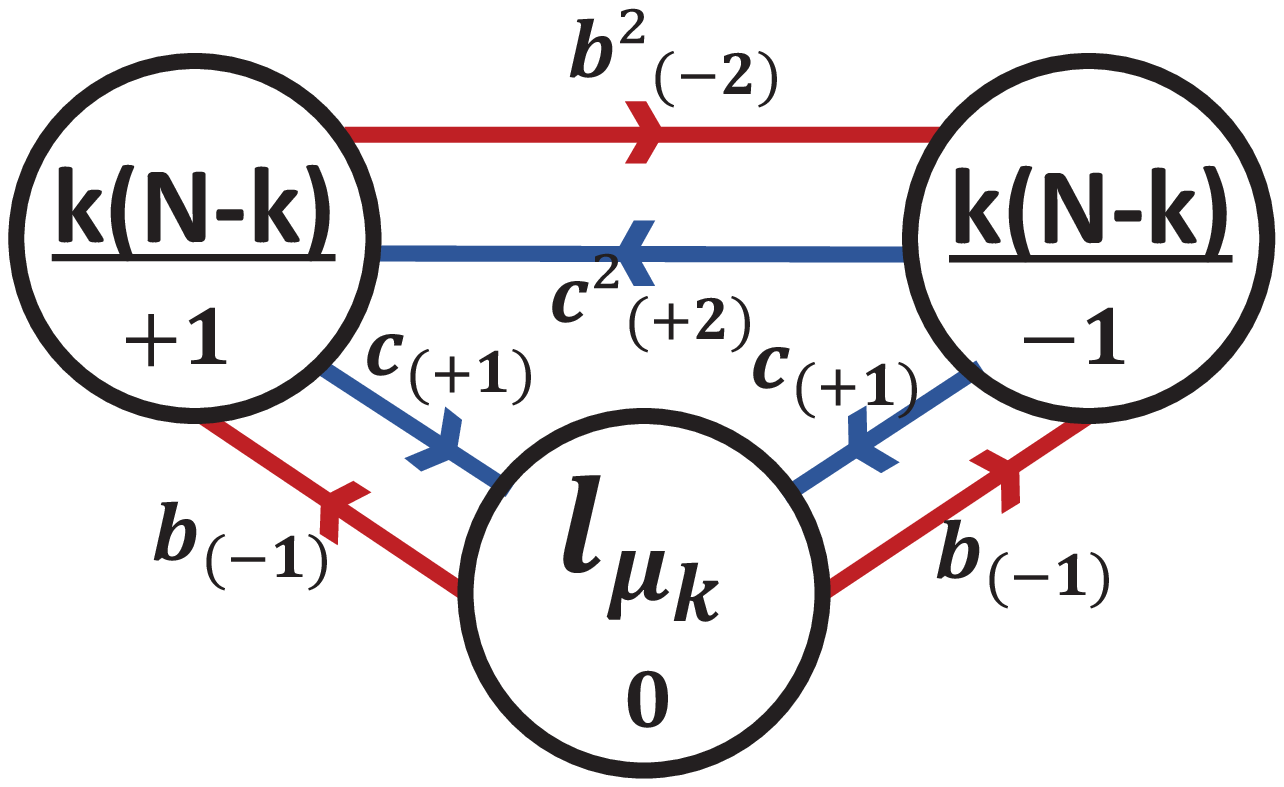}
\end{center}
\par
\vspace{-0.5cm}
\caption{The topological quiver Q$_{\boldsymbol{adj}\left( sl_{N}\right) }^{%
\mathbf{\protect\mu }_{k}}$ for the adjoint representation of $sl_{N}$. It
has three nodes $\mathcal{N}_{0}=\boldsymbol{l}_{\protect\mu _{k}}$ and $%
\boldsymbol{n}_{\pm }=\boldsymbol{k}\left( \boldsymbol{N-k}\right) _{\pm }$.
}
\label{adjs}
\end{figure}
\newline
\end{description}

\section{Vector 't Hooft lines of D$_{N}$- type}

In this section, we study the class of vector-like L-operators $\mathcal{L}%
_{so_{2N}}^{vect}$ in the 4D Chern-Simons theory with $SO_{2N}$ gauge
symmetry. This is a sub-family of the family of D- type Lax operators which
contains moreover the Lax operators $\mathcal{L}_{so_{2N}}^{spin}$ of the
spinorial class to be studied in the next section. Because $%
SO_{4}=SU_{2}\times SU_{2}$ and $SO_{6}\sim D_{3}$ is isomorphic to $SL_{4}$%
, we assume that $N\geq 4$ so that the first element of the $D_{N}$ series
is given by $SO_{8}$. \newline
Notice that the general aspects of the present construction are similar to
those introduced in the previous sections. The 't Hooft line tH$_{\mathrm{%
\gamma }_{0}}^{\mu }$ is taken as the horizontal x-axis of $\mathbb{R}^{2}$
and the $W_{\mathrm{\xi }_{z}}^{\boldsymbol{R}}$ is chosen as the vertical
y-axis; the $z$ is a generic position in the holomorphic line $\mathbb{CP}%
^{1}$, and $\boldsymbol{R}$\ is a given representation of $so_{2N}$.
Moreover, most of the features associated to the derivation of Lax operators
from 4D CS with $SO_{2N}$ gauge symmetry have been considered in \textrm{%
\cite{34A,54A}}. Therefore, we focus here on analysing the internal
algebraic structure of this theory allowing to illustrate the key elements
of the quiver gauge Q$_{so_{2N}}^{{\small vect}}$ associated to $\mathcal{L}%
_{so_{2N}}^{{\small vect}}$. This quiver constitutes a necessary part in the
unified theory chain in the sense that it links the A-type symmetries to the
exceptional ones, and allows to indirectly include the B-type symmetries
thanks to its similarity with the minuscule coweight of the $so_{2N+1}$ Lie
algebra.

\subsection{Vector lines tH$_{\mathrm{\protect\gamma }_{0}}^{\protect\mu %
_{1}}$ and their L-operators}

We begin by recalling that minuscule 't Hooft lines within the $D_{N}$
family of 4D CS theory are magnetically charged with magnetic charge given
by the minuscule coweights $\mu $ of $D_{N}$. Because there are three
minuscule coweights in the D$_{N}$ Lie algebras given by $\mu _{1},\mu
_{N-1},\mu _{N}$ (see the Figures \textbf{\ref{DV} }and\textbf{\ \ref{ds}}),
we distinguish three types of 't Hooft lines tH$_{\mathrm{\gamma }_{0}}^{\mu
}$ in the 4D Chern-Simons theory with orthogonal gauge symmetry $SO_{2N}$
that we can refer to as
\begin{equation}
\begin{tabular}{lllll}
tH$_{\mathrm{\gamma }_{0}}^{\mu _{1}}=$tH$_{\mathrm{\gamma }_{0}}^{vect}$ & ,
& tH$_{\mathrm{\gamma }_{0}}^{\mu _{N}}=$tH$_{\mathrm{\gamma }_{0}}^{\text{%
spin}}$ & , & tH$_{\mathrm{\gamma }_{0}}^{\mu _{N-1}}=$tH$_{\mathrm{\gamma }%
_{0}}^{\text{cospin}}$%
\end{tabular}%
\end{equation}%
The coweights $\mu _{1},\mu _{N-1},\mu _{N}$ are respectively dual to the
vector representation $2N$, the spinor representation 2$_{L}^{N-1}$ and the
cospinor representation 2$_{R}^{N-1}$.
\begin{figure}[h]
\begin{center}
\includegraphics[width=12cm]{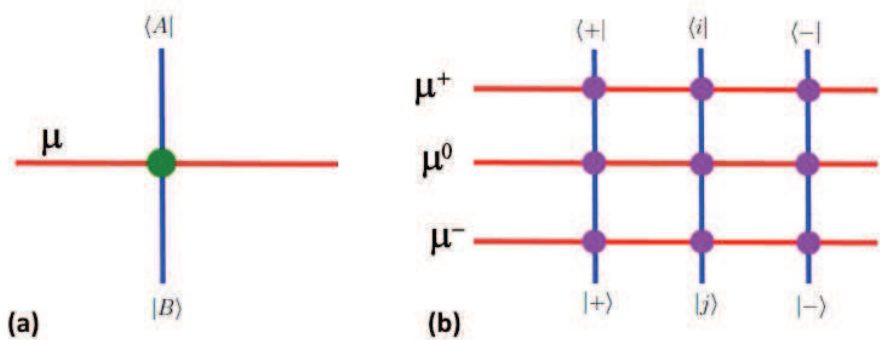}
\end{center}
\par
\vspace{-0.5cm}
\caption{(\textbf{a}) A horizontal vector-like 't Hooft line with magnetic
charge $\protect\mu _{1}$ crossing an electrically charged vertical Wilson
line in fundamental vector representation $\boldsymbol{R=2N}$. The green dot
refers to the coupling between the two lines; it is given by the Lax
operator $\mathcal{L}_{\boldsymbol{2N}}^{\protect\mu _{1}}$. (\textbf{b})
Intrinsic structure of the Lax operator which will be interpreted as a
topological gauge quiver with three nodes and 6 links.}
\label{2N}
\end{figure}
Here, we first focus on the coupling of the vector-like tH$_{\mathrm{\gamma }%
_{0}}^{\mu _{1}}$ with the Wilson line in fundamental $\boldsymbol{R=2N}$;
then we move in the next section to the study of tH$_{\mathrm{\gamma }%
_{0}}^{\mu _{N-1}}$ and tH$_{\mathrm{\gamma }_{0}}^{\mu _{N}}$. To fix the
ideas, we illustrate in the Figure \textbf{\ref{2N}} the Levi splitting
characterising tH$_{\mathrm{\gamma }_{0}}^{vect}$. This intrinsic structure
will be derived and commented later on.

\subsubsection{Vectorial tH$_{\mathrm{\protect\gamma }_{0}}^{vect}$ line:
magnetic charge}

\begin{figure}[h]
\begin{center}
\includegraphics[width=12cm]{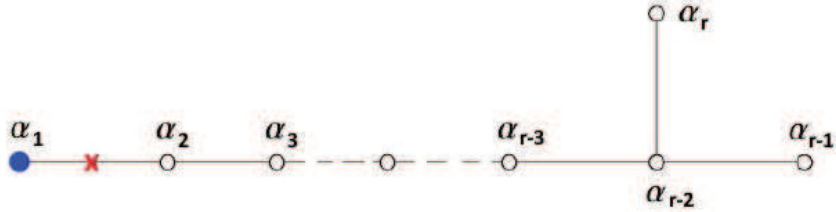}
\end{center}
\par
\vspace{-0.5cm}
\caption{Dynkin diagram of $D_{N}$ Lie algebras where the $N$ simple roots $%
\protect\alpha _{i}$ are exhibited. The Levi decomposition of $%
so_{2N}\rightarrow so_{2}\oplus so_{2N-2}$ using the vector coweight is
given by cutting the simple root $\protect\alpha _{1}$. }
\label{DV}
\end{figure}

The fundamental coweight $\mu _{1}$ is the dual to the simple root $\alpha
_{1}$ of the $so_{2N}$ Lie algebra. By taking the $N$ simple roots of $%
SO_{2N}$ as $\alpha _{i}=e_{i}-e_{i+1}$ for $i\in \left[ 1,N-1\right] $ and $%
\alpha _{N}=e_{N-1}+e_{N};$ it follows that the value of the minuscule
coweight constrained as $\mu _{1}\alpha _{i}=\delta _{i1}$ can be solved
like $\mu _{1}=e_{1}.$ In terms of the simple roots, we have
\begin{equation}
\mu _{1}=\alpha _{1}+...+\alpha _{N-2}+\frac{1}{2}\left( \alpha
_{N-1}+\alpha _{N}\right)  \label{e1}
\end{equation}%
Notice that by setting N=3 in this relation, the resulting $\mu _{1}$ takes
the value $\alpha _{1}+\frac{1}{2}\left( \alpha _{2}+\alpha _{3}\right) $
which can be compared with the fundamental weight $\tilde{\mu}_{2}=\tilde{%
\alpha}_{2}+\frac{1}{2}(\tilde{\alpha}_{1}+\tilde{\alpha}_{3})$ of the $%
sl_{4}$ Lie algebra which is isomorphic to $so_{6}.$ Here, the $\tilde{\alpha%
}_{i}$'s stand for the simple roots of $sl_{4}$. \newline
From the Dynkin diagram of the D$_{N}$ Lie algebras given in Figure \textbf{%
\ref{DV}}, we can see that the Levi decomposition $\boldsymbol{l}_{\mu
_{1}}\oplus n_{+}\oplus n_{-}$ of $so_{2N}$ with respect to the vectorial
coweight $\mu _{1}$ is given by $\boldsymbol{l}_{\mu _{1}}=so_{2}\oplus
so_{2N-2}$ and $n_{\pm }=\left( 2N-2\right) _{\pm }$ with the charge
symmetry $so_{2}\sim sl_{1}$. As such, the dimensions of the $so_{2N}$ and
its vector $\boldsymbol{2N}$ split as follows%
\begin{equation}
\begin{tabular}{lll}
$N\left( 2N-1\right) $ & $=$ & $1_{0}+\left( N-1\right) \left( 2N-3\right)
_{0}+\left( 2N-2\right) _{+}+\left( 2N-2\right) _{-}$ \\
$\mathbf{\ \ \ \ \ \ \ \ \ \ \ 2N}$ & $\mathbf{=}$ & $\mathbf{2}_{0}\mathbf{+%
}\left( \mathbf{2N-2}\right) _{0}$%
\end{tabular}%
\end{equation}%
where we have also exhibited the charge of $so_{2}$. To construct the
L-operator of the tH$_{\mathrm{\gamma }_{0}}^{vect}$ line represented
graphically by the Figure \textbf{\ref{2N}-(a), }we need the adjoint action
of the coweight $\mu _{1}$ and the explicit expressions of the generators of
the nilpotent subalgebras $n_{\pm }$.
\newline
The $2N-2$ generators of $n_{+}$ are denoted by $X_{i}$ and their homologues
generating $n_{-}$ are denoted like $Y^{i}$, their realisation should solve
the Levi decomposition constraint $\left[ \mathbf{\mu }_{1},n_{\pm }\right]
=\pm n_{\pm }$ and $\left[ n_{q},n_{q}\right] =0$ with $q=\pm $. \newline
To get this solution, we consider $\left( \mathbf{i}\right) $ an electric
vertical Wilson line as in the Figure \textbf{\ref{2N}-(a)}
\begin{equation}
W_{\mathrm{\xi }_{z}}^{\boldsymbol{R}=\boldsymbol{2N}}\qquad ,\qquad \mathrm{%
\xi }_{z}=\left \{ \left( x,y\right) |x=0;-\infty <y<\infty \right \}
\end{equation}%
with incoming vector-like states $\left \langle A\right \vert $ (A=1,...,
2N) and outgoing $\left \vert B\right \rangle $ ones propagating along the
line $\mathrm{\xi }_{z}.$\ $\left( \mathbf{ii}\right) $ a horizontal 't
Hooft line defect tH$_{\mathrm{\gamma }_{0}}^{vect}$ with the magnetic
charge $\mu _{1}; $%
\begin{equation}
tH_{\mathrm{\gamma }_{0}}^{vect}\qquad ,\qquad \mathrm{\gamma }_{0}=\mathbb{R%
}_{y\leq 0}^{2}\cap \mathbb{R}_{y\geq 0}^{2}
\end{equation}%
In this case, we can split the vector representation $\left \vert
B\right
\rangle $ of SO$_{2N}$ as a direct sum $\left \vert \beta
\right
\rangle \oplus \left \vert j\right \rangle $ where $\left \vert
\beta \right
\rangle $ is a vector of $so_{2}$ and $\left \vert
j\right
\rangle $ a vector of $so_{2N-2}$. Moreover, we use the isomorphism
$so_{2}\sim sl_{1}$ to split $\left \vert \beta \right \rangle $ as $%
\left
\vert +\right \rangle $ and $\left \vert -\right \rangle .$
Eventually, we have
\begin{equation}
\left \vert B\right \rangle =\left(
\begin{array}{c}
\left \vert 0\right \rangle \\
\left \vert j\right \rangle \\
\left \vert \bar{0}\right \rangle%
\end{array}%
\right) \equiv \left(
\begin{array}{c}
\left \vert +\right \rangle \\
\left \vert j\right \rangle \\
\left \vert -\right \rangle%
\end{array}%
\right) ,\qquad 1\leq j\leq M  \label{vb}
\end{equation}%
where we have set $M=2N-2$ and considered the splitting of the $\mathbf{2N}$
vector as $\mathbf{1}_{+}\oplus \left( \mathbf{2N-2}\right) _{0}\oplus
\mathbf{1}_{-}$ such that\ the Levi subalgebra is $sl_{1}\oplus so_{2N-2}.$
In this vector states basis (\ref{vb}), the operators $X_{i}$ and $Y^{i}$
generating the nilpotent subalgebras are given by
\begin{equation}
\begin{tabular}{lll}
$X_{i}$ & $=$ & $\left \vert +\right \rangle \left \langle i\right \vert
-\left \vert i\right \rangle \left \langle -\right \vert $ \\
$Y^{i}$ & $=$ & $\left \vert i\right \rangle \left \langle +\right \vert
-\left \vert -\right \rangle \left \langle i\right \vert $%
\end{tabular}
\label{r1}
\end{equation}%
The action of these operators X$_{i}$ and Y$^{i}$ on the vector
representation of $so_{2N}$ can be visualized in the the Figure \textbf{\ref%
{14}} describing the splitting of the \textbf{2N} vector. As for the adjoint
action of the minuscule coweight, it is given by a particular linear
combination of projectors $\varrho _{R}$ on the irreducible representations $%
\mathbf{1}_{\pm }$ and $\left( \mathbf{2N-2}\right) _{0}$\ of the $%
so_{2}\oplus so_{2N-2}$ Levi subalgebra as follows
\begin{equation}
\mathbf{\mu }_{1}=\varrho _{+}-\varrho _{-}  \label{r2}
\end{equation}%
with $\varrho _{+}=\left \vert +\right \rangle \left \langle +\right \vert $
and $\varrho _{-}=\left \vert -\right \rangle \left \langle -\right \vert .$
\begin{figure}[h]
\begin{center}
\includegraphics[width=10cm]{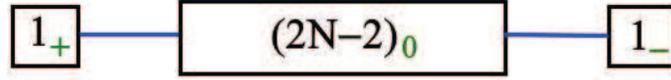}
\end{center}
\par
\vspace{-0.5cm}
\caption{A graphic representation of the splitting of the vector 2N
representation under the vectorial Levi decomposition. The projectors on
these three blocks are $\protect\varrho _{+}=\left \vert +\right \rangle
\left \langle +\right \vert ,$ $\sum \protect\varrho _{i}=\sum \left \vert
i\right \rangle \left \langle i\right \vert $ and $\protect\varrho %
_{-}=\left \vert -\right \rangle \left \langle -\right \vert .$}
\label{14}
\end{figure}
\newline
Because of the vanishing $so_{2}$ charge of $\left( \mathbf{2N-2}\right)
_{0} $, the minuscule coweight has no dependence on the projector
\begin{equation}
\Pi _{0}=\sum_{i}\varrho _{i}
\end{equation}%
with $\varrho _{i}=\left \vert i\right \rangle \left \langle i\right \vert $%
. Notice that $X_{i}$ and $Y^{j}$ satisfy some characteristic relations like
for example $X_{i}Y^{j}=\delta _{j}^{j}\varrho _{+}+\left \vert
i\right
\rangle \left \langle j\right \vert $ indicating that
\begin{equation}
Tr\left( X_{i}Y^{i}\right) =2\delta _{j}^{j}
\end{equation}%
From the realisation of eqs(\ref{r1}-\ref{r2}) we can deduce that $\left[
\mathbf{\mu }_{1},X_{i}\right] =+X_{i}$, $\left[ \mathbf{\mu }_{1},Y^{i}%
\right] =-Y^{i}$ and $\left[ X_{i},Y^{i}\right] =\mathbf{\mu }_{1}$. Other
useful and simplifying relations are listed below%
\begin{equation}
\begin{tabular}{lllllll}
$X_{i}X_{j}$ & $=$ & $-\delta _{ij}\left \vert +\right \rangle \left \langle
-\right \vert $ & $\qquad ,\qquad $ & $X_{i}X_{j}X_{l}$ & $=$ & $0$ \\
$Y^{i}Y^{j}$ & $=$ & $-\delta ^{ij}\left \vert -\right \rangle \left \langle
+\right \vert $ & $\qquad ,\qquad $ & $Y^{i}Y^{j}Y^{l}$ & $=$ & $0$%
\end{tabular}%
\end{equation}%
and%
\begin{equation}
\begin{tabular}{lllllll}
$\varrho _{-}X_{i}$ & $=$ & $0$ & $\qquad ,\qquad $ & $X_{i}\varrho _{+}$ & $%
=$ & $0$ \\
$\varrho _{+}Y^{i}$ & $=$ & $0$ & $\qquad ,\qquad $ & $Y^{i}\varrho _{-}$ & $%
=$ & $0$%
\end{tabular}
\label{ct}
\end{equation}%
as well as%
\begin{equation}
\begin{tabular}{lllllll}
$X_{i}\varrho _{-}$ & $=$ & $-\left \vert i\right \rangle \left \langle
-\right \vert $ & $\qquad ,\qquad $ & $\varrho _{+}X_{i}$ & $=$ & $\left
\vert +\right \rangle \left \langle i\right \vert $ \\
$\varrho _{-}Y^{i}$ & $=$ & $-\left \vert -\right \rangle \left \langle
i\right \vert $ & $\qquad ,\qquad $ & $Y^{i}\varrho _{+}$ & $=$ & $\left
\vert i\right \rangle \left \langle +\right \vert $%
\end{tabular}
\label{y}
\end{equation}%
By considering the linear combinations%
\begin{equation}
X=b^{i}X_{i}\in \mathbf{n}_{+}\qquad ,\qquad Y=c_{i}Y^{i}\in \mathbf{n}_{-}
\end{equation}%
where $b^{i}$ and $c_{i}$ are the phase space coordinates, we can calculate
their powers $X^{n}$ and $Y^{n};$ and then $e^{X}$ and $e^{Y}$. We find that
$X^{2}=-\mathbf{b}^{2}E$, $Y^{2}=-\mathbf{c}^{2}F$ and $X^{3}=Y^{3}=0$ where
we have set $\mathbf{b}^{2}=b^{i}\delta _{ij}b^{j}$ and $\mathbf{c}%
^{2}=c_{i}\delta ^{ij}c_{j}$ as well as $E=\left \vert +\right \rangle
\left
\langle -\right \vert $ and $F=\left \vert -\right \rangle
\left
\langle +\right \vert $ satisfying $\left[ E,F\right] =\mathbf{\mu }%
_{1}$ and $Tr\left( EF\right) =1$. We also have
\begin{equation}
\begin{tabular}{lllllll}
$b^{i}$ & $=$ & $\frac{1}{2}Tr\left( XY^{i}\right) $ & $\qquad ,\qquad $ & $%
\mathbf{b}^{2}$ & $=$ & $-Tr\left( X^{2}F\right) $ \\
$c_{i}$ & $=$ & $\frac{1}{2}Tr\left( X_{i}Y\right) $ & $\qquad ,\qquad $ & $%
\mathbf{c}^{2}$ & $=$ & $-Tr\left( Y^{2}E\right) $%
\end{tabular}
\label{2x}
\end{equation}%
Moreover, we have%
\begin{equation}
\begin{tabular}{lllllllllll}
$\varrho _{-}X$ & $=$ & $0$ & $,\qquad $ & $\varrho _{+}X$ & $=$ & $%
b^{i}\left \vert +\right \rangle \left \langle i\right \vert $ & , & $%
X\varrho _{-}$ & $=$ & $-b^{i}\left \vert i\right \rangle \left \langle
-\right \vert $ \\
$\varrho _{+}Y$ & $=$ & $0$ & $,\qquad $ & $\varrho _{-}Y$ & $=$ & $%
-c_{i}\left \vert -\right \rangle \left \langle i\right \vert $ & , & $%
\varrho _{-}Y$ & $=$ & $-c_{i}\left \vert -\right \rangle \left \langle
i\right \vert $%
\end{tabular}
\label{x}
\end{equation}

\subsubsection{Vector- like tH$_{\mathrm{\protect\gamma }_{0}}^{vect}$ line:
building the L-operator}

Using the properties $X^{3}=Y^{3}=0$\textbf{\ }indicating that $e^{X}=I+X+%
\frac{1}{2}X^{2}$ and equivalently for $e^{Y};$ then putting back into the
expression of the L-operator namely $\mathcal{L}=e^{X}z^{\mathbf{\mu }%
_{1}}e^{Y},$ we obtain%
\begin{equation}
\begin{tabular}{lll}
$\mathcal{L}_{\boldsymbol{2N}}^{{\small vect}}$ & $=$ & $z^{\mathbf{\mu }%
_{1}}+Xz^{\mathbf{\mu }_{1}}+z^{\mathbf{\mu }_{1}}Y+$ \\
&  & $\frac{1}{2}z^{\mathbf{\mu }_{1}}Y^{2}+\frac{1}{2}X^{2}z^{\mathbf{\mu }%
_{1}}+Xz^{\mathbf{\mu }_{1}}Y$ \\
&  & $+\frac{1}{2}Xz^{\mathbf{\mu }_{1}}Y^{2}+\frac{1}{2}X^{2}z^{\mathbf{\mu
}_{1}}Y+\frac{1}{4}X^{2}z^{\mathbf{\mu }_{1}}Y^{2}$%
\end{tabular}%
\end{equation}%
with higher monomial given by $X^{2}z^{\mathbf{\mu }_{1}}Y^{2}$. Replacing $%
z^{\mathbf{\mu }_{1}}=z\varrho _{+}+z^{-1}\varrho _{-}$ and using eq.(\ref%
{ct}) indicating that
\begin{equation}
Xz^{\mathbf{\mu }_{1}}=z^{-1}X\varrho _{-},\qquad z^{\mathbf{\mu }%
_{1}}Y=z^{-1}\varrho _{-}Y
\end{equation}%
the above L-operator reads as follows%
\begin{equation}
\begin{tabular}{lll}
$\mathcal{L}_{\boldsymbol{2N}}^{{\small vect}}$ & $=$ & $z\varrho
_{+}+z^{-1}\varrho _{-}+z^{-1}X\varrho _{-}+z^{-1}\varrho _{-}Y+$ \\
&  & $\frac{1}{2}z^{-1}\varrho _{-}Y^{2}+\frac{1}{2}z^{-1}X^{2}\varrho
_{-}+z^{-1}X\varrho _{-}Y$ \\
&  & $+\frac{1}{2}z^{-1}X\varrho _{-}Y^{2}+\frac{1}{2}z^{-1}X^{2}\varrho
_{-}Y+\frac{1}{4}z^{-1}X^{2}\varrho _{-}Y^{2}$%
\end{tabular}
\label{lv}
\end{equation}%
This operator has a remarkable dependence on the projector $\varrho _{-}$.
Using the non vanishing $\varrho _{+}X_{i}X_{j}\varrho _{-}=-\delta _{ij}E$
and $\varrho _{-}Y^{i}Y^{j}\varrho _{+}=-\delta ^{ij}F$ as well as $\varrho
_{+}X_{i}X_{j}\varrho _{-}Y^{k}Y^{l}\varrho _{+}=\delta _{ij}\delta
^{kl}\varrho _{+}$, we have%
\begin{equation}
\begin{tabular}{lll}
$\varrho _{+}\mathcal{L}\varrho _{+}$ & $=$ & $z\varrho _{+}+\frac{1}{4}%
z^{-1}\varrho _{+}X^{2}\varrho _{-}Y^{2}\varrho _{+}$ \\
$\varrho _{+}\mathcal{L}\Pi _{0}$ & $=$ & $\frac{1}{2}z^{-1}\varrho
_{+}X^{2}\varrho _{-}Y\Pi _{0}$ \\
$\varrho _{+}\mathcal{L}\varrho _{-}$ & $=$ & $\frac{1}{2}z^{-1}\varrho
_{+}X^{2}\varrho _{-}$%
\end{tabular}%
\end{equation}%
and%
\begin{equation}
\begin{tabular}{lll}
$\Pi _{0}\mathcal{L}\varrho _{+}$ & $=$ & $\frac{1}{2}z^{-1}\Pi _{0}X\varrho
_{-}Y^{2}\varrho _{+}$ \\
$\Pi _{0}\mathcal{L}\Pi _{0}$ & $=$ & $z^{-1}\Pi _{0}X\varrho _{-}Y\Pi
_{0}=z^{-1}b^{i}\mathrm{E}_{i}^{j}c_{j}$ \\
$\Pi _{0}\mathcal{L}\varrho _{-}$ & $=$ & $z^{-1}\Pi _{0}X\varrho _{-}$%
\end{tabular}%
\end{equation}%
with E$_{i}^{j}=\left \vert i\right \rangle \left \langle j\right \vert $,
and
\begin{equation}
\begin{tabular}{lll}
$\varrho _{-}\mathcal{L}\varrho _{+}$ & $=$ & $\frac{1}{2}z^{-1}\varrho
_{-}Y^{2}\varrho _{+}$ \\
$\varrho _{-}\mathcal{L}\Pi _{0}$ & $=$ & $z^{-1}\varrho _{-}Y\Pi _{0}$ \\
$\varrho _{-}\mathcal{L}\varrho _{-}$ & $=$ & $z^{-1}\varrho _{-}$%
\end{tabular}%
\end{equation}%
Substituting $X\varrho _{-}=-b^{i}x_{i}$ and $\varrho _{-}Y=-c_{i}y^{i}$ as
well as $X^{2}\varrho _{-}=-\mathbf{b}^{2}E$ and $\varrho _{-}Y^{2}=-\mathbf{%
c}^{2}F$ by help of \textrm{eqs.(\ref{y},\ref{2x},\ref{x}}), we obtain%
\begin{equation}
\begin{tabular}{lll}
$\mathcal{L}_{\boldsymbol{2N}}^{{\small vect}}$ & $=$ & $\left( z+\frac{1}{4}%
z^{-1}\mathbf{b}^{2}\mathbf{c}^{2}\right) \varrho _{+}+z^{-1}\varrho
_{-}-z^{-1}b^{i}x_{i}-z^{-1}c_{i}y^{i}+$ \\
&  & $-\frac{1}{2}z^{-1}\mathbf{c}^{2}F-\frac{1}{2}z^{-1}\mathbf{b}%
^{2}E+z^{-1}\left( b^{i}c_{j}\right) x_{i}y^{j}$ \\
&  & $+\frac{1}{2}z^{-1}\left( b^{i}\mathbf{c}^{2}\right) x_{i}F+\frac{1}{2}%
z^{-1}\mathbf{b}^{2}c_{i}Ey^{i}$%
\end{tabular}
\label{L1}
\end{equation}

\subsection{Topological quiver Q$_{\boldsymbol{2N}}^{{\protect\small vect}}:$
case of the vector 't Hooft line}

From the realisation eqs.(\ref{r1}-\ref{r2}) and the diagram of the Figure
\textbf{\ref{2N}}, we learn that the Lax operator $\mathcal{L}_{\boldsymbol{%
2N}}^{{\small vect}}$ has an intrinsic structure that can be represented by
a topological gauge quiver Q$_{\boldsymbol{2N}}^{{\small vect}}$. To draw
this topological quiver diagram, we use the projectors $\varrho _{+},\varrho
_{-}$ and $\Pi _{0}=\sum \varrho _{i},$ singling out the representations of
the Levi subgroup $SO_{2}\times SO_{2N-2}$ of the orthogonal symmetry $%
SO_{2N},$ to cast eq(\ref{lv}) as follows
\begin{equation}
\mathcal{L}_{\boldsymbol{2N}}^{{\small vect}}=\left(
\begin{array}{ccc}
\varrho _{+}\mathcal{L}\varrho _{+} & \varrho _{+}\mathcal{L}\Pi _{0} &
\varrho _{+}\mathcal{L}\varrho _{-} \\
\Pi _{0}\mathcal{L}\varrho _{+} & \Pi _{0}\mathcal{L}\Pi _{0} & \Pi _{0}%
\mathcal{L}\varrho _{-} \\
\varrho _{-}\mathcal{L}\varrho _{+} & \varrho _{-}\mathcal{L}\Pi _{0} &
\varrho _{-}\mathcal{L}\varrho _{-}%
\end{array}%
\right)
\end{equation}%
In this decomposition, we have used the relation $\varrho _{+}+\Pi
_{0}+\varrho _{-}=I_{id}$ and $\varrho _{\pm }\Pi _{0}=\varrho _{+}\varrho
_{-}=0$. Finally, we recover the matrix representation in agreement with
\cite{FrB}
\begin{equation}
\mathcal{L}_{\boldsymbol{2N}}^{{\small vect}}=z^{-1}\left(
\begin{array}{lll}
z^{2}+\frac{1}{4}\mathbf{b}^{2}\mathbf{c}^{2} & \frac{1}{2}\mathbf{b}%
^{2}c_{i} & -\frac{1}{2}\mathbf{b}^{2} \\
\frac{1}{2}b^{j}\mathbf{c}^{2} & b^{j}c_{i} & -b^{j} \\
-\frac{1}{2}\mathbf{c}^{2} & -c_{i} & 1%
\end{array}%
\right)  \label{L2}
\end{equation}%
\begin{figure}[h]
\begin{center}
\includegraphics[width=10cm]{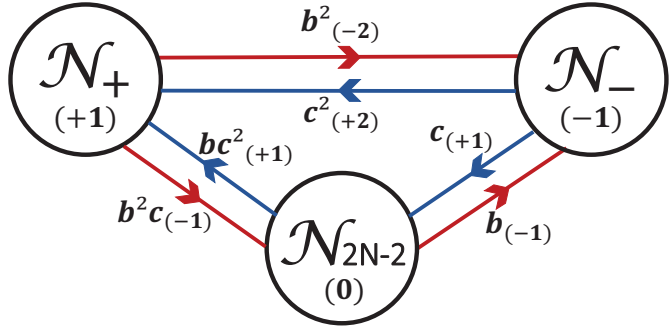}
\end{center}
\par
\vspace{-0.5cm}
\caption{The topological quiver representing $\mathcal{L}_{\boldsymbol{2N}}^{%
{\protect\small vect}}$. It has three nodes and 6 links. The nodes describe
self-dual topological matter and the links describe topological bi-matter$.$}
\label{fig4}
\end{figure}
\newline
The topological gauge quiver Q$_{\boldsymbol{2N}}^{{\small vect}}$
representing the above vector like $\mathcal{L}_{\boldsymbol{2N}}^{{\small %
vect}}$ is given by the Figure \textbf{\ref{fig4}}. The Q$_{\boldsymbol{2N}%
}^{{\small vect}}$ has three nodes $\mathcal{N}_{+},$ $\mathcal{N}_{2N-2}$
and $\mathcal{N}_{-}$ given by%
\begin{equation}
\mathcal{N}_{+}\equiv \left \langle \varrho _{+}\mathcal{L}\varrho
_{+}\right \rangle \qquad ,\qquad \mathcal{N}_{2N-2}\equiv \left \langle \Pi
_{0}\mathcal{L}\Pi _{0}\right \rangle \qquad ,\qquad \mathcal{N}_{-}\equiv
\left \langle \varrho _{-}\mathcal{L}\varrho _{-}\right \rangle
\end{equation}%
It has $3+3$ links $L_{ij}$ with $i,j=0,\pm $ interpreted as topological
\emph{bi-fundamental matter }$SO_{2}\times SO_{2N-2}$ reading as
\begin{equation}
\begin{tabular}{lllll}
$L_{+0}=\left \langle \varrho _{+}\mathcal{L}\Pi _{0}\right \rangle $ & , & $%
L_{0+}=\left \langle \Pi _{0}\mathcal{L}\varrho _{+}\right \rangle $ & , & $%
L_{-+}=\left \langle \varrho _{-}\mathcal{L}\varrho _{+}\right \rangle $ \\
$L_{+-}=\left \langle \varrho _{+}\mathcal{L}\varrho _{-}\right \rangle $ & ,
& $L_{0-}=\left \langle \Pi _{0}\mathcal{L}\varrho _{-}\right \rangle $ & ,
& $L_{-0}=\left \langle \varrho _{-}\mathcal{L}\Pi _{0}\right \rangle $%
\end{tabular}%
\end{equation}%
Notice that The Darboux coordinates can be expressed in terms of the
operator $\mathcal{L}_{\boldsymbol{2N}}^{{\small vect}}$ and the generators $%
X_{i},$ $Y^{i}$ and the minuscule coweight $\mathbf{\mu }_{1}$ as follows :%
\begin{equation}
b^{i}=zTr\left( \mathbf{\mu }_{1}Y^{i}\mathcal{L}_{\boldsymbol{2N}}^{{\small %
vect}}\right) ,\qquad c_{i}=zTr\left( \mathcal{L}_{\boldsymbol{2N}}^{{\small %
vect}}X_{i}\mathbf{\mu }_{1}\right)
\end{equation}%
While $b^{i}$ and $c_{i}$ sit respectively in the vector representation of $%
SO_{2N-2}$ and its transpose, they carry opposite unit charges under the
minuscule coweight $\mu _{1}$. \newline
As for the Darboux, we also have their composites that appear in the
expression of the L-operator, they are scalars of $SO_{2N-2}$ and carry non
trivial $SO_{2}$ charges. They are given by
\begin{equation}
\mathbf{b}^{2}=-2zTr\left( F\mathcal{L}_{\boldsymbol{2N}}^{{\small vect}%
}\right) ,\qquad \mathbf{c}^{2}=-2zTr\left( E\mathcal{L}_{\boldsymbol{2N}}^{%
{\small vect}}\right)
\end{equation}%
where E and F are related to the minuscule coweight operator as $\left[ E,F%
\right] =\mathbf{\mu }_{1}$. Interesting composites of the Darboux
coordinates that transform non trivially under $SO_{2}$ are given by
\begin{equation}
\mathbf{b}^{2}c_{i}=2zTr\left( \mathbf{\mu }_{1}F\mathcal{L}_{\boldsymbol{2N}%
}^{{\small vect}}X_{i}\right) ,\qquad b^{i}\mathbf{c}^{2}=2zTr\left( \mathbf{%
\mu }_{1}Y^{i}\mathcal{L}_{\boldsymbol{2N}}^{{\small vect}}E\right)
\end{equation}

\section{Spinorial 't Hooft lines of $D_{N}$- type}

This section is a continuation to the previous one, it concerns the
operators $\mathcal{L}_{\boldsymbol{2N}}^{{\small spin}}$. Here, we
introduce the two spinorial like 't Hooft lines of $D_{N}$ type denoted as tH%
$_{\mathrm{\gamma }_{0}}^{\mathbf{\mu }_{N-1}}$ and tH$_{\mathrm{\gamma }%
_{0}}^{\mathbf{\mu }_{N}}$ and construct associated Lax operators. We cast
their special properties in the associated topological quivers Q$_{%
\boldsymbol{R}}^{{\small spin}}$. We also treat exotic cases where the
electric charges are given by representations beyond the (anti)fundamental
of the $so_{2N}$ Lie algebra.

\subsection{'t Hooft line with magnetic charges $\protect\mu _{N-1}$ and $%
\protect\mu _{N}$}

\begin{figure}[h]
\begin{center}
\includegraphics[width=10cm]{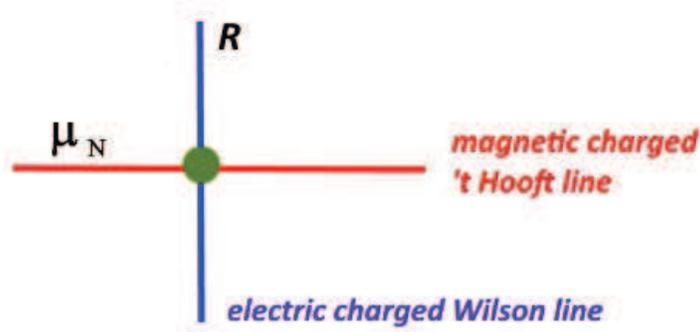}
\end{center}
\par
\vspace{-0.5cm}
\caption{A horizontal 't Hooft line of D- type with spinor-like magnetic
charge given by the minuscule coweight $\protect\mu _{N}$ of $SO_{2N}$
couples to a vertical Wilson line characterized by a representation $%
\boldsymbol{R}$ of $so_{2N}$.}
\label{hw}
\end{figure}

\begin{figure}[h]
\begin{center}
\includegraphics[width=10cm]{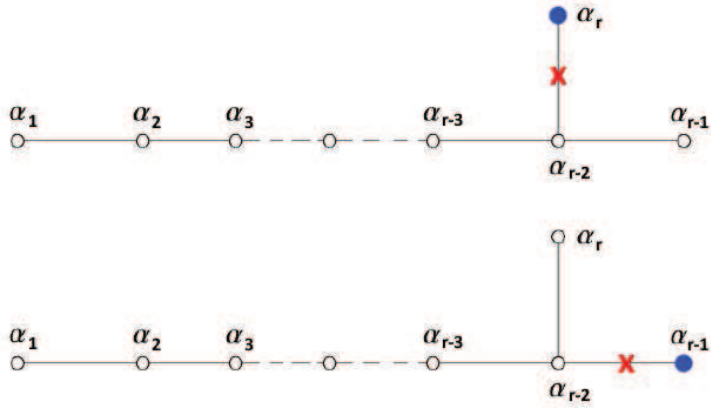}
\end{center}
\par
\vspace{-0.5cm}
\caption{Dynkin diagram of D$_{N}$ Lie algebras where the two Levi
decompositions with respect to the spinorial coweights are illustrated by : $%
(a)$ removing the simple root $\protect\alpha _{N}$ for the minuscule
coweight $\protect\mu _{N}.$ $\left( b\right) $ removing the simple root $%
\protect\alpha _{N-1}$ for the minuscule coweight $\protect\mu _{N-1}.$}
\label{ds}
\end{figure}

Besides the vectorial $\mu _{1}=e_{1}$ given by eq.(\ref{e1}), the $SO_{2N}$
has moreover two other minuscule coweights $\mu _{N-1}$ and $\mu _{N}.$
These coweights yield the magnetic charges of the two spinorial-like 't
Hooft lines :%
\begin{equation*}
\begin{tabular}{lll}
tH$_{\mathrm{\gamma }_{0}}^{\mathbf{\mu }_{N-1}}$ & , & tH$_{\mathrm{\gamma }%
_{0}}^{\mathbf{\mu }_{N}}$%
\end{tabular}%
\end{equation*}%
These line defects are represented similarly to the vector-like line tH$_{%
\mathrm{\gamma }_{0}}^{\mathbf{\mu }_{1}}$ of previous section as depicted
in Figure \textbf{\ref{hw}} where the tH$_{\mathrm{\gamma }_{0}}^{\mathbf{%
\mu }_{N}}$ couples to a vertical Wilson line $W_{\mathrm{\xi }_{z}}^{%
\boldsymbol{R}}$ carrying internal states $\left \vert A\right \rangle $
belonging to some representation $\boldsymbol{R}$ of $so_{2N}.$ Interesting
candidates for $\boldsymbol{R}$ are given by the vectorial and the
spinorials, namely%
\begin{equation}
\boldsymbol{R}=2N\qquad ,\qquad \boldsymbol{R}=2_{L}^{N-1}\qquad ,\qquad
\boldsymbol{R}=2_{R}^{N-1}\qquad ,\qquad \boldsymbol{R}=2^{N}  \label{R}
\end{equation}%
\newline
So depending on the electric charge of the Wilson line $W_{\mathrm{\xi }%
_{z}}^{\boldsymbol{R}}$, one distinguishes various kinds of $\mathcal{L}$%
-operators that generally speaking, can be labeled as follows
\begin{equation}
\mathcal{L}_{\boldsymbol{R}}^{\mathbf{\mu }_{s}}=e^{X_{\boldsymbol{R}%
}}z^{\mu _{s}}e^{Y_{\boldsymbol{R}}}  \label{lr}
\end{equation}%
with a spinor-like minuscule coweight $\mu _{s}$ of $SO_{2N}$. For an
electric representation $\boldsymbol{R}$, we have a Lax operator $\mathcal{L}%
_{\boldsymbol{R}}^{\mathbf{\mu }}$ described by a $\dim _{\boldsymbol{R}%
}\times \dim _{\boldsymbol{R}}$ matrix whose entries are functions of the
Darboux coordinates. These phase space coordinates labeled as $\left( b^{%
\left[ ij\right] },c_{\left[ ij\right] }\right) $ appear in the expression
of the $X_{\boldsymbol{R}}$ and $Y_{\boldsymbol{R}}$ as follows
\begin{equation}
X_{\boldsymbol{R}}=b^{\left[ ij\right] }X_{\left[ ij\right] }^{\boldsymbol{R}%
}\qquad ,\qquad Y_{\boldsymbol{R}}=c_{\left[ ij\right] }Y_{\boldsymbol{R}}^{%
\left[ ij\right] }
\end{equation}%
where the $X_{\left[ ij\right] }^{\boldsymbol{R}}$ and $Y_{\boldsymbol{R}}^{%
\left[ ij\right] }$ are generators of the nilpotent subalgebras $\mathbf{n}%
_{\pm }^{\boldsymbol{R}}$ issued from the Levi decomposition of $so_{2N}$.
In fact, for the spinor-like coweights $\mu _{s}=\mu _{N-1}$ or $\mu _{N}$,
we have the following Levi decomposition of $so_{2N}$

\begin{equation}
so_{2N}\rightarrow \boldsymbol{l}_{\mu _{s}}\oplus \boldsymbol{n}_{+}\oplus
\boldsymbol{n}_{-}
\end{equation}%
with $\boldsymbol{l}_{\mu _{s}}=gl_{N}$. This can be directly read from the
Figure \textbf{\ref{ds}} where we see that the fundamental coweight $\mu
_{N-1}$ is the dual of the simple root $\alpha _{N-1}=e_{N-1}-e_{N}$, while $%
\mu _{N}$ is the dual of $\alpha _{N}=e_{N-1}+e_{N}$.
\newline
Notice that by cutting the root $\alpha _{N-1}$ from \textbf{\ref{ds}-a}, we
end up with the Dynkin diagram of an $sl_{N}$ Lie algebra with the following
simple roots :%
\begin{equation}
\alpha _{1},...,\alpha _{N-2};\alpha _{N}
\end{equation}%
And if instead, we cut the root $\alpha _{N}$ as in \textbf{\ref{ds}-b}, we
also end up with the Dynkin diagram of an $sl_{N}^{\prime }$ Lie algebra
having the simple roots :%
\begin{equation}
\alpha _{1},...,\alpha _{N-2};\alpha _{N-1}
\end{equation}%
The two $sl_{N}$ and $sl_{N}^{\prime }$ are isomorphic, they are related by
the exchange $\alpha _{N}\leftrightarrow \alpha _{N-1}$. We can therefore
focus our analysis on the minuscule tH$_{\mathrm{\gamma }_{0}}^{\mu _{N}}$
since the calculations are similar for tH$_{\mathrm{\gamma }_{0}}^{\mu
_{N-1}}$. Notice however that the expressions of the coweights in terms of
the $e_{i}$ weight vector basis are given by
\begin{equation}
\begin{tabular}{rrr}
$\mu _{N-1}$ & $=$ & $\frac{1}{2}\left( e_{1}+...+e_{N-1}-e_{N}\right) $ \\
$\mu _{N}$ & $=$ & $\frac{1}{2}\left( e_{1}+...+e_{N-1}+e_{N}\right) $%
\end{tabular}
\label{m10}
\end{equation}

\subsection{ Magnetic charge $\protect\mu _{N}$ and the link between $%
SO_{2N} $ and $SL_{N}$}

Here, we study the Levi decomposition of $so_{2N}$ with respect to $\mu _{N}$
in order to explore intrinsic aspects of the coupling between the minuscule
tH$_{\mathrm{\gamma }_{0}}^{\mu _{N}}$ and the Wilson line in a
representation $\boldsymbol{R}$ of $so_{2N}$ that is usually taken as the
vectorial $\boldsymbol{2N}$. Particularly, we extend the results here for
Wilson lines in the spinorial representation $\boldsymbol{2}^{N}$ where we
build the graphic representation of their remarkable coupling with 't Hooft
lines; see Figure \textbf{\ref{ths}-(a)}.

\subsubsection{Spinorial 't Hooft line tH$_{\mathrm{\protect\gamma }_{0}}^{%
\protect\mu _{N}}$}

As shown by the Figure \textbf{\ref{ds}-a }without $\alpha _{N}$, there is a
close relationship between $SO_{2N}$ and $SL_{N}.$ It is given by the Levi
decomposition $so_{2N}\rightarrow \boldsymbol{l}_{\mu _{N}}\oplus
\boldsymbol{n}_{+}\oplus \boldsymbol{n}_{-}$ with respect to the coweight $%
\mu _{N}$ of the $SO_{2N}$ gauge symmetry of the CS theory. In this
decomposition, we have the following dimension splitting%
\begin{equation}
N\left( 2N-1\right) =N^{2}+\frac{1}{2}N\left( N-1\right) +\frac{1}{2}N\left(
N-1\right)
\end{equation}%
and the subalgebra structures%
\begin{equation}
\begin{tabular}{lll}
$\boldsymbol{l}_{\mu _{N}}$ & $=$ & $sl_{1}\oplus sl_{N}$ \\
$\boldsymbol{n}_{+}$ & $\boldsymbol{=}$ & $\boldsymbol{N}_{+\frac{1}{2}%
}\wedge \boldsymbol{N}_{+\frac{1}{2}}$ \\
$\boldsymbol{n}_{-}$ & $\boldsymbol{=}$ & $\boldsymbol{N}_{-\frac{1}{2}%
}\wedge \boldsymbol{N}_{-\frac{1}{2}}$%
\end{tabular}%
\end{equation}%
with $sl_{1}\oplus sl_{N}\sim gl_{N}$ and $\left[ sl_{1},\boldsymbol{n}_{\pm
}\right] =\boldsymbol{n}_{\pm }$ indicating that
\begin{equation}
\left[ sl_{1},\boldsymbol{N}_{\pm \frac{1}{2}}\right] =\pm \frac{1}{2}%
\boldsymbol{N}_{\pm \frac{1}{2}}
\end{equation}%
We also have the $\boldsymbol{R}_{so_{2N}}$ representations' splitting%
\begin{equation}
\begin{tabular}{|l|l|}
\hline
repres $\boldsymbol{R}_{so_{2N}}$ & repres $\boldsymbol{R}_{gl_{N}}$ \\
\hline\hline
$\mathbf{2}\boldsymbol{N}$ & $\boldsymbol{N_{+\frac{1}{2}}}\oplus
\boldsymbol{N}_{-\frac{1}{2}}\boldsymbol{\ \ }$ \\ \hline
$\mathbf{2}\boldsymbol{N}\wedge \mathbf{2}\boldsymbol{N}$ & $adj_{0}\oplus
\left( \boldsymbol{N_{+\frac{1}{2}}}\wedge \boldsymbol{N_{+\frac{1}{2}}}%
\right) \oplus \left( \boldsymbol{N}_{-\frac{1}{2}}\wedge \boldsymbol{N}_{-%
\frac{1}{2}}\right) \boldsymbol{\ \ }$ \\ \hline
$\mathbf{2}\boldsymbol{N}\vee \mathbf{2}\boldsymbol{N}$ & $adj_{0}\oplus
\left( \boldsymbol{N_{+\frac{1}{2}}}\vee \boldsymbol{N_{+\frac{1}{2}}}%
\right) \oplus \left( \boldsymbol{N}_{-\frac{1}{2}}\vee \boldsymbol{N_{+%
\frac{1}{2}}}\right) $ \\ \hline
$\mathbf{2}^{N}$ & $\boldsymbol{\oplus }_{k=0}^{N}\boldsymbol{N}%
_{q_{k}}^{\wedge k}$ \\ \hline\hline
\end{tabular}
\label{red}
\end{equation}%
where $\mathbf{2}\boldsymbol{N}$ describes vector- like states, $\mathbf{2}%
\boldsymbol{N}\wedge \mathbf{2}\boldsymbol{N}$ the antisymmetric (adjoint)
and $\mathbf{2}\boldsymbol{N}\vee \mathbf{2}\boldsymbol{N}$ the symmetric.
The $\mathbf{2}^{N}$ states describe a Dirac-type spinor reducible into left
handed and right handed Weyl spinors as follows
\begin{equation}
\mathbf{2}^{N}\mathbf{=2}_{L}^{N-1}\oplus \mathbf{2}_{R}^{N-1}  \label{2s}
\end{equation}%
Notice that the wedge product $\wedge ^{k}\boldsymbol{N}$ is the k-th
anti-symmetrisation order (for short $\boldsymbol{N}^{\wedge k}$) of the
tensor product of k representation $\boldsymbol{N}\mathbf{.}$ Its dimension
is equal to $\frac{N!}{\left( N-k\right) !k!}.$ As illustrating examples of
the degrees of freedom described by such wedge products, we give below the
reductions associated with the leading gauge symmetry groups%
\begin{equation}
\begin{tabular}{|l|l|l|l|l|}
\hline
$so_{2N}$ & $\mathbf{2}\boldsymbol{N}$ & $\mathbf{2}^{N}$ & $\mathbf{2}%
_{L}^{N-1}$ & $\mathbf{2}_{R}^{N-1}$ \\ \hline\hline
$so_{6}$ & $\mathbf{6}$ & $\mathbf{8}$ & $\mathbf{4}_{L}$ & $\mathbf{4}_{R}$
\\ \hline
$so_{8}$ & $\mathbf{8}$ & $\mathbf{16}$ & $\mathbf{8}_{L}$ & $\mathbf{8}_{R}$
\\ \hline
$so_{10}$ & $\mathbf{10}$ & $\mathbf{32}$ & $\mathbf{16}_{L}$ & $\mathbf{16}%
_{R}$ \\ \hline
$so_{12}$ & $\mathbf{12}$ & $\mathbf{64}$ & $\mathbf{32}_{L}$ & $\mathbf{32}%
_{R}$ \\ \hline\hline
\end{tabular}%
\end{equation}%
where we have also given the $so_{6}$ which is isomorphic to $sl_{4}$ with
no Levi charge operator $sl_{1}$. The Levi decompositions with respect to $%
\mu _{N}$ of the above spinorial representations $\mathbf{2}^{N}$ are given
by the sum of two blocks: $\left( i\right) $ the first block involving the
even powers $\boldsymbol{N}^{\wedge 2l}$, it corresponds to Weyl spinor; say
$\mathbf{2}_{L}^{N-1}$. $\left( ii\right) $ the second block having the odd
powers $\boldsymbol{N}^{\wedge 2l+1}$ and corresponding to $\mathbf{2}%
_{R}^{N-1}.$ So, we have:%
\begin{equation}
\begin{tabular}{|l|l|l|}
\hline
$so_{2N}$ & $\ \ \ \ \ \ \ \ \ \ \ \ \ \ \ \mathbf{2}_{L}^{N-1}$ & $\ \ \ \
\ \ \ \ \ \ \ \ \ \ \ \mathbf{2}_{R}^{N-1}$ \\ \hline\hline
$so_{6}$ & $\mathbf{4}_{L}=1+3^{\wedge 2}$ & $\mathbf{4}_{R}=3^{\wedge
1}+3^{\wedge 3}$ \\ \hline
$so_{8}$ & $\mathbf{8}_{L}=1+4^{\wedge 2}+4^{\wedge 4}$ & $\mathbf{8}%
_{R}=4^{\wedge 1}+4^{\wedge 3}$ \\ \hline
$so_{10}$ & $\mathbf{16}_{L}=1+5^{\wedge 2}+5^{\wedge 4}$ & $\mathbf{16}%
_{R}=5^{\wedge 1}+5^{\wedge 3}+5^{\wedge 5}$ \\ \hline
$so_{12}$ & $\mathbf{32}_{L}=1+6^{\wedge 2}+6^{\wedge 4}+6^{\wedge 6}$ & $%
\mathbf{32}_{R}=6^{\wedge 1}+6^{\wedge 3}+6^{\wedge 5}$ \\ \hline\hline
\end{tabular}%
\end{equation}%
By assuming the $\mathbf{2}_{L}^{N-1}$ and the $\mathbf{2}_{R}^{N-1}$ as
traceless, we can exhibit the Levi charges in the above relations leading to
\begin{equation}
\begin{tabular}{|l|l|l|l|l|}
\hline
$so_{2N}$ & \multicolumn{2}{|l}{$\ \ \ \ \ \ \ \ \ \ \ \ \ \ \ \mathbf{2}%
_{L}^{N-1}$} & \multicolumn{2}{|l|}{$\ \ \ \ \ \ \ \ \ \ \ \ \ \ \ \mathbf{2}%
_{R}^{N-1}$} \\ \hline\hline
$so_{6}$ & $\mathbf{4}_{L}$ & ${\small =1}_{+3/4}{\small +3}_{-1/4}$ & $%
\mathbf{4}_{R}$ & ${\small =3}_{+1/4}{\small +1}_{-3/4}$ \\ \hline
$so_{8}$ & $\mathbf{8}_{L}$ & ${\small =1}_{+1}{\small +6}_{0}{\small +1}%
_{-1}$ & $\mathbf{8}_{R}$ & ${\small =4}_{+1/2}{\small +4}_{-1/2}$ \\ \hline
$so_{10}$ & $\mathbf{16}_{L}$ & ${\small =1}_{5/4}{\small +10}_{+1/4}{\small %
+5}_{-3/4}$ & $\mathbf{16}_{R}$ & ${\small =5}_{+3/4}{\small +10}_{-1/4}%
{\small +1}_{-5/4}$ \\ \hline
$so_{12}$ & $\mathbf{32}_{L}$ & ${\small =1}_{+3/2}{\small +15}_{+1/2}%
{\small +15}_{-1/2}{\small +1}_{-3/2}$ & $\mathbf{32}_{R}$ & ${\small =6}%
_{+1}{\small +20}_{0}{\small +6}_{-1}$ \\ \hline\hline
\end{tabular}
\label{rf}
\end{equation}%
Thanks to the reduction of $so_{2N}$ representations in terms of gl$_{N}$
ones like in eqs.(\ref{red}), one can construct various kinds of
spinorial-like Lax couplings depending on the electric representation $%
\boldsymbol{R}$ hosted by the Wilson line $W_{\mathrm{\gamma }_{z}}^{%
\boldsymbol{R}}$ crossing the tH$_{\mathrm{\gamma }_{0}}^{\mu _{N}}$ line.
Two of such couplings are studied here below:

$\bullet $ \emph{Case of electric }$\boldsymbol{R}_{s}=\mathbf{2}^{N}$%
\begin{figure}[h]
\begin{center}
\includegraphics[width=12cm]{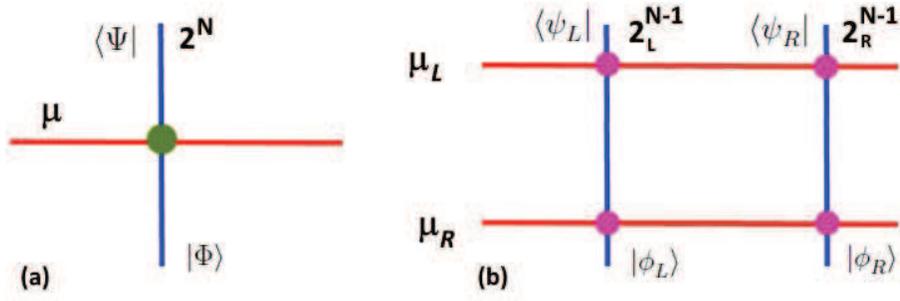}
\end{center}
\par
\vspace{-0.5cm}
\caption{On the left, a horizontal spinorial like 't Hooft line crossing a
vertical Wilson line carrying internal fermionic states $\Psi =\left(
\protect\psi _{L},\protect\psi _{R}\right) .$ On the right, the structure of
the coupling under the Levi decomposition showing chiral and antichiral Weyl
states traveling along vertical lines.}
\label{ths}
\end{figure}
\newline
In this case, the coupling is given by the interaction between the spinorial
tH$_{\mathrm{\gamma }_{0}}^{\mu _{N}}$ and a Wilson $W_{\mathrm{\gamma }%
_{z}}^{\boldsymbol{R}}$ line with electric representation $\boldsymbol{R}%
_{s}=\mathbf{2}^{N}$ as illustrated by the Figure \textbf{\ref{ths}-(a)}.
The quantum states propagating along the vertical Wilson line form a Dirac
spinor $\Psi =\Psi _{L}\oplus \Psi _{R}.$ By using the projector $\Pi _{L}$
on the left handed spinor and the projector $\Pi _{R}$ on the right handed
one, we can use the properties $\Pi _{L}+\Pi _{R}=I_{id}$\ and $\Pi _{L}\Pi
_{R}=0$ to decompose the action of the minuscule coweight on $\mathbf{2}^{N}$
like
\begin{equation}
\mathbf{\mu }=\Pi _{L}\mathbf{\mu }+\Pi _{R}\mathbf{\mu }\qquad
\leftrightarrow \qquad \mathbf{\mu }=\mathbf{\mu }_{L}+\mathbf{\mu }_{R}
\end{equation}%
This splitting is illustrated by the Figure \textbf{\ref{ths}-(b)} where the
states propagating in the two vertical Wilson lines are given by the left
handed $\Psi _{L}$ and the right handed $\Psi _{R}$ Weyl spinors. In this
case, the L-operator decomposes into four blocks as follows%
\begin{equation}
\mathcal{L}_{\boldsymbol{R}_{s}}^{\mathbf{\mu }_{N}}=\left(
\begin{array}{cc}
\Pi _{L}\mathcal{L}\Pi _{L} & \Pi _{L}\mathcal{L}\Pi _{R} \\
\Pi _{R}\mathcal{L}\Pi _{L} & \Pi _{R}\mathcal{L}\Pi _{R}%
\end{array}%
\right)
\end{equation}%
Notice that in this expression of $\mathcal{L}_{\boldsymbol{R}_{s}}^{\mathbf{%
\mu }_{N}}$, we have not yet implemented the Levi decomposition; we have
only exhibited the chiral and anti-chiral structure of the Dirac spinor. To
implement the effect of the Levi decomposition, we introduce other types of
projectors
\begin{equation}
P_{\mathbf{N}^{\wedge k}}=\left \vert \mathbf{N}^{\wedge k}\right \rangle
\left \langle \mathbf{N}^{\wedge k}\right \vert
\end{equation}%
that give the reduction $\mathbf{2}^{N}=\oplus _{k}\mathbf{N}^{\wedge k}$
and eqs.(\ref{red}-\ref{rf}). \textrm{This leads to a more complicated
structure of this specific type of coupling; we will come back to this case
later for further development.}

$\bullet $ \emph{Case of electric} $\boldsymbol{R}_{v}=\mathbf{2}N$%
\begin{figure}[h]
\begin{center}
\includegraphics[width=12cm]{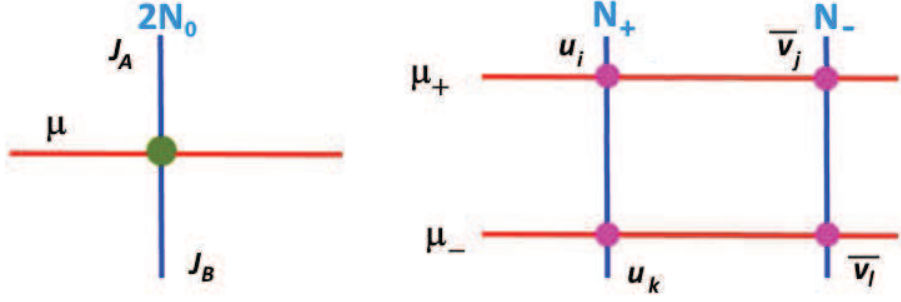}
\end{center}
\par
\vspace{-0.5cm}
\caption{On the left, a horizontal spinorial like 't Hooft line crossing a
vertical Wilson line carrying a bosonic current $J^{A}=\Psi \Gamma ^{A}\Psi
. $ On the right, The splitting of the current into two currents $u_{i}=\Psi
\Upsilon _{i}\Phi $ and $\bar{v}^{i}=\Psi \bar{\Upsilon}^{i}\Phi $ traveling
along the vertical lines.}
\label{th2s}
\end{figure}
\newline
In this case, the spinorial tH$_{\mathrm{\gamma }_{0}}^{\mu _{N}}$ crosses a
Wilson $W_{\mathrm{\xi }_{z}}^{\boldsymbol{R}}$ line with electric
representation $\boldsymbol{R}=\mathbf{2}N$ as shown by the Figure \textbf{%
\ref{th2s}-(a).} This representation can be related to the previous $%
\boldsymbol{R}_{s}=\mathbf{2}^{N}$ because it can also be viewed as an $%
so_{2N}$ electric current $J_{A}$ given by the Dirac bi-linear like
\begin{equation}
J_{A}=\left \langle \Psi \right \vert \Gamma _{A}\left \vert \Phi \right
\rangle  \label{vp}
\end{equation}%
where the 2N Gammas $\Gamma _{A}$ are $2^{N}\times 2^{N}$ Dirac matrices. To
deal with this $so_{2N}$ Wilson lines, it is interesting to use the new basis%
\begin{equation}
\Upsilon _{l}=\frac{1}{\sqrt{2}}\left( \Gamma _{l}+i\Gamma _{N+1}\right)
,\qquad \bar{\Upsilon}^{l}=\frac{1}{\sqrt{2}}\left( \Gamma _{l}-i\Gamma
_{N+1}\right)
\end{equation}%
Then by putting back into (\ref{vp}), we find that the $so_{2N}$ electric
current $J_{A}$ decomposes as two (covariant and contravariant) $gl_{N}$
currents given by%
\begin{equation}
u_{i}=\Psi \Upsilon _{i}\Phi \qquad ,\qquad \bar{v}^{i}=\Psi \bar{\Upsilon}%
^{i}\Phi  \label{58}
\end{equation}%
The $u_{i}$ transforms in the fundamental $\boldsymbol{N}_{+}$ of the Levi
subalgebra $gl_{N}$; and the $\bar{v}^{i}$ transforms in the anti-
fundamental $\boldsymbol{N}_{-}.$ Using the projector $\varrho _{+}$ on the $%
\boldsymbol{N}_{+}$ and the projector $\varrho _{-}$ on the $\boldsymbol{N}%
_{-},$ we can express the L-operator as follows, see also Figure \textbf{\ref%
{th2s}-(b).}%
\begin{equation}
\mathcal{L}_{\boldsymbol{2N}}^{\mathbf{\mu }_{N}}=\left(
\begin{array}{cc}
\varrho _{+}\mathcal{L}\varrho _{+} & \varrho _{+}\mathcal{L}\varrho _{-} \\
\varrho _{-}\mathcal{L}\varrho _{+} & \varrho _{-}\mathcal{L}\varrho _{-}%
\end{array}%
\right)  \label{mve}
\end{equation}

\subsubsection{Levi and nilpotent subalgebras within $so_{2N}$}

To model properties of the spinorial 't Hooft lines in 4D Chern-Simons
theory with $SO_{2N}$ symmetry characterized by the following Levi
decomposition with respect to $\mu _{n}$
\begin{equation}
so_{2N}\rightarrow \left( \boldsymbol{N}_{-}\wedge \boldsymbol{N}_{-}\right)
\oplus gl_{N}\oplus \left( \boldsymbol{N}_{+}\wedge \boldsymbol{N}_{+}\right)
\label{so2n}
\end{equation}%
where $\boldsymbol{N}_{\pm }$ stand for $\boldsymbol{N}_{\pm 1/2}$, it is
interesting to recall some useful tools concerning the euclidian Dirac
spinors in higher dimensions and the algebra of Gamma matrices.\newline
In even 2N dimensions, the Dirac spinor $\left \vert \psi _{{\small Dirac}%
}\right \rangle $ has 2$^{N}$ components and decomposes as a sum of two Weyl
spinors like $\left \vert \psi _{L}\right \rangle +\left \vert \psi
_{R}\right \rangle $ where
\begin{equation}
\begin{tabular}{lll}
$\left \vert \psi _{L}\right \rangle $ & $=$ & $\Pi _{L}\left \vert \psi _{%
{\small Dirac}}\right \rangle $ \\
$\left \vert \psi _{R}\right \rangle $ & $=$ & $\Pi _{R}\left \vert \psi _{%
{\small Dirac}}\right \rangle $%
\end{tabular}%
\end{equation}%
and%
\begin{equation}
\begin{tabular}{lll}
$\Pi _{L}$ & $=$ & $\frac{1}{2}\left( I+\Gamma _{2N+1}\right) $ \\
$\Pi _{r}$ & $=$ & $\frac{1}{2}\left( I-\Gamma _{2N+1}\right) $%
\end{tabular}%
\end{equation}%
The $\psi _{L}$ and $\psi _{R}$ are Weyl spinors transforming in $%
2_{L}^{N-1} $ and $2_{R}^{N-1}$ while the $\Pi _{L}$ and the $\Pi _{R}$\ are
the spin projectors encountered earlier reading as follows
\begin{equation}
\Gamma _{L}=\left(
\begin{array}{cc}
\boldsymbol{I} & \boldsymbol{0} \\
\boldsymbol{0} & \boldsymbol{0}%
\end{array}%
\right) ,\qquad \Gamma _{R}=\left(
\begin{array}{cc}
\boldsymbol{0} & \boldsymbol{0} \\
\boldsymbol{0} & \boldsymbol{I}%
\end{array}%
\right)
\end{equation}%
The identity and the zeros appearing in these matrices live in $2^{N-1}$
dimensions. The $\Gamma _{2N+1}$ is the chiral operator given by
\begin{equation}
\Gamma _{A_{1}}\Gamma _{A_{2}}....\Gamma _{A_{2N}}=\left( i\right)
^{N}\varepsilon _{A_{1}.....A_{2N}}\Gamma _{2N+1}  \label{5}
\end{equation}%
where $\varepsilon _{A_{1}.....A_{2N}}$ is the completely antisymmetric
tensor with $\varepsilon _{1...2N}=1$ and $\Gamma _{A}$ obeying the Clifford
algebra of a 2N dimension euclidian space$.$%
\begin{equation}
\Gamma _{A}\Gamma _{B}+\Gamma _{B}\Gamma _{A}=2\delta _{AB}
\end{equation}%
The relations (\ref{so2n}) and (\ref{5}) allow to split the 2N Gamma
matrices $\Gamma _{A}$ into two subsets that will be used later to construct
a new basis for the Gammas that is compatible with $gl_{N}$,%
\begin{equation}
\left.
\begin{array}{l}
\Gamma _{i} \\
\Gamma _{N+i}%
\end{array}%
\right. ,\qquad i=1,...,N
\end{equation}%
Recall also that the generators $J_{\left[ AB\right] }$ of the $so_{2N}$
spinor representation are defined by the commutators
\begin{equation}
\Gamma _{AB}=\frac{1}{2i}\left[ \Gamma _{A},\Gamma _{B}\right]
\end{equation}%
As for $sl_{1}\oplus sl_{N}$, the $so_{2N}$ algebra also has N commuting
diagonal generators $H_{l}$ realised in terms of the Gamma matrices as
\begin{equation}
H_{l}=\frac{1}{2i}\left[ \Gamma _{l},\Gamma _{N+l}\right] =-i\Gamma
_{l}\Gamma _{N+l}\text{\qquad },\qquad l=1,...,N
\end{equation}%
To exhibit the realisation of the $sl_{1}\oplus sl_{N}$ representations
within the $so_{2N}$ orthogonal symmetry group, we substitute the spliting $%
\Gamma _{A}=(\Gamma _{i},\Gamma _{N+l})$ into the $N\left( 2N-1\right) $
generators $\Gamma _{AB}$ of $so_{2N}$ and we obtain the following
antisymmetric $2\times 2$ block matrix%
\begin{equation}
\Gamma _{AB}=\left(
\begin{array}{cc}
\Gamma _{\left[ ij\right] } & \hat{\Gamma}_{i}^{j} \\
-\hat{\Gamma}_{j}^{i} & \tilde{\Gamma}^{\left[ ij\right] }%
\end{array}%
\right)  \label{gdec}
\end{equation}%
This decomposition contains: \newline
$\left( \mathbf{a}\right) $ the $N^{2}$ operators $\hat{\Gamma}_{i}^{j}$
generating $N_{+}\otimes N_{-}$ of the Levi subalgebra $sl_{1}\oplus sl_{N}.$%
\newline
$\left( \mathbf{b}\right) $ the $\frac{1}{2}N\left( N-1\right) $ operators $%
\Gamma _{\left[ ij\right] }$ generating the $N_{+}\wedge N_{+}$ nilpotent
subalgebras.\newline
$\left( \mathbf{c}\right) $ the $\frac{1}{2}N\left( N-1\right) $ operators $%
\bar{\Gamma}^{\left[ ij\right] }$ generating the $N_{-}\wedge N_{-}$ dual
nilpotent subalgebra.

\subsection{Nilpotent subalgebras and L-operator}

In order to explicitly realise the generators $\Gamma _{\left[ ij\right] },%
\tilde{\Gamma}^{\left[ ij\right] }$ and $\hat{\Gamma}_{i}^{j}$ appearing in
the decomposition (\ref{gdec}) and consequently the generators $X_{\left[ ij%
\right] }$ and $Y^{\left[ kl\right] }$ of the nilpotent subalgebras $\mathbf{%
n}_{\pm }$, we first think of the set of the $2N$ Dirac matrices $\Gamma
_{A}=(\Gamma _{i},\Gamma _{N+1})$ as follows,%
\begin{equation}
\Upsilon _{l}=\frac{1}{\sqrt{2}}\left( \Gamma _{l}+i\Gamma _{N+1}\right)
,\qquad \bar{\Upsilon}^{l}=\frac{1}{\sqrt{2}}\left( \Gamma _{l}-i\Gamma
_{N+1}\right)
\end{equation}%
This new Gamma matrix basis satisfy the Clifford algebra
\begin{equation}
\begin{tabular}{lll}
$\Upsilon _{i}\bar{\Upsilon}^{j}+\bar{\Upsilon}^{j}\Upsilon _{i}$ & $=$ & $%
2\delta _{i}^{j}$ \\
$\Upsilon _{i}\Upsilon _{j}+\Upsilon _{j}\Upsilon _{i}$ & $=$ & $0$ \\
$\bar{\Upsilon}^{k}\bar{\Upsilon}^{l}+\bar{\Upsilon}^{l}\bar{\Upsilon}^{k}$
& $=$ & $0$%
\end{tabular}%
\end{equation}%
Then, we consider the two $gl_{N}$ vector currents $u_{i}=\left \langle \xi
|\Upsilon _{i}|\psi \right \rangle $ and $\bar{v}^{i}=\left \langle \psi |%
\bar{\Upsilon}^{i}|\xi \right \rangle $ of eq(\ref{58}) constructed out of
bilinears of the Dirac fermions and use them to construct $\Gamma _{\left[ ij%
\right] },\tilde{\Gamma}^{\left[ ij\right] }$ and $\hat{\Gamma}_{i}^{j}$.
These two currents transform in the $\boldsymbol{N}_{+}$ and $\boldsymbol{N}%
_{-}$ representation of $sl_{1}\oplus sl_{N}$.

\subsubsection{Realising the nilpotent generators of\emph{\ }$\boldsymbol{n}%
_{\pm }$}

First, using the $N+N$ complex variables $u_{i}$ and $\bar{v}^{i},$ we build
the translation operators $\bar{\partial}^{i}=\partial /\partial u_{i}$ and $%
\partial _{i}=\partial /\partial \bar{v}^{i}$ as well as the rotations
\begin{equation}
\begin{tabular}{rrrrrrr}
$X_{\left[ ij\right] }$ & $=$ & $u_{i}\partial _{j}-u_{j}\partial _{i}$ & ,
& $Z_{i}^{l}$ & $=$ & $u_{i}\bar{\partial}^{l}-\bar{v}^{l}\partial _{i}$ \\
$Y^{\left[ ij\right] }$ & $=$ & $\bar{v}^{i}\bar{\partial}^{j}-\bar{v}^{j}%
\bar{\partial}^{i}$ & , & $H$ & $=$ & $\frac{1}{2}Tr\left( Z_{i}^{l}\right) $%
\ \ \ \ \
\end{tabular}
\label{op}
\end{equation}%
In these relations, the operator
\begin{equation}
H=\frac{1}{2}\sum_{i}\left( u_{i}\bar{\partial}^{i}-\bar{v}^{i}\partial
_{i}\right)  \label{po}
\end{equation}%
is the charge generator of $sl_{1}$. It acts on the complex variables like
\begin{equation}
Hu_{i}=+\frac{1}{2}u_{i},\qquad H\bar{v}^{i}=-\frac{1}{2}\bar{v}^{i}
\end{equation}%
We also have $X_{\left[ ij\right] }\bar{v}^{l}=\delta _{j}^{l}u_{i}-\delta
_{i}^{l}u_{j}$ and $Y^{\left[ ij\right] }u_{l}=\delta _{l}^{j}\bar{v}%
^{i}-\delta _{l}^{i}\bar{v}^{j}$ as well as $Z_{i}^{j}u_{k}=u_{i}\delta
_{k}^{j}$ and $Z_{i}^{j}\bar{v}^{l}=-\bar{v}^{j}\delta _{i}^{l}$. The above
operators (\ref{op}-\ref{po}) obey interesting commutation relations such as
\begin{equation}
\begin{tabular}{lll}
$\left[ X_{\left[ ij\right] },Y^{\left[ kl\right] }\right] $ & $=$ & $\left(
\delta _{j}^{k}Z_{i}^{l}-\delta _{i}^{k}Z_{j}^{l}\right) -\left( \delta
_{j}^{l}Z_{i}^{k}-\delta _{i}^{l}Z_{j}^{k}\right) $ \\
$\left[ X_{\left[ ij\right] },X_{\left[ kl\right] }\right] $ & $=$ & $0$ \\
$\left[ Y^{\left[ ij\right] },Y^{\left[ kl\right] }\right] $ & $=$ & $0$%
\end{tabular}%
\end{equation}%
For particular values of the labels, we obtain
\begin{equation}
\begin{tabular}{rrr}
$\left[ X_{\left[ ij\right] },Y^{\left[ jl\right] }\right] $ & $=$ & $\left(
N-2\right) Z_{i}^{l}+2\delta _{i}^{l}H$ \\
$\left[ Z_{i}^{j},X_{kl}\right] $ & $=$ & $+\left( \delta _{k}^{j}X_{\left[
il\right] }-\delta _{l}^{j}X_{\left[ ik\right] }\right) $ \\
$\left[ Z_{i}^{j},Y^{\left[ kl\right] }\right] $ & $=$ & $-\left( \delta
_{i}^{k}Y^{\left[ jl\right] }-\delta _{i}^{l}Y^{\left[ jk\right] }\right) $\
\end{tabular}%
\end{equation}%
and%
\begin{equation}
\begin{tabular}{rrr}
$\left[ H,X_{\left[ kl\right] }\right] $ & $=$ & $+X_{\left[ kl\right] }$ \\
$\left[ H,Y^{\left[ kl\right] }\right] $ & $=$ & $-Y^{\left[ kl\right] }$\
\end{tabular}%
\end{equation}%
In order to introduce similar notations to the ones used in the previous
sections, we associate to the variables $u_{i}$ and $\bar{v}^{i}$ the kets
\begin{equation}
u_{i}\rightarrow \left \vert +\frac{1}{2},i\right \rangle \qquad ,\qquad
\bar{v}^{i}\rightarrow \left \vert -\frac{1}{2},i\right \rangle
\end{equation}%
and to the translation operators $\bar{\partial}^{i}=\partial /\partial
u_{i} $ and $\partial _{i}=\partial /\partial \bar{v}^{i}$ the following
bras
\begin{equation}
\bar{\partial}^{i}\rightarrow \left \langle -\frac{1}{2},i\right \vert
\qquad ,\qquad \partial _{i}\rightarrow \left \langle +\frac{1}{2},i\right
\vert
\end{equation}%
We use moreover the following notation
\begin{equation}
\begin{tabular}{lllllll}
$\left \langle -,j|+,i\right \rangle $ & $=$ & $\delta _{i}^{j}$ & , & $%
\left \langle +,j|+,i\right \rangle $ & $=$ & $0$ \\
$\left \langle +,j|-,i\right \rangle $ & $=$ & $\delta _{i}^{j}$ & , & $%
\left \langle -,j|-,i\right \rangle $ & $=$ & $0$%
\end{tabular}
\label{ort}
\end{equation}%
to realise the operators $X_{\left[ ij\right] },Y^{\left[ kl\right] }$ and $%
Z_{i}^{l}$ as
\begin{equation}
\begin{tabular}{rrr}
$X_{\left[ ij\right] }$ & $=$ & $\left \vert +,i\right \rangle \left \langle
+,j\right \vert -\left \vert +,j\right \rangle \left \langle +,i\right \vert
$ \\
$Y^{\left[ kl\right] }$ & $=$ & $\left \vert -,k\right \rangle \left \langle
-,l\right \vert -\left \vert -,l\right \rangle \left \langle -,k\right \vert
$ \\
$Z_{i}^{l}$ & $=$ & $\left \vert +,i\right \rangle \left \langle -,l\right
\vert -\left \vert -,l\right \rangle \left \langle +,i\right \vert $%
\end{tabular}%
\end{equation}%
We also have $X_{\left[ ij\right] }Y^{\left[ kl\right] }=U_{\left[ ij\right]
}^{\left[ kl\right] }$ with%
\begin{equation}
U_{\left[ ij\right] }^{\left[ kl\right] }=\delta _{j}^{k}\left \vert
+,i\right \rangle \left \langle -,l\right \vert -\delta _{i}^{k}\left \vert
+,j\right \rangle \left \langle -,l\right \vert -\delta _{j}^{l}\left \vert
+,i\right \rangle \left \langle -,k\right \vert +\delta _{i}^{l}\left \vert
+,j\right \rangle \left \langle -,k\right \vert
\end{equation}%
as well as
\begin{equation}
H=\frac{1}{2}\varrho ^{+}-\frac{1}{2}\varrho ^{-}
\end{equation}%
where%
\begin{equation}
\begin{tabular}{lllllll}
$\Pi ^{+}$ & $=$ & $\sum_{i}\varrho _{i}^{+}$ & $\qquad ,\qquad $ & $\varrho
_{i}^{+}$ & $=$ & $\left \vert +,i\right \rangle \left \langle -,i\right
\vert $ \\
$\Pi ^{-}$ & $=$ & $\sum_{i}\varrho _{i}^{-}$ & $\qquad ,\qquad $ & $\varrho
_{i}^{-}$ & $=$ & $\left \vert -,i\right \rangle \left \langle +,i\right
\vert $%
\end{tabular}
\label{pr}
\end{equation}%
with the properties $\Pi ^{+}X_{\left[ ij\right] }=X_{\left[ ij\right] }$
and $Y^{\left[ kl\right] }\Pi ^{+}=Y^{\left[ kl\right] }.$ Notice also that
using (\ref{ort}), we have%
\begin{equation}
X_{ij}X_{kl}=0\qquad ,\qquad Y^{\left[ ij\right] }Y^{\left[ kl\right] }=0
\label{xij}
\end{equation}%
and%
\begin{equation}
X_{\left[ ij\right] }Y^{\left[ jl\right] }=\left \vert +,i\right \rangle
\left \langle -,l\right \vert +\delta _{i}^{l}\Pi ^{+}
\end{equation}

\subsubsection{Building the Lax operator $\mathcal{L}_{\boldsymbol{2N}}^{%
\mathbf{\protect\mu }_{N}}$}

Now, we are finally able to explicitly calculate the expression of the
spinorial Lax operator of the 4D CS theory with $SO_{2N}$ gauge symmetry.
This operator $\mathcal{L}_{\boldsymbol{2N}}^{\mathbf{\mu }_{N}}$ describing
the coupling of Figure \textbf{\ref{th2s}-(b)} is generally given by%
\begin{equation}
\mathcal{L}_{\boldsymbol{2N}}^{\mathbf{\mu }_{N}}=e^{X_{vect}}z^{\mathbf{\mu
}_{N}}e^{Y_{vect}}  \label{mn}
\end{equation}%
where the $X_{vect}$ and $Y_{vect}$ are $2N\times 2N$ matrices given by the
following linear combinations
\begin{equation}
X_{vect}=b^{\left[ ij\right] }X_{\left[ ij\right] }^{vect},\qquad
Y_{vect}=c_{\left[ ij\right] }Y_{vect}^{\left[ ij\right] }  \label{xiy}
\end{equation}%
such that the antisymmetric $b^{\left[ ij\right] }$ and $c_{\left[ ij\right]
}$ are Darboux coordinates satisfying the Poisson Bracket%
\begin{equation}
\left \{ b^{\left[ ij\right] },c_{\left[ kl\right] }\right \} _{PB}=\delta
_{k}^{i}\delta _{l}^{j}-\delta _{l}^{i}\delta _{k}^{j}
\end{equation}%
The adjoint form $\mathbf{\mu }_{N}$ of the minuscule coweight in (\ref{mn})
is given by
\begin{equation}
\mathbf{\mu }_{N}=\frac{1}{2}\Pi ^{+}-\frac{1}{2}\Pi ^{-}
\end{equation}%
where the projectors $\Pi ^{\pm }$ are as given in (\ref{pr}) with the
properties $\Pi ^{+}+\Pi ^{-}=I_{id}$ and $\Pi ^{+}\Pi ^{-}=0$. This allows
us to write\
\begin{equation}
z^{\mathbf{\mu }_{N}}=z^{\frac{1}{2}}\Pi ^{+}+z^{-\frac{1}{2}}\Pi ^{-}
\label{mz}
\end{equation}%
Moreover, because of the properties (\ref{xij}), the matrices $X$ and $Y$ (%
\ref{xiy}) are nilpotent with degree $2$, that is $X^{2}=Y^{2}=0$.
Therefore, the L-operator expands as
\begin{equation}
\mathcal{L}_{\boldsymbol{2N}}^{\mathbf{\mu }_{N}}=z^{\mathbf{\mu }_{N}}+Xz^{%
\mathbf{\mu }_{N}}+z^{\mathbf{\mu }_{N}}Y+Xz^{\mathbf{\mu }_{N}}Y
\label{xmy}
\end{equation}%
By substituting $z^{\mathbf{\mu }_{N}}$ by its expression (\ref{mz}) and
using the properties $X\Pi ^{+}=0$ and $\Pi ^{+}Y=0,$ we end up with%
\begin{equation}
\mathcal{L}_{\boldsymbol{2N}}^{\mathbf{\mu }_{N}}=z^{\frac{1}{2}}\Pi
^{+}+z^{-\frac{1}{2}}\Pi ^{-}+z^{-\frac{1}{2}}X\Pi ^{-}+z^{-\frac{1}{2}}\Pi
^{-}Y+z^{-\frac{1}{2}}X\Pi ^{-}Y  \label{xny}
\end{equation}%
And by putting $X=b^{\left[ ij\right] }X_{\left[ ij\right] }$ and $Y=c_{%
\left[ kl\right] }Y^{\left[ kl\right] }$, this operator can be also
expressed like%
\begin{equation}
\begin{tabular}{lll}
$\mathcal{L}_{\boldsymbol{2N}}^{\mathbf{\mu }_{N}}$ & $=$ & $z^{\frac{1}{2}%
}\Pi ^{+}+z^{-\frac{1}{2}}\Pi ^{-}+8z^{-\frac{1}{2}}(b^{\left[ ik\right]
}E_{i}^{j}c_{\left[ kj\right] })$ \\
&  & $+(2z^{-\frac{1}{2}}b^{\left[ ij\right] })X_{\left[ ij\right] }+(2z^{-%
\frac{1}{2}}c_{\left[ kl\right] })Y^{\left[ kl\right] }$%
\end{tabular}%
\end{equation}%
where $E_{i}^{k}=\left \vert +,i\right \rangle \left \langle -,k\right \vert
.$ Moreover, using%
\begin{equation}
\begin{tabular}{lll}
$Tr\left( X_{\left[ ij\right] }Y^{\left[ kl\right] }\right) $ & $=$ & $%
2\left( \delta _{i}^{l}\delta _{j}^{k}-\delta _{j}^{l}\delta _{i}^{k}\right)
$ \\
$Tr\left( XY^{\left[ ij\right] }\right) $ & $=$ & $-2b^{\left[ ij\right] }$
\\
$Tr\left( X_{\left[ ij\right] }Y\right) $ & $=$ & $-2c_{\left[ ij\right] }$%
\end{tabular}%
\end{equation}%
we have%
\begin{equation}
b^{\left[ ij\right] }=-\frac{1}{4}z^{\frac{1}{2}}Tr\left( Y^{\left[ ij\right]
}\mathcal{L}^{\mathbf{\mu }_{N}}\right) ,\qquad c_{\left[ ij\right] }=-\frac{%
1}{4}z^{\frac{1}{2}}Tr\left( X_{\left[ ij\right] }\mathcal{L}^{\mathbf{\mu }%
_{N}}\right)
\end{equation}%
The expression of the L-operator in the basis $\left \vert +,i\right \rangle
,\left \vert -,j\right \rangle $ defined in eq(\ref{gdec}) reads as follows%
\begin{equation}
\mathcal{L}_{\boldsymbol{2N}}^{\mathbf{\mu }_{N}}=z^{-\frac{1}{2}}\left(
\begin{array}{cc}
2c_{\left[ ij\right] } & z\delta _{j}^{i}+8b^{\left[ ik\right] }c_{\left[ kj%
\right] } \\
\delta _{j}^{i} & 2b^{\left[ ij\right] }%
\end{array}%
\right)
\end{equation}%
This is equivalent to spinor solutions in \cite{FrB} by change of basis.

\subsection{Topological quiver Q$_{\boldsymbol{2N}}^{\mathbf{\protect\mu }%
_{N}}$ of $\mathcal{L}_{\boldsymbol{2N}}^{\mathbf{\protect\mu }_{N}}$}

In order to construct the topological gauge quiver Q$_{\boldsymbol{2N}}^{%
\mathbf{\mu }_{N}}$ associated to the spinor coweight and the fundamental
representation of the D-type symmetry, we begin by rewriting the $\mathcal{L}%
_{\boldsymbol{2N}}^{\mathbf{\mu }_{N}}$ in the projector basis $\left( \Pi
^{+},\Pi ^{-}\right) $ of the representation $2N=N_{+}\oplus N_{-}.$ \newline
Using the properties of the $gl_{N}$ projectors on $N_{+}\oplus N_{-}$, in
particular $\left( \Pi ^{+}\right) ^{2}=\Pi ^{+},$ $\left( \Pi ^{-}\right)
^{2}=\Pi ^{-}$ and
\begin{equation}
\Pi ^{+}+\Pi ^{-}=I_{id}\qquad ,\qquad \Pi ^{+}\Pi ^{-}=0
\end{equation}%
as well as $\Pi ^{+}X=X$ and $Y\Pi ^{+}=Y,$ we can rewrite the Lax operator (%
\ref{xny}) as follows%
\begin{equation}
\mathcal{L}_{\boldsymbol{2N}}^{\mathbf{\mu }_{N}}=\left(
\begin{array}{ll}
z^{\frac{1}{2}}\Pi ^{+}+z^{-\frac{1}{2}}\Pi ^{+}X\Pi ^{-}Y\Pi ^{+} & z^{-%
\frac{1}{2}}X\Pi ^{-} \\
z^{-\frac{1}{2}}\Pi ^{-}Y & z^{-\frac{1}{2}}\Pi ^{-}%
\end{array}%
\right)  \label{xpy}
\end{equation}%
Moreover, by using the remarkable properties $X\Pi ^{-}=X$ and $\Pi ^{-}Y=Y$
that can be checked with the explicit realisations $X_{\left[ ij\right]
}=\left \vert +,i\right \rangle \left \langle +,j\right \vert -\left \vert
+,j\right \rangle \left \langle +,i\right \vert $ and $Y^{\left[ kl\right]
}=\left \vert -,k\right \rangle \left \langle -,l\right \vert -\left \vert
-,l\right \rangle \left \langle -,k\right \vert $, the term $X\Pi ^{-}Y\Pi
^{+}$ reduces to $XY\Pi ^{+}$ and the eq(\ref{xpy}) becomes%
\begin{equation}
\mathcal{L}_{\boldsymbol{2N}}^{\mathbf{\mu }_{N}}=z^{-\frac{1}{2}}\left(
\begin{array}{ll}
\Pi ^{+}(z+XY)\Pi ^{+} & X\Pi ^{-} \\
\Pi ^{-}Y & \Pi ^{-}%
\end{array}%
\right)  \label{sp}
\end{equation}%
\begin{figure}[h]
\begin{center}
\includegraphics[width=12cm]{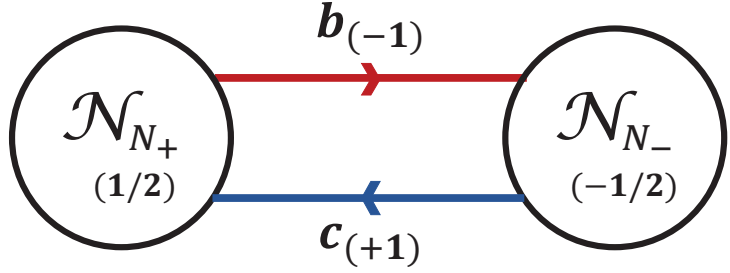}
\end{center}
\par
\vspace{-0.5cm}
\caption{The topological quiver Q$_{\boldsymbol{R}_{v}}^{\mathbf{\protect\mu
}_{N}}$ representing the operator $\mathcal{L}_{\boldsymbol{R}_{v}}^{\mathbf{%
\protect\mu }_{N}}$. It has 2 nodes $\mathcal{N}_{1}$, $\mathcal{N}_{2}$;
and 2 links $L_{12},L_{21}$. The nodes describe self-dual topological matter
and the links describe topological bi-matter$.$}
\label{mr}
\end{figure}
\newline
The nodes $\mathcal{N}_{1}$ and $\mathcal{N}_{2}$ of the topological gauge
quiver Q$_{\boldsymbol{2N}}^{\mathbf{\mu }_{N}}$ representing $\mathcal{L}_{%
\boldsymbol{2N}}^{\mathbf{\mu }_{N}}$ as depicted in Figure \textbf{\ref{mr}}
are given by the diagonal entries of the matrix (\ref{sp})%
\begin{equation}
\mathcal{N}_{1}\equiv \Pi _{+}\mathcal{L}\Pi _{+}\qquad ,\qquad \mathcal{N}%
_{2}\equiv \Pi _{-}\mathcal{L}\Pi _{-}
\end{equation}%
They are interpreted in terms of topological self-dual matter in the sense
that they have no $sl_{1}$ Levi charge. This feature is manifestly exhibited
by their dependence into the monomials $b^{\left[ ik\right] }c_{\left[ kj%
\right] }$ that are neutral under $sl_{1}$ because the Darboux coordinates $%
b^{\left[ ik\right] }$ and $c_{\left[ kj\right] }$ have opposite charges. On
the other hand, the two links are given by%
\begin{equation}
L_{1\rightarrow 2}\equiv \Pi _{+}\mathcal{L}\Pi _{-}\qquad ,\qquad
L_{2\rightarrow 1}\equiv \Pi _{-}\mathcal{L}\Pi _{+}
\end{equation}%
They are remarkably equivalent to the Darboux coordinates $b^{\left[ ij%
\right] }$ and $c_{\left[ ij\right] }$ and are interpreted in terms of
topological bi-fundamental matter of $sl_{1}\oplus sl_{N}$. The $sl_{1}$
charges data for the Q$_{\boldsymbol{2N}}^{\mathbf{\mu }_{N}}$ is collected
in the following table
\begin{equation}
\begin{tabular}{|l|l|l|l|l|}
\hline
{\small Quiver} & $\mathcal{N}_{1}$ & $\mathcal{N}_{2}$ & $L_{1\rightarrow
2} $ & $L_{2\rightarrow 1}$ \\ \hline
$\ \ sl_{1}$ & $+\frac{1}{2}$ & $-\frac{1}{2}$ & $-1$ & $+1$ \\ \hline
\end{tabular}%
\end{equation}%
where we remark that the transition from the topological quiver node $%
\mathcal{N}_{1}$ to the $\mathcal{N}_{2}$ is given by the link $%
L_{1\rightarrow 2}$ carrying a Levi charge $-1$; while the reverse
transition is given by the link $L_{2\rightarrow 1}$ with Levi charge $+1$.

\section{Exceptional E$_{6}$ 't Hooft lines}

This section is dedicated to the 4D Chern-Simons having as gauge symmetry
the E$_{6}$ group. This case is characterized by two minuscule 't Hooft
lines tH$_{\mathrm{\gamma }_{0}}^{\mu _{1}}$ and tH$_{\mathrm{\gamma }%
_{0}}^{\mu _{5}}$, and therefore two types of minuscule Lax operators $%
\mathcal{L}_{\boldsymbol{R_{e_{6}}}}^{\mu _{1}}$ and $\mathcal{L}_{%
\boldsymbol{R_{e_{6}}}}^{\mu _{5}}$ that we need to study in order to build
the associated topological gauge quivers. In particular, we focus here on $%
\boldsymbol{R_{e_{6}}=27}$; other possibilities are considered in the
conclusion section (\ref{TE}).

\subsection{ Minuscule coweights and Levi subalgebras of E$_{6}$}

\begin{figure}[h]
\begin{center}
\includegraphics[width=8cm]{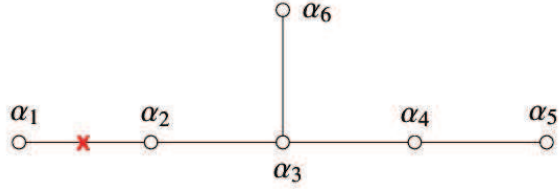}
\end{center}
\par
\vspace{-0.5cm}
\caption{The Dynkin Diagram of e$_{6}$ having six nodes labeled by the
simple roots $\protect\alpha _{i}$. The cross $\left( \times \right) $
indicates the cutted node in the Levi decomposition with respect to $\protect%
\mu_{1}$, the Levi subalgebra in this case is given by $so(10)\oplus so(2).$}
\label{E6}
\end{figure}

We begin by describing the interesting properties of the finite dimensional
exceptional Lie algebra $\boldsymbol{e}_{6}$ that are useful for our
construction. This is a simply laced Lie algebra with dimension 78 and rank
6; its algebraic properties are described by the root system $\Phi _{e_{6}}$
generated by six simple roots $\alpha _{i}$. The intersection between these
simple roots is represented in the Dynkin diagram $\mathcal{D}_{e_{6}}$
depicted in the Figure \textbf{\ref{E6} } and having the symmetric Cartan
matrix $K_{e_{6}}=\alpha _{i}.\alpha _{j}$ given by :%
\begin{equation}
K_{e_{6}}=\left(
\begin{array}{cccccc}
{\small 2} & {\small -1} & {\small 0} & {\small 0} & {\small 0} & {\small 0}
\\
{\small -1} & {\small 2} & {\small -1} & {\small 0} & {\small 0} & {\small 0}
\\
{\small 0} & {\small -1} & {\small 2} & {\small -1} & {\small 0} & {\small -1%
} \\
{\small 0} & {\small 0} & {\small -1} & {\small 2} & {\small -1} & {\small 0}
\\
{\small 0} & {\small 0} & {\small 0} & {\small -1} & {\small 2} & {\small 0}
\\
{\small 0} & {\small 0} & {\small -1} & {\small 0} & {\small 0} & {\small 2}%
\end{array}%
\right)
\end{equation}%
\textbf{\ }
\newline
The root system $\Phi _{e_{6}}$ contains 72 roots generated by the simple
root basis $\left \{ \alpha _{i}\right \} _{1\leq i\leq 6}$, it has 36
positive roots $\alpha \in \Phi _{e_{6}}^{+}$ and 36 negative ones $-\alpha
\in \Phi _{e_{6}}^{-}$. All of these roots have length $\alpha ^{2}=2$ and
are realised in the Euclidean $\mathbb{R}^{8}$ generated by the unit vector
basis $\left \{ \epsilon _{i}\right \} _{1\leq i\leq 8}$ as follows
\begin{equation}
\begin{tabular}{lllll}
E$_{6}$ & $:$ & $\alpha _{1}$ & $=$ & $\frac{1}{2}\left( \epsilon
_{1}-\epsilon _{2}-\epsilon _{3}-\epsilon _{4}-\epsilon _{5}-\epsilon
_{6}-\epsilon _{7}+\epsilon _{8}\right) $ \\
&  & $\alpha _{i}$ & $=$ & $\epsilon _{i}-\epsilon _{i-1}\qquad ,\qquad
i=1,2,3,4,5$ \\
&  & $\alpha _{6}$ & $=$ & $\epsilon _{1}+\epsilon _{2}$%
\end{tabular}
\label{al1}
\end{equation}%
From the Figure \textbf{\ref{E6}}, we learn that the Dynkin diagram $%
\mathcal{D}_{e_{6}}$ is invariant under a manifest $\mathbb{Z}_{2}^{aut}$
outer- automorphism symmetry exchanging four simple roots and leaving
invariant $\alpha _{3}$\ and $\alpha _{6}$. It acts like $\alpha
_{i}\rightarrow \alpha _{6-i}$ with $i=1,...,5$, by exchanging $\alpha _{2}$%
\ with $\alpha _{4}$ and $\alpha _{1}$\ with $\alpha _{5}$. In permutation
symmetry language, the $\mathbb{Z}_{2}^{aut}$ is generated by the double
transposition $\left( 15\right) \left( 24\right)$, i.e:%
\begin{equation}
\mathbb{Z}_{2}^{aut}=\left \{ I_{id},\left( 15\right) \left( 24\right)
\right \}
\end{equation}%
The 36+36 roots $\alpha $ of the e$_{6}$ Lie algebra can be organised as
follows%
\begin{equation}
\begin{tabular}{|l|l|l|l|}
\hline
{\small root} & {\small \ \ \ \ \ \ \ \ realisation} & {\small \ \ \ \ \
labels} & {\small number} \\ \hline
$\mathrm{\beta }_{ij}^{+}$ & $+\epsilon _{i}+\epsilon _{j}$ & $1\leq j<i\leq
5$ & $20$ \\ \hline
$\mathrm{\beta }_{ij}^{-}$ & $-\epsilon _{i}-\epsilon _{j}$ & $2\leq j<i\leq
5$ & $20$ \\ \hline
$\mathrm{\gamma }_{q_{i}}^{+}$ & $+\frac{1}{2}\left( q_{i}\epsilon
_{i}-\epsilon _{6}-\epsilon _{7}+\epsilon _{8}\right) $ & $\Pi
_{i=1}^{5}q_{i}=1$ & $16$ \\ \hline
$\mathrm{\gamma }_{q_{i}}^{-}$ & $-\frac{1}{2}\left( q_{i}\epsilon
_{i}-\epsilon _{6}-\epsilon _{7}+\epsilon _{8}\right) $ & $\Pi
_{i=1}^{5}q_{i}=1$ & $16$ \\ \hline
\end{tabular}%
\end{equation}%
\ where the five $q_{i}$ can take the values $\pm 1$ with the constraint $%
\Pi q_{i}=1.$\newline
Regarding the fundamental coweights $\omega _{i}$ of the six fundamental
representations of the Lie algebra of E$_{6},$ they are given by the duality
relation $\omega ^{i}.\alpha _{j}=\delta _{j}^{i}$; this equation can either
be solved in terms of roots, or by using the weight unit vectors $\epsilon
_{l}$. The $\omega _{i}$ read in terms of the simple roots as follows%
\begin{equation}
\begin{tabular}{|l|l|l|l|}
\hline
{\small fund- }$\omega _{i}$ & {\small \ \ \ \ \ \ \ \ \ \ \ \ in terms of
roots} & {\small height} & {\small Repres} \\ \hline
$\omega _{1}$ & $\frac{4}{3}\alpha _{1}+\frac{5}{3}\alpha _{2}+2\alpha _{3}+%
\frac{4}{3}\alpha _{4}+\frac{2}{3}\alpha _{5}+\alpha _{6}$ & {\small 8} & $%
27_{+}$ \\ \hline
$\omega _{2}$ & $\frac{5}{3}\alpha _{1}+\frac{10}{3}\alpha _{2}+4\alpha _{3}+%
\frac{8}{3}\alpha _{4}+\frac{4}{3}\alpha _{5}+2\alpha _{6}$ & {\small 15} & $%
351_{+}$ \\ \hline
$\omega _{3}$ & $2\alpha _{1}+4\alpha _{2}+6\alpha _{3}+4\alpha _{4}+2\alpha
_{5}+3\alpha _{6}$ & {\small 21} & $2925_{0}$ \\ \hline
$\omega _{4}$ & $\frac{4}{3}\alpha _{1}+\frac{8}{3}\alpha _{2}+4\alpha _{3}+%
\frac{10}{3}\alpha _{4}+\frac{5}{3}\alpha _{5}+2\alpha _{6}$ & {\small 15} &
$351_{-}$ \\ \hline
$\omega _{5}$ & $\frac{2}{3}\alpha _{1}+\frac{4}{3}\alpha _{2}+2\alpha _{3}+%
\frac{5}{3}\alpha _{4}+\frac{4}{3}\alpha _{5}+\alpha _{6}$ & {\small 8} & $%
27_{-}$ \\ \hline
$\omega _{6}$ & $\alpha _{1}+2\alpha _{2}+3\alpha _{3}+2\alpha _{4}+\alpha
_{5}+2\alpha _{6}$ & {\small 11} & $78_{0}$ \\ \hline
\end{tabular}
\label{tab1}
\end{equation}%
\begin{equation*}
\end{equation*}%
From these expressions, we see that the outer-automorphism symmetry $\mathbb{%
Z}_{2}^{aut}$ discussed above can be manifestly exhibited as follows,%
\begin{equation}
\begin{tabular}{lll}
$\omega _{1}+\omega _{5}$ & $=$ & $2\left( \alpha _{1}+\alpha _{5}\right)
+3\left( \alpha _{2}+\alpha _{4}\right) +4\alpha _{3}+2\alpha _{6}$ \\
$\omega _{2}+\omega _{4}$ & $=$ & $3\left( \alpha _{1}+\alpha _{5}\right)
+6\left( \alpha _{2}+\alpha _{4}\right) +8\alpha _{3}+4\alpha _{6}$ \\
$\omega _{3}$ & $=$ & $2\left( \alpha _{1}+\alpha _{5}\right) +4\left(
\alpha _{2}+\alpha _{4}\right) +6\alpha _{3}+3\alpha _{6}$ \\
$\omega _{6}$ & $=$ & $\left( \alpha _{1}+\alpha _{5}\right) +2\left( \alpha
_{2}+\alpha _{4}\right) +3\alpha _{3}+2\alpha _{6}$%
\end{tabular}%
\end{equation}%
Moreover, by using (\ref{al1}) and $\alpha _{i}\rightarrow \alpha _{6-i}$
with $\alpha _{0}\equiv \alpha _{6}$, one can write down the action of the
outer-automorphism symmetry $\mathbb{Z}_{2}^{aut}$ on the weight vector
basis $\epsilon _{i}$. In what follows, we will be particularly interested
into: $\left( 1\right) $ the representation 78$_{0},$ associated with the
simple root $\alpha _{6}$, and $\left( 2\right) $ the 27$_{\pm }$ associated
with $\alpha _{1}$ and $\alpha _{5}.$ \newline
The two minuscule coweights $\mu _{1}$ and $\mu _{5}$ that are dual to the $%
\alpha _{1}$ and $\alpha _{5}$ of the $e_{6}$ are respectively associated
with the fundamentals $27_{+}$ and $27_{-}$ as shown in table (\ref{tab1}).
Being related by $\mathbb{Z}_{2}^{aut}$, we focus below on one of the two
minuscule coweights, say $\mu =\omega _{1}$; Similar results can be derived
for $\mu _{5}$.

\subsubsection{The e$_{6}$ algebra and the representation \textbf{78}}

There are different ways to decompose the root system of the $e_{6}$ Lie
algebra. The interesting Levi decomposition with respect to charges of the
minuscule coweight $\mu =\mu _{1}$ considered here reads as follows
\begin{equation}
e_{6}\rightarrow so_{2}\oplus so_{10}\oplus \boldsymbol{16}_{+}\oplus
\boldsymbol{16}_{-}  \label{e6d}
\end{equation}%
From this splitting, we learn that the Levi subalgebra $\boldsymbol{l}_{\mu
}=so_{2}\oplus so_{10}$ and the nilpotent subalgebras $\boldsymbol{n}_{\pm }=%
\boldsymbol{16}_{\pm }$. The root system $\Phi _{e_{6}}$ containing the 72
roots of $e_{6}$ is therefore decomposed in terms of two subsets: a subset $%
\Phi _{so_{10}}$, and a subset given by the complement $\Phi
_{e_{6}}\backslash \Phi _{so_{10}}$; they are described here below as they
play an important role in the construction of the Lax operator $\mathcal{L}%
_{e_{6}}^{\mathbf{\mu }}$.

$\bullet $ \emph{Roots within} $\Phi _{so_{10}}$\newline
The subset $\Phi _{so_{10}}$ contains 40 roots $\beta _{so_{10}}$, 20
positive and 20 negative; they define the step operators $Z_{\pm \beta
_{so_{10}}}$ generating $so_{10}$ within $\boldsymbol{e}_{6}$. It is
generated by the simple roots
\begin{equation}
\begin{tabular}{lllll}
$\alpha _{2},$ & $\alpha _{3},$ & $\alpha _{4},$ & $\alpha _{5},$ & $\alpha
_{6}$%
\end{tabular}%
\end{equation}%
and has the usual symmetry properties of the root system of $so_{10}$. The
root subsystem $\Phi _{so_{10}}\subset \Phi _{e_{6}}$ can be defined as
containing the roots $\beta _{so_{10}}$ with no dependence into $\alpha _{1}$%
, formally
\begin{equation}
\frac{\delta \beta _{so_{10}}}{\delta \alpha _{1}}=0
\end{equation}%
This can be noticed b\textrm{y cutting the node} $\alpha _{1}$ in the Dynkin
diagram of the Figure \textbf{\ref{E6}}, where we recover the Dynkin diagram
of $so_{10}$ and a free node $\alpha _{1}$ associated with the $so_{10}$
spinor representations $16_{\pm }$ charged under $so_{2}$.

$\bullet $ \emph{Roots outside} $\Phi _{so_{10}}$\newline
This is the complementary subset of $\Phi _{so_{10}}$ within $\Phi _{e_{6}}$%
; it is given by $\Phi _{e_{6}}\backslash \Phi _{so_{10}}$ and reads
directly from the root system of $e_{6}$ by considering only the roots $%
\beta _{e_{6}}$ with a dependence into $\alpha _{1}$ :
\begin{equation}
\delta \beta _{so_{10}}/\delta \alpha _{1}\neq 0  \label{n}
\end{equation}%
This subset contains 32 roots of spinorial type as they linearly depend on
the simple root $\alpha _{1}$ which is spinorial-like. The importance of
these roots is that they define the 16 step operators $X_{+\beta }$
generating the nilpotent $\boldsymbol{16}_{+}$ and 16 step operators $%
X_{-\beta }=Y^{\beta }$ generating the $\boldsymbol{16}_{-}$.

\subsubsection{Decomposing the representation 27}

\begin{figure}[h]
\begin{center}
\includegraphics[width=16cm]{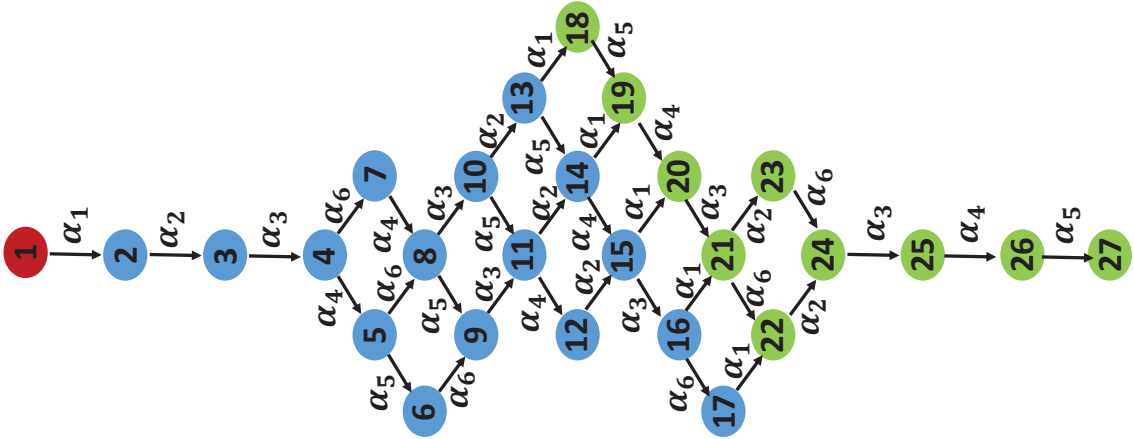}
\end{center}
\par
\vspace{-0.5cm}
\caption{The weight diagram of the representation 27 of the exceptional Lie
algebra $e_{6}$ where every state $\left \vert \protect\xi %
_{A}\right
\rangle $ is simply represented by the node carrying its number.
The states 1+16+10 of the $SO_{10}$ sub-representations of $e_{6}$ are
represented by different colors.}
\label{27e6}
\end{figure}

As for the adjoint representation of $e_{6}$, the fundamental representation
also decomposes in terms of representations of $so_{2}\oplus so_{10}$. This
representation is interesting in our study as it will be taking as the
electric charge of the Wilson line $W_{\mathrm{\xi }_{z}}^{\boldsymbol{R}}$
where $\boldsymbol{R=27}_{\pm }$. Generally speaking, given a representation
$\boldsymbol{R}_{e_{6}}$ of the algebra $e_{6}$, it can be decomposed into a
direct sum of representations of $so_{2}\oplus so_{10}.$ such as
\begin{equation}
\boldsymbol{R}_{e_{6}}=\sum_{l}n_{l}\left( \boldsymbol{R}_{l}^{so_{10}},%
\boldsymbol{R}_{l}^{so_{2}}\right)
\end{equation}%
where $n_{l}$ are some positive integers. In the case of $\boldsymbol{R}%
_{e_{6}}=\boldsymbol{27}$, we have the following reduction \textrm{\cite{59}}
\begin{equation}
\mathbf{27}=(\mathbf{1},-\frac{4}{3})+(\mathbf{10},+\frac{2}{3})+(\mathbf{16}%
,-\frac{1}{3})  \label{27}
\end{equation}%
that we can simply write as $27=1_{-4/3}+10_{2/3}+16_{-1/3}$. Notice that by
cutting the simple root $\alpha _{1}$ in the Dynkin diagram, the $SO_{10}$
representations get charged under $SO_{2}$; these charges play the role of a
"glue" between these representations within the 27. This property is
manifested by the constraint that the sum (or the trace) of the charges of
the 27 states with respect to $SO_{2}\sim E_{6}/SO_{10}$ must vanish. Notice
moreover that these charges can be also observed in the following relation%
\begin{equation}
\omega _{1}-\omega _{5}=\frac{2}{3}\alpha _{1}+\frac{1}{3}\alpha _{2}-\frac{1%
}{3}\alpha _{4}-\frac{2}{3}\alpha _{5}
\end{equation}%
where $\alpha _{2}$ stands for the spinorial of $SO_{10}$ and $\alpha _{5}$
for the vectorial.\newline
In order to understand the structure of the 27 states in the fundamental
representation of $e_{6},$ we refer to the weight diagram of the Figure
\textbf{\ref{27e6}} where we have a top state $\left \vert \xi
_{1}\right
\rangle $ with weight $\xi _{1}=\omega _{1}$ and a bottom state $%
\xi _{27}=-\omega _{5}$. The other 25 states in between can be generated
either by starting from the $\left \vert \xi _{1}\right \rangle $ and
successively acting on it by the step operators $\left( E_{\beta }\right)
^{\dagger }=E_{-\beta }$ where $\beta $\ a positive root of e$_{6}$, or by
acting on the bottom state $\left \vert \xi _{27}\right \rangle $ with \emph{%
\ }$\left( E_{-\beta }\right) ^{\dagger }=E_{\beta }$.\emph{\ }\newline
The subspaces of the $\boldsymbol{27}$ representation correspond in the
figure \textbf{\ref{27e6}} to :
\begin{equation}
\begin{tabular}{llll}
& $\left \vert \mathbf{1}\right \rangle $ & $\left \vert \xi _{1}\right
\rangle _{-4/3}$ & $=\left \vert \omega _{1}\right \rangle $ \\
& $\ \downarrow $ &  &  \\
& $\left \vert \mathbf{16}\right \rangle $ & $\left \vert \xi _{\alpha
}\right \rangle _{+1/3}$ &  \\
& $\ \downarrow $ &  &  \\
& $\left \vert \mathbf{10}\right \rangle $ & $\left \vert \xi _{i}\right
\rangle _{-2/3}$ &
\end{tabular}
\label{st}
\end{equation}%
such that the top state $\left \vert \xi _{1}\right \rangle $ is an $SO_{10}$
singlet, the 16 states $\left \vert \xi _{2}\right \rangle ,...,\left \vert
\xi _{17}\right \rangle $ constitute a chiral spinor of $SO_{10}$, and the
ten states $\left \vert \xi _{18}\right \rangle ,...,\left \vert \xi
_{27}\right \rangle $ form a vector of $SO_{10}$.

\subsection{Minuscule E$_{6}$ 't Hooft operator}

We can now use the collected mathematical tools concerning the exceptional
Lie algebra $e_{6}$ to calculate the Lax operator $\mathcal{L}_{e_{6}}^{%
\mathbf{\mu }}$ describing the coupling of an exceptional minuscule 't Hooft
line tH$_{\mathrm{\gamma }_{0}}^{\mathbf{\mu }}$ with magnetic charge $\mu
=\mu_{1} $ interacting with a Wilson line $W_{\mathrm{\xi }_{z}}^{%
\boldsymbol{R}}$ with electric charge $\boldsymbol{R=27}$.

\subsubsection{Realizing the generators of the nilpotent subalgebras}

\begin{figure}[h]
\begin{center}
\includegraphics[width=8cm]{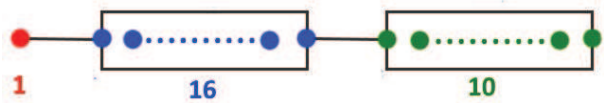}
\end{center}
\par
\vspace{-0.5cm}
\caption{A graphical illustration of the Levi decomposition of the
representation 27 of e$_{6}$ in terms of representations of $so_{10}.$}
\label{270}
\end{figure}

To construct the 't Hoof line operator $\mathcal{L}_{\boldsymbol{27}}^{%
\mathbf{\mu }}$ of the exceptional E$_{6}$ Chern-Simons theory in 4D, we
begin by building the generators of the nilpotent subalgebras that appear in
the Levi factorisation -based formula \textrm{\cite{34A}} where $\mathbf{\mu
=\omega }_{1}$ and the nilpotent matrix operators are given by
\begin{equation}
X=\sum_{\beta =1}^{16}b^{^{\beta }}X_{\beta }\qquad ,\qquad Y=\sum_{\beta
=1}^{16}c_{\beta }Y^{\beta }  \label{lc}
\end{equation}%
In these expansions, the sixteen $b^{\beta }$ and the sixteen $c_{\beta }$
are the 16+16 Darboux coordinates of the phase space of the exceptional E$%
_{6}$ 't Hooft line tH$_{\mathrm{\gamma }_{0}}^{\mathbf{\mu }}$. They
satisfy the Poisson bracket $\left \{ b^{\gamma },c_{\beta }\right \}
=\delta _{\beta }^{\gamma }$ that must be promoted to a commutator in the
study of interacting quantum lines. $X_{\beta }$ and $\ Y^{\beta }$ are the
generators of the nilpotent subalgebras $\mathbf{16}_{+}$ and $\mathbf{16}%
_{-}.$ The charge operator $\mathbf{\mu }$ of the Levi subalgebra associated
with the minuscule coweight can be presented as
\begin{equation}
\mathbf{\mu }=-\frac{4}{3}\varrho _{\underline{\mathbf{1}}}+\frac{2}{3}%
\varrho _{\underline{\mathbf{10}}}-\frac{1}{3}\varrho _{\underline{\mathbf{16%
}}}  \label{mu}
\end{equation}%
where $\varrho _{\underline{\mathbf{1}}},$ $\varrho _{\underline{\mathbf{10}}%
}$ and $\varrho _{\underline{\mathbf{16}}}$ are projectors on the $%
so_{2}\oplus so_{10}$ representation subspaces making the $\boldsymbol{27}$
fundamental of E$_{6}$ as given by eq.(\ref{27}). By denoting the 27 states $%
\left \vert \xi _{A}\right \rangle $ of this representation as
\begin{equation}
\begin{tabular}{|l|l|l|l|l|}
\hline
{\small Groups} & E$_{6}$ & \multicolumn{3}{|l|}{\ $SO_{10}\times SO_{2}$}
\\ \hline
{\small States} & $\left \vert \xi _{A}\right \rangle $ & $\left \vert
\upsilon _{i}\right \rangle $ & $\left \vert s_{\alpha }\right \rangle $ & $%
\left \vert \varphi \right \rangle $ \\ \hline
{\small Repres} & $\mathbf{27}_{0}$ & $\mathbf{10}_{+2/3}$ & $\mathbf{16}%
_{-1/3}$ & $\mathbf{1}_{-4/3}$ \\ \hline
\end{tabular}%
\end{equation}%
following the splitting formally represented in the picture \textbf{\ref{270}%
},
\newline
we can write the projectors $\varrho _{\underline{\mathbf{R}}}$ on the
fundamental representation of $e_{6}$ as
\begin{equation}
\varrho _{\underline{\mathbf{10}}}=\sum_{l=1}^{10}\left \vert v_{l}\right
\rangle \left \langle v^{l}\right \vert ,\qquad \varrho _{\underline{\mathbf{%
16}}}=\sum_{\beta =1}^{16}\left \vert s_{\beta }\right \rangle \left \langle
s^{\beta }\right \vert ,\qquad \varrho _{\underline{\mathbf{1}}}=\left \vert
\varphi \right \rangle \left \langle \varphi \right \vert
\end{equation}%
Using the state basis kets $\left \vert v_{l}\right \rangle ,$ $\left \vert
s_{\beta }\right \rangle $ and $\left \vert \varphi \right \rangle $
satisfying the orthogonality properties $\left \langle \varphi
|v_{l}\right
\rangle =\left \langle \varphi |s_{\beta }\right \rangle
=\left \langle v_{l}|s_{\beta }\right \rangle =0$, we realise the generators
$X_{\beta }$ and $Y^{\beta }$ of the nilpotent subalgebras like
\begin{equation}
\begin{tabular}{lll}
$X_{\beta }$ & $=$ & $\left \vert v_{i}\right \rangle \left( \Gamma
^{i}\right) _{\beta \gamma }\left \langle s^{\gamma }\right \vert +\left
\vert s_{\beta }\right \rangle \left \langle \varphi \right \vert $ \\
$Y^{\beta }$ & $=$ & $\left \vert \varphi \right \rangle \left \langle
s^{\beta }\right \vert +\left \vert s_{\gamma }\right \rangle \left( \Gamma
_{i}\right) ^{\beta \gamma }\left \langle v^{i}\right \vert $%
\end{tabular}
\label{rxy}
\end{equation}%
where the $\Gamma _{i}$'s are Gamma matrices satisfying the usual Clifford
algebra in ten dimensional space, namely $\Gamma _{i}\Gamma _{j}+\Gamma
_{j}\Gamma _{i}=2\delta _{ij}$. Moreover, if we adopt the short notations $%
\left \vert \mathbf{1}\right \rangle ,$ $\left \vert \mathbf{10}%
\right
\rangle $ and $\left \vert \mathbf{16}\right \rangle $ to refer to
the singlet state $\left \vert \varphi \right \rangle ,$ the vector $%
\left
\vert v_{l}\right \rangle $ and the spinor $\left \vert s_{\beta
}\right
\rangle $, we can express the projectors more simply like $\varrho
_{\underline{\mathbf{1}}}=\left \vert \mathbf{1}\right \rangle \left \langle
\mathbf{1}\right \vert $, and $\varrho _{\underline{\mathbf{10}}%
}=\left
\vert \mathbf{10}\right \rangle \left \langle \mathbf{10}%
\right
\vert $ as well as $\varrho _{\underline{\mathbf{16}}}=\left \vert
\mathbf{16}\right
\rangle \left \langle \mathbf{16}\right \vert $. Then, we
also end up with the following expressions for the nilpotent generators (\ref%
{rxy}) :
\begin{equation}
\begin{tabular}{lll}
$X_{\beta }$ & $=$ & $\left \vert \mathbf{10}\right \rangle \left \langle
\mathbf{16}\right \vert +\left \vert \mathbf{16}\right \rangle \left \langle
\mathbf{1}\right \vert $ \\
$Y^{\beta }$ & $=$ & $\left \vert \mathbf{1}\right \rangle \left \langle
\mathbf{16}\right \vert +\left \vert \mathbf{16}\right \rangle \left \langle
\mathbf{10}\right \vert $ \\
$\mathbf{\mu }$ & $=$ & $\frac{2}{3}\left \vert \mathbf{10}\right \rangle
\left \langle \mathbf{10}\right \vert -\frac{1}{3}\left \vert \mathbf{16}%
\right \rangle \left \langle \mathbf{16}\right \vert -\frac{4}{3}\left \vert
\mathbf{1}\right \rangle \left \langle \mathbf{1}\right \vert $%
\end{tabular}
\label{16}
\end{equation}%
We can check that this realisation solves the Levi decomposition
constraints, namely%
\begin{equation}
\left[ \mathbf{\mu },X_{\beta }\right] =X_{\beta }\qquad ,\qquad \left[ \mu
,Y^{\beta }\right] =-Y^{\beta }
\end{equation}%
We have for example $\mathbf{\mu }X_{\beta }=\frac{2}{3}\left \vert \mathbf{%
10}\right \rangle \left \langle \mathbf{16}\right \vert -\frac{1}{3}%
\left
\vert \mathbf{16}\right \rangle \left \langle \mathbf{1}\right \vert $
and $X_{\beta }\mathbf{\mu }=-\frac{1}{3}\left \vert \mathbf{10}%
\right
\rangle \left \langle \mathbf{16}\right \vert -\frac{4}{3}%
\left
\vert \mathbf{16}\right \rangle \left \langle \mathbf{1}\right \vert $%
, thus leading to $\left[ \mathbf{\mu },X_{\beta }\right] =X_{\beta }$.
Notice that this realisation leads to
\begin{equation}
\begin{tabular}{lll}
$X_{\alpha }X_{\beta }$ & $=$ & $\left \vert v_{i}\right \rangle \left(
\Gamma ^{i}\right) _{\alpha \beta }\left \langle \varphi \right \vert $ \\
$Y^{\alpha }Y^{\beta }$ & $=$ & $\left \vert \varphi \right \rangle \left(
\Gamma _{i}\right) ^{\beta \alpha }\left \langle v^{i}\right \vert $%
\end{tabular}%
\end{equation}%
and%
\begin{equation}
X_{\alpha }X_{\beta }X_{\gamma }=0\qquad ,\qquad Y^{\alpha }Y^{\beta
}Y^{\gamma }=0
\end{equation}%
We also have as interesting properties $X_{\beta }\varrho _{\underline{%
\mathbf{10}}}=0$ and $\varrho _{\underline{\mathbf{10}}}Y^{\beta }=0,$ as
well as%
\begin{equation}
\begin{tabular}{lllllll}
$X_{\beta }\varrho _{\underline{\mathbf{1}}}$ & $=$ & $X_{\beta }$ & $%
,\qquad $ & $\varrho _{\underline{\mathbf{1}}}Y^{\beta }$ & $=$ & $Y^{\beta
} $ \\
$X_{\beta }\varrho _{\underline{\mathbf{16}}}$ & $=$ & $X_{\beta }$ & $%
,\qquad $ & $\varrho _{\underline{\mathbf{16}}}Y^{\beta }$ & $=$ & $Y^{\beta
}$%
\end{tabular}%
\end{equation}%
From these relations and the linear combinations $X=b^{^{\beta }}X_{\beta }$
and $Y=c_{\beta }Y^{\beta }$ given by (\ref{lc}), we learn that $%
X^{3}=Y^{3}=0$ while
\begin{equation}
X^{2}=2V^{i}\left \vert v_{i}\right \rangle \left \langle 0\right \vert
\qquad ,\qquad Y^{2}=2W_{i}\left \vert 0\right \rangle \left \langle
v^{i}\right \vert  \label{nl}
\end{equation}%
where we have set
\begin{equation}
V^{i}=\frac{1}{2}b^{\alpha }\left( \Gamma ^{i}\right) _{\alpha \beta
}b^{\beta }\qquad ,\qquad W_{i}=\frac{1}{2}c_{\alpha }\left( \Gamma
_{i}\right) ^{\alpha \beta }c_{\beta }
\end{equation}%
In terms of the short notations, we have $X_{\alpha }X_{\beta }\sim
\left
\vert \mathbf{10}\right \rangle \left \langle \mathbf{1}\right \vert $
and $Y^{\alpha }Y^{\beta }\sim \left \vert \mathbf{1}\right \rangle
\left
\langle \mathbf{10}\right \vert $ as well as $X^{2}=2\mathbf{V}%
\left
\vert \mathbf{10}\right \rangle \left \langle \mathbf{1}\right \vert $
and $Y^{2}=2\mathbf{W}\left \vert \mathbf{1}\right \rangle \left \langle
\mathbf{10}\right \vert $ where $\mathbf{V}$ and $\mathbf{W}$ are the
vectors appearing in (\ref{nl}).

\subsubsection{Constructing the operator $\mathcal{L}_{e_{6}}^{\mathbf{%
\protect\mu }}$}

For the final step, we use the nilpotency feature of X and Y yielding the
finite expansions $e^{X}=I+X+\frac{1}{2}X^{2}$ and $e^{Y}=I+Y+\frac{1}{2}%
Y^{2}$ as well as $z^{\mathbf{\mu }}e^{Y}=z^{\mathbf{\mu }}+z^{\mathbf{\mu }%
}Y+\frac{1}{2}z^{\mathbf{\mu }}Y^{2}.$ Moreover, by replacing with
\begin{equation}
z^{\mathbf{\mu }}=z^{-\frac{4}{3}}\varrho _{\underline{\mathbf{1}}}+z^{\frac{%
2}{3}}\varrho _{\underline{\mathbf{10}}}+z^{-\frac{1}{3}}\varrho _{%
\underline{\mathbf{16}}}  \label{mmm}
\end{equation}%
and $\varrho _{\underline{\mathbf{10}}}Y=0$, we obtain%
\begin{equation}
\begin{tabular}{lll}
$z^{\mathbf{\mu }}e^{Y}$ & $=$ & $z^{-4/3}\varrho _{\underline{\mathbf{1}}%
}+z^{-1/3}\varrho _{\underline{\mathbf{16}}}+z^{2/3}\varrho _{\underline{%
\mathbf{10}}}$ \\
&  & $+z^{-4/3}\varrho _{\underline{\mathbf{1}}}Y+z^{-1/3}\varrho _{%
\underline{\mathbf{16}}}Y+\frac{1}{2}z^{-4/3}\varrho _{\underline{\mathbf{1}}%
}Y^{2}$%
\end{tabular}%
\end{equation}%
Substituting this into $e^{X}z^{\mathbf{\mu }}e^{Y}$ and using the property $%
X\varrho _{\underline{\mathbf{10}}}=0$, we finally find the expression of
the L-operator we are looking for :
\begin{equation}
\begin{tabular}{lll}
$\mathcal{L}_{\boldsymbol{27}}^{\mathbf{\mu }}$ & $=$ & $z^{-\frac{4}{3}%
}\varrho _{\underline{\mathbf{1}}}+z^{-1/3}\varrho _{\underline{\mathbf{16}}%
}+z^{2/3}\varrho _{\underline{\mathbf{10}}}+z^{-4/3}\varrho _{\underline{%
\mathbf{1}}}Y+z^{-1/3}\varrho _{\underline{\mathbf{16}}}Y+$ \\
&  & $z^{-\frac{4}{3}}X\varrho _{\underline{\mathbf{1}}}+z^{-1/3}X\varrho _{%
\underline{\mathbf{16}}}+z^{-\frac{4}{3}}X\varrho _{\underline{\mathbf{1}}%
}Y+ $ \\
&  & $\frac{1}{2}z^{-\frac{4}{3}}\varrho _{\underline{\mathbf{1}}%
}Y^{2}+z^{-1/3}X\varrho _{\underline{\mathbf{16}}}Y+\frac{1}{2}z^{-\frac{4}{3%
}}X\varrho _{\underline{\mathbf{1}}}Y^{2}+$ \\
&  & $\frac{1}{2}z^{-\frac{4}{3}}X^{2}\varrho _{\underline{\mathbf{1}}}+%
\frac{1}{2}z^{-\frac{4}{3}}X^{2}\varrho _{\underline{\mathbf{1}}}Y+\frac{1}{4%
}z^{-\frac{4}{3}}X^{2}\varrho _{\underline{\mathbf{1}}}Y^{2}$%
\end{tabular}
\label{le6}
\end{equation}%
Notice that each one of the $z^{\mathbf{\mu }}$, $e^{X}$ and $e^{Y}$ has 3
monomials leading in general to 81 monomials for the $\mathcal{L}_{%
\boldsymbol{27}}^{\mathbf{\mu }}$. However, The above expression was
simplified thanks to useful properties such as $X\varrho _{\underline{%
\mathbf{10}}}=0$ and $\varrho _{\underline{\mathbf{10}}}Y=0$ and the other
ones mentioned above. It can be further expressed in terms of Darboux
ccordinates by substituting the following relations%
\begin{equation}
\begin{tabular}{lllllll}
$X\varrho _{\underline{\mathbf{1}}}$ & $=$ & $b^{\beta }$ & ,\qquad \qquad &
$X^{2}\varrho _{\underline{\mathbf{1}}}$ & $=$ & $b^{\beta }\Gamma _{\beta
\gamma }^{i}b^{\gamma }$ \\
$\varrho _{\underline{\mathbf{1}}}Y$ & $=$ & $c_{\alpha }$ & ,\qquad \qquad
& $X\varrho _{\underline{\mathbf{1}}}Y^{2}$ & $=$ & $b^{\alpha }c_{\beta
}\Gamma _{i}^{\beta \gamma }c$ \\
$X\varrho _{\underline{\mathbf{1}}}Y$ & $=$ & $b^{\beta }c_{\alpha }$ &
,\qquad \qquad & $X^{2}\varrho _{\underline{\mathbf{1}}}Y^{2}$ & $=$ & $%
b^{\beta }\Gamma _{\beta \gamma }^{i}b^{\gamma }c_{\beta }\Gamma _{i}^{\beta
\gamma }c_{\gamma }$%
\end{tabular}%
\end{equation}%
and%
\begin{equation}
\begin{tabular}{lll}
$X\varrho _{\underline{\mathbf{16}}}$ & $=$ & $b^{\gamma }\Gamma _{\gamma
\beta }^{i}$ \\
$\varrho _{\underline{\mathbf{16}}}Y$ & $=$ & $\Gamma _{i}^{\gamma \beta
}c_{\gamma }$ \\
$X\varrho _{\underline{\mathbf{16}}}Y$ & $=$ & $b^{\gamma }\Gamma _{\gamma
\beta }^{i}\Gamma _{i}^{\gamma \beta }c_{\gamma }$ \\
$\varrho _{\underline{\mathbf{1}}}Y^{2}$ & $=$ & $c_{\beta }\Gamma
_{i}^{\beta \gamma }c_{\gamma }$%
\end{tabular}%
\end{equation}%
and
\begin{equation}
X\varrho _{\underline{\mathbf{16}}}Y^{2}=0\qquad ,\qquad X^{2}\varrho _{%
\underline{\mathbf{16}}}Y^{2}=0
\end{equation}

\subsection{Topological gauge quiver for E$_{6}$}

\begin{figure}[h]
\begin{center}
\includegraphics[width=10cm]{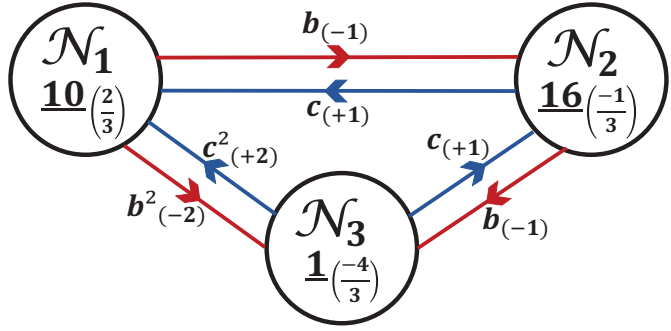}
\end{center}
\par
\vspace{-0.5cm}
\caption{$\mathcal{L}_{\boldsymbol{27}}^{\mathbf{\protect\mu }}$ as a
topological quiver with 3 nodes and $6$ links. The nodes are given by the
self-dual $R_{i}\otimes \bar{R}_{i}$ and the links by bi-matter $%
R_{i}\otimes \bar{R}_{j}.$ In addition to $SO_{10}$ representations, the
Darboux coordinates $b^{\protect\alpha },$ $c_{\protect\alpha }$ carry $%
SO_{2}$ charges given by $q=\pm 1$. The fundamental vector-like matter $%
V^{i} $ and $W_{i}$ carry $-2$ and $+2$.}
\label{Q6}
\end{figure}

In this subsection, we construct the topological gauge quiver Q$_{%
\boldsymbol{27}}^{\mathbf{\mu }}$ associated with the operator $\mathcal{L}_{%
\boldsymbol{27}}^{\mathbf{\mu }}$ (\ref{le6}). First, we give the matrix
form of the L-operator in terms of the phase variables $b^{\beta }$ and $%
c_{\beta }$ to underline their field theory interpretation in terms of
topological bi-matter. Then, we derive the quiver representation Q$_{%
\boldsymbol{27}}^{\mathbf{\mu }}$ using the projectors $\varrho _{\underline{%
\mathbf{1}}},$ $\varrho _{\underline{\mathbf{10}}}$ and $\varrho _{%
\underline{\mathbf{16}}}$ on the sub-representations of $so_{10}$ within the
27 of E$_{6}$.\newline
By ordering the above mentioned projectors like $\left( \varrho _{\underline{%
\mathbf{10}}},\varrho _{\underline{\mathbf{16}}},\varrho _{\underline{%
\mathbf{1}}}\right) $ and thinking of them as representing the sub-blocks of
the matrix; the operator $\mathcal{L}_{\boldsymbol{27}}^{\mathbf{\mu }}$ is
put as follows%
\begin{equation}
\mathcal{L}_{\boldsymbol{27}}^{\mathbf{\mu }}=\left(
\begin{array}{lll}
z^{\frac{2}{3}}\varrho _{\underline{\mathbf{10}}}+z^{-\frac{1}{3}}X\varrho _{%
\underline{\mathbf{16}}}Y+\frac{1}{4}z^{-\frac{4}{3}}X^{2}\varrho _{%
\underline{\mathbf{1}}}Y^{2} & z^{-\frac{1}{3}}X\varrho _{\underline{\mathbf{%
16}}}+\frac{1}{2}z^{-\frac{4}{3}}X^{2}\varrho _{\underline{\mathbf{1}}}Y &
\frac{1}{2}z^{-\frac{4}{3}}X^{2}\varrho _{\underline{\mathbf{1}}} \\
z^{-\frac{1}{3}}\varrho _{\underline{\mathbf{16}}}Y+\frac{1}{2}z^{-\frac{4}{3%
}}X\varrho _{\underline{\mathbf{1}}}Y^{2} & z^{-\frac{1}{3}}\varrho _{%
\underline{\mathbf{16}}}+z^{-\frac{4}{3}}X\varrho _{\underline{\mathbf{1}}}Y
& z^{-\frac{4}{3}}X\varrho _{\underline{\mathbf{1}}} \\
\frac{1}{2}z^{-\frac{4}{3}}\varrho _{\underline{\mathbf{1}}}Y^{2} & z^{-%
\frac{4}{3}}\varrho _{\underline{\mathbf{1}}}Y & z^{-\frac{4}{3}}\varrho _{%
\underline{\mathbf{1}}}%
\end{array}%
\right)
\end{equation}%
which is also obtained in \cite{54}. By substituting eqs(\ref{rxy}) and (\ref%
{nl}) into the expansions $X=b^{^{\beta }}X_{\beta }$ and $Y=c_{\beta
}Y^{\beta }$ as well as into their squares $X^{2}$ and $Y^{2}$, we obtain
\begin{equation}
\mathcal{L}_{\boldsymbol{27}}^{\mathbf{\mu }}=\left(
\begin{array}{lll}
z^{\frac{2}{3}}+z^{-\frac{1}{3}}b^{\beta }c_{\beta }+\frac{1}{4}z^{-\frac{4}{%
3}}V^{i}W_{i} & z^{-\frac{1}{3}}b^{\beta }\Gamma _{\beta \gamma }^{i}+\frac{1%
}{2}z^{-\frac{4}{3}}V^{i}c_{\beta } & \frac{1}{2}z^{-\frac{4}{3}}V^{i} \\
z^{-\frac{1}{3}}c_{\beta }\Gamma _{i}^{\beta \gamma }+\frac{1}{2}z^{-\frac{4%
}{3}}b^{\beta }W_{i} & z^{-\frac{1}{3}}+z^{-\frac{4}{3}}b^{\beta }c_{\beta }
& z^{-\frac{4}{3}}b^{\beta } \\
\frac{1}{2}z^{-\frac{4}{3}}W_{i} & z^{-\frac{4}{3}}c_{\beta } & z^{-\frac{4}{%
3}}%
\end{array}%
\right)  \label{zm}
\end{equation}%
where $V^{i}=\frac{1}{2}\mathbf{b}\Gamma ^{i}\mathbf{b}$ and $W_{i}=\frac{1}{%
2}\mathbf{c}\Gamma _{i}\mathbf{c}.$ This is the most convenient expression
of the coupling between 'tH$_{\mathrm{\gamma }_{0}}^{\mu _{e_{6}}}$ and $W_{%
\mathrm{\xi }_{z}}^{27}$ in the E$_{6}$ CS theory allowing to derive the
associated topological quiver Q$_{\boldsymbol{27}}^{\mathbf{\mu }}$. In
fact, by writing the L-operator like $\left \langle \varrho _{\underline{%
\boldsymbol{R}}_{i}}|\mathcal{L}^{\mathbf{\mu }}|\varrho _{\underline{%
\boldsymbol{R}}_{j}}\right \rangle $, which is
\begin{equation}
\mathcal{L}_{ij}^{\mathbf{\mu }}=\varrho _{\underline{\boldsymbol{R}}_{i}}%
\mathcal{L}^{\mathbf{\mu }}\varrho _{\underline{\boldsymbol{R}}_{j}}
\end{equation}%
We have in terms of the projectors :
\begin{equation}
\mathcal{L}_{\boldsymbol{27}}^{\mathbf{\mu }}=\left(
\begin{array}{ccc}
\varrho _{\underline{\mathbf{10}}}\mathcal{L}^{\mathbf{\mu }}\varrho _{%
\underline{\mathbf{10}}} & \varrho _{\underline{\mathbf{10}}}\mathcal{L}^{%
\mathbf{\mu }}\varrho _{\underline{\mathbf{16}}} & \varrho _{\underline{%
\mathbf{10}}}\mathcal{L}^{\mathbf{\mu }}\varrho _{\underline{\mathbf{1}}} \\
\varrho _{\underline{\mathbf{16}}}\mathcal{L}^{\mathbf{\mu }}\varrho _{%
\underline{\mathbf{10}}} & \varrho _{\underline{\mathbf{16}}}\mathcal{L}^{%
\mathbf{\mu }}\varrho _{\underline{\mathbf{16}}} & \varrho _{\underline{%
\mathbf{16}}}\mathcal{L}^{\mathbf{\mu }}\varrho _{\underline{\mathbf{1}}} \\
\varrho _{\underline{\mathbf{1}}}\mathcal{L}^{\mathbf{\mu }}\varrho _{%
\underline{\mathbf{10}}} & \varrho _{\underline{\mathbf{1}}}\mathcal{L}^{%
\mathbf{\mu }}\varrho _{\underline{\mathbf{16}}} & \varrho _{\underline{%
\mathbf{1}}}\mathcal{L}^{\mathbf{\mu }}\varrho _{\underline{\mathbf{1}}}%
\end{array}%
\right)  \label{ent}
\end{equation}%
\begin{equation*}
\text{ \ \ }
\end{equation*}%
This directly indicates that the topological gauge quiver Q$_{\boldsymbol{27}%
}^{\mathbf{\mu }}$ has three nodes $\mathcal{N}_{1}$, $\mathcal{N}_{2}$, $%
\mathcal{N}_{3}$ and six links, three L$_{ij}$ and three L$_{ji}$ with $%
i>j=1,2,3$, as depicted by the Figure \textbf{\ref{Q6}}.
\newline
The $\mathcal{N}_{i}$ nodes are associated with the diagonal enties of (\ref%
{ent}), namely
\begin{equation}
\mathcal{N}_{1}\equiv \varrho _{\underline{\mathbf{10}}}\mathcal{L}^{\mathbf{%
\mu }}\varrho _{\underline{\mathbf{10}}}\qquad ,\qquad \mathcal{N}_{2}\equiv
\varrho _{\underline{\mathbf{16}}}\mathcal{L}^{\mathbf{\mu }}\varrho _{%
\underline{\mathbf{16}}}\qquad ,\qquad \mathcal{N}_{3}\equiv \varrho _{%
\underline{\mathbf{1}}}\mathcal{L}^{\mathbf{\mu }}\varrho _{\underline{%
\mathbf{1}}}
\end{equation}%
We will refer to them in terms of the $SO_{2}\times SO_{10}$ representations
as follows%
\begin{equation}
\begin{tabular}{lll}
$\mathcal{N}_{1}$ & : & $\mathbf{10}_{+2/3}$ \\
$\mathcal{N}_{2}$ & : & $\mathbf{16}_{-1/3}$ \\
$\mathcal{N}_{3}$ & : & $\mathbf{1}_{-4/3}$%
\end{tabular}%
\end{equation}%
The L$_{ij}$ links of the quiver Q$_{\boldsymbol{27}}^{\mathbf{\mu }}$ are
given by the off diagonal terms $\varrho _{\underline{\boldsymbol{R}}_{i}}%
\mathcal{L}^{\mathbf{\mu }}\varrho _{\underline{\boldsymbol{R}}_{j}}$ with $%
i\neq j$. These links transform in the fundamental representations of $%
SO_{2}\times SO_{10}$ knowing that $\mathbf{10}$ and $\mathbf{16}$ and their
duals are fundamental representations of $SO_{10}$. The explicit expressions
of these links are given in the following table\
\begin{equation}
\begin{tabular}{|l|l|l||l|l|l|}
\hline
{\small link} & {\small \ \ \ Repres} & {\small bi-matter} & {\small link} &
{\small \ \ \ Repres} & {\small bi-matter} \\ \hline
$L_{1\rightarrow 2}$ & $\mathbf{16}_{-\frac{1}{3}}\times \mathbf{10}_{-\frac{%
2}{3}}$ & \textbf{\ }$\ \mathbf{b,}$ $\mathbf{b}^{2}c$ & $L_{2\rightarrow 1}$
& $\mathbf{10}_{\frac{2}{3}}\times \mathbf{16}_{\frac{1}{3}}$ & \textbf{\ }$%
\ \mathbf{c,}$ $\mathbf{bc}^{2}$ \\ \hline
$L_{2\rightarrow 3}$ & $\ \mathbf{1}_{-\frac{4}{3}}\times \mathbf{16}_{+%
\frac{1}{3}}$ & \textbf{\ }$\ \ \ \ \mathbf{b}$ & $L_{2\rightarrow 1}$ & $%
\mathbf{16}_{-\frac{1}{3}}\times \boldsymbol{1}_{\frac{4}{3}}$ & \textbf{\ }$%
\ \ \ \ \mathbf{c}$ \\ \hline
$L_{1\rightarrow 3}$ & $\ \mathbf{1}_{-\frac{4}{3}}\times \mathbf{10}_{-%
\frac{2}{3}}$ & \textbf{\ }$\ \ \ \ \mathbf{b}^{2}$ & $L_{3\rightarrow 1}$ &
$\mathbf{10}_{\frac{2}{3}}\times \boldsymbol{1}_{+\frac{4}{3}}$ & \textbf{\ }%
$\ \ \ \ \mathbf{c}^{2}$ \\ \hline
\end{tabular}%
\end{equation}

\section{Minuscule line defects in E$_{7}$ CS theory}

In this section, we complete the study undertaken in this paper regarding
the minuscule L-operators of ADE type by investigating the case of 4D Chern
Simons theory with exceptional E$_{7}$ gage symmetry. Just as before, we
treat this theory by studying the properties of interacting minuscule 't
Hooft and Wilson lines, and construct the Lax operators $\mathcal{L}_{%
\boldsymbol{R}_{\boldsymbol{e}_{7}}}^{\mu }$ and the associated topological
gauge quivers Q$_{\boldsymbol{R}_{\boldsymbol{e}_{7}}}^{\mu }$ by focusing
on the fundamental $\boldsymbol{R}_{\boldsymbol{e}_{7}}=\mathbf{56}$.

\subsection{Levi subalgebra of E$_{7}$ and weights of the \textbf{56}$_{%
\mathbf{e}_{7}}$}

\begin{figure}[h]
\begin{center}
\includegraphics[width=8cm]{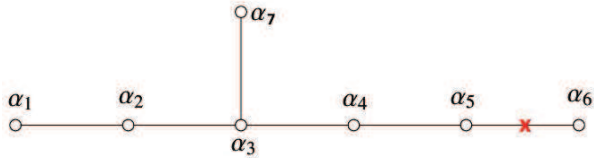}
\end{center}
\par
\vspace{-0.5cm}
\caption{Dynkin Diagram of E$_{7}$ having seven nodes labeled by the simple
roots $\protect\alpha _{i}$. The cross $\left( \times \right) $ indicates
the root cut by the Levi decomposition where the Levi subgroup is $%
SO_{2}\times E_{6}.$}
\label{E7}
\end{figure}

First, we begin by recalling the useful aspects of the e$_{7}$ Lie algebra
that will play an important role in our construction. In particular, the
root system $\Phi _{\boldsymbol{e}_{7}}$ containing 126 roots is generated
by seven simple roots $\alpha _{i}$ realised as follows%
\begin{equation}
\begin{tabular}{lllll}
E$_{7}$ & $:$ & $\alpha _{1}$ & $=$ & $\frac{1}{2}\left( \epsilon
_{1}-\epsilon _{2}-\epsilon _{3}-\epsilon _{4}-\epsilon _{5}-\epsilon
_{6}-\epsilon _{7}+\epsilon _{8}\right) $ \\
&  & $\alpha _{i}$ & $=$ & $\epsilon _{i}-\epsilon _{i-1}\qquad ,\qquad
i=2,3,4,6$ \\
&  & $\alpha _{7}$ & $=$ & $\epsilon _{1}+\epsilon _{2}$%
\end{tabular}
\label{a}
\end{equation}%
The Dynkin diagram underlying the gauge symmetry of the 4D CS theory with E$%
_{7}$ symmetry is given by the Figure \ref{E7} where the seven simple roots $%
\alpha _{i}$ are exhibited.
\newline
The associated Cartan matrix $K_{e_{7}}$ reads as%
\begin{equation}
K_{e_{7}}=\left(
\begin{array}{ccccccc}
{\small 2} & {\small -1} & {\small 0} & {\small 0} & {\small 0} & {\small 0}
& {\small 0} \\
{\small -1} & {\small 2} & {\small -1} & {\small 0} & {\small 0} & {\small 0}
& {\small 0} \\
{\small 0} & {\small -1} & {\small 2} & {\small -1} & {\small 0} & {\small 0}
& {\small -1} \\
{\small 0} & {\small 0} & {\small -1} & {\small 2} & {\small -1} & {\small 0}
& {\small 0} \\
{\small 0} & {\small 0} & {\small 0} & {\small -1} & {\small 2} & {\small 0}
& {\small 0} \\
{\small 0} & {\small 0} & {\small 0} & {\small 0} & {\small 0} & {\small 2}
& {\small 0} \\
{\small 0} & {\small 0} & {\small -1} & {\small 0} & {\small 0} & {\small 0}
& {\small 2}%
\end{array}%
\right)
\end{equation}%
It describes the intersection matrix $\alpha _{i}.\alpha _{j}$ while its
inverse gives the fundamental coweights of E$_{7}.$ One of these coweights
is particularly interesting for our present study; the $\mu $ dual to $%
\alpha _{6}$ is the only minuscule coweight of $e_{7}$.

\subsubsection{Minuscule coweight of E$_{7}$}

From the Cartan matrix $K_{e_{7}}$, we can learn useful informations
regarding the Lie algebra $e_{7}$ and its representations, in particular the
expressions of fundamental weights $\omega _{i}$ in terms of simple roots :
\begin{equation}
\begin{tabular}{|l|l|}
\hline
{\small fund- }$\omega _{i}$ & {\small \ \ \ \ \ \ \ \ \ \ \ \ in terms of
roots} \\ \hline
$\omega _{1}$ & $2\alpha _{1}+3\alpha _{2}+4\alpha _{3}+3\alpha _{4}+2\alpha
_{5}+\alpha _{6}+2\alpha _{7}$ \\ \hline
$\omega _{2}$ & $3\alpha _{1}+6\alpha _{2}+8\alpha _{3}+6\alpha _{4}+4\alpha
_{5}+2\alpha _{6}+4\alpha _{7}$ \\ \hline
$\omega _{3}$ & $4\alpha _{1}+8\alpha _{2}+12\alpha _{3}+9\alpha
_{4}+6\alpha _{5}+3\alpha _{6}+6\alpha _{7}$ \\ \hline
$\omega _{4}$ & $3\alpha _{1}+6\alpha _{2}+9\alpha _{3}+\frac{15}{2}\alpha
_{4}+5\alpha _{5}+\frac{5}{2}\alpha _{6}+\frac{9}{2}\alpha _{7}$ \\ \hline
$\omega _{5}$ & $2\alpha _{1}+4\alpha _{2}+6\alpha _{3}+5\alpha _{4}+4\alpha
_{5}+2\alpha _{6}+3\alpha _{7}$ \\ \hline
$\omega _{6}$ & $\alpha _{1}+2\alpha _{2}+3\alpha _{3}+\frac{5}{2}\alpha
_{4}+2\alpha _{5}+\frac{3}{2}\alpha _{6}+\frac{3}{2}\alpha _{7}$ \\ \hline
$\omega _{7}$ & $2\alpha _{1}+4\alpha _{2}+6\alpha _{3}+\frac{9}{2}\alpha
_{4}+3\alpha _{5}+\frac{3}{2}\alpha _{6}+\frac{7}{2}\alpha _{7}$ \\ \hline
\end{tabular}%
\end{equation}%
\begin{equation*}
\text{ \ \ }
\end{equation*}%
The exceptional Lie algebra e$_{7}$ has one minuscule coweight $\mu $ given
by $\omega _{6}$, thus the corresponding Levi decomposition $\boldsymbol{n}%
_{-}\oplus \boldsymbol{l}_{\mu }\oplus \boldsymbol{n}_{+}$ for this algebra
is given by%
\begin{equation}
\boldsymbol{l}_{\mu }=so_{2}\oplus e_{6}\qquad ,\qquad \boldsymbol{n}_{\pm
}=27_{\pm }
\end{equation}%
The dimensions of $\boldsymbol{n}_{\pm }$ can be calculated by dispatching
the algebraic dimensions of $e_{7}$ with respect to $so_{2}\oplus e_{6}$, in
fact we have $133=1+78+27+27.$ This Levi decomposition with respect to the
minuscule coweight $\mu $ requires the following adjoint actions%
\begin{equation}
\left[ \mathbf{\mu },\boldsymbol{n}_{\pm }\right] =\pm \boldsymbol{n}_{\pm
}\qquad ,\qquad \left[ \boldsymbol{n}_{+},\boldsymbol{n}_{-}\right] =0
\label{lct}
\end{equation}%
These constraints show that the $27$ generators X$_{\beta }$ of the
nilpotent algebra $\boldsymbol{n}_{+}$ and the $27$ generators Y$^{\beta }$
of the algebra $\boldsymbol{n}_{-}$ have opposite $so_{2}$ charges $\pm 1$,
which is important to consider when realising the action of X$_{\beta }$ and
Y$^{\beta }$ on the electrically charged quantum states $\left \vert
A\right
\rangle $ that we take in the fundamental representation of E$_{7}$.

\subsubsection{Representation $\mathbf{56}$ of the e$_{7}$ Lie algebra}

\begin{figure}[ph]
\begin{center}
\includegraphics[width=9cm]{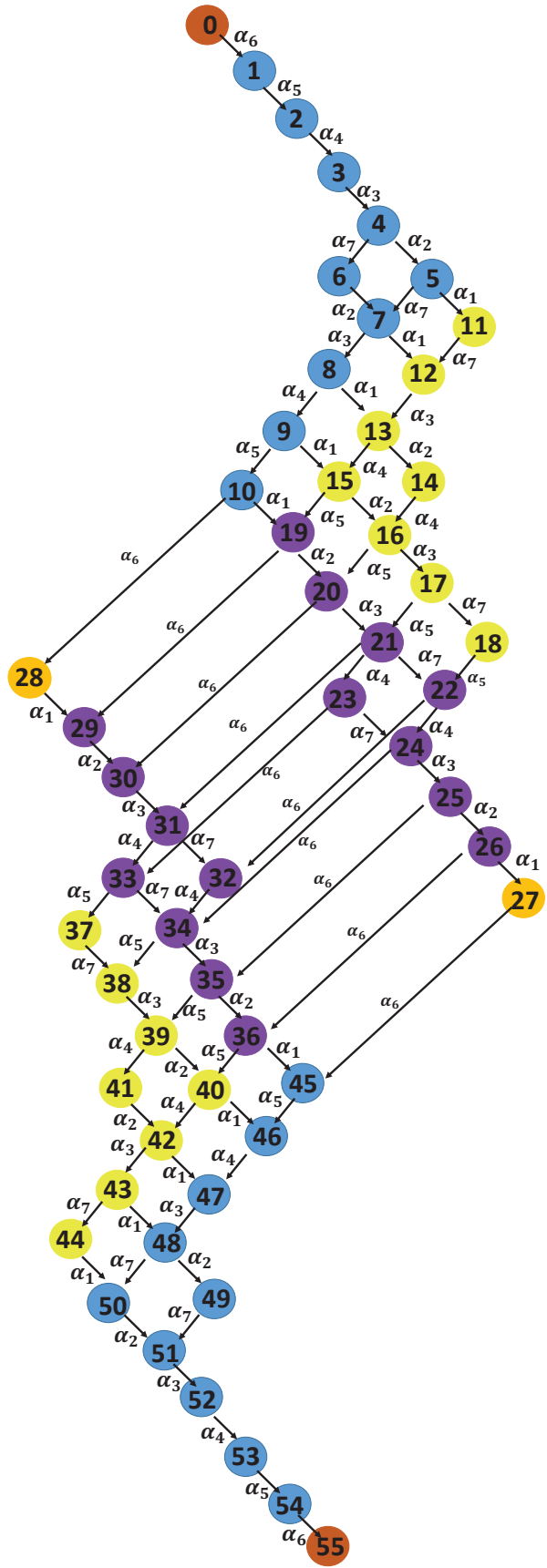}
\end{center}
\par
\vspace{-0.5cm}
\caption{The decomposition of the $\mathbf{56}$ representation of $e_{7}$ in
terms of representations of $e_{6}$. We have $\mathbf{56}=\mathbf{28}%
_{+}\oplus \mathbf{28}_{-}$ where $\mathbf{28}_{\pm }$ are reducible like $%
\mathbf{1}_{\pm 3/2}\oplus \mathbf{27}_{\pm 1/2}.$}
\label{56}
\end{figure}

\begin{figure}[h]
\begin{center}
\includegraphics[width=10cm]{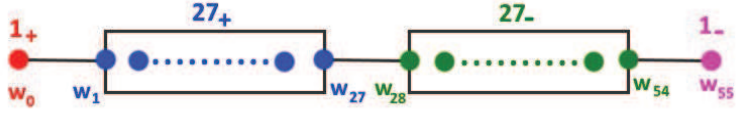}
\end{center}
\par
\vspace{-0.5cm}
\caption{The decomposition of the $\mathbf{56}$ representation of E$_{7}$ in
terms of representations of E$_{6}$. We have $\mathbf{56}=\mathbf{28}%
_{+}\oplus \mathbf{28}_{-}$ with $\mathbf{28}_{\pm }$ reducible like $%
\mathbf{1}_{\pm 3/2}\oplus \mathbf{27}_{\pm 1/2}.$}
\label{127}
\end{figure}

The fundamental representation of the $\boldsymbol{e}_{7}$ algebra has 56
dimensions, it is self dual and pseudo-real \textrm{\cite{60}}. Its weight
diagram is given by the Figure\textbf{\ \ref{56}} where the weight $\xi _{0}$
of the top state $\left \vert \xi _{0}\right \rangle $ corresponds to the
minuscule coweight $\omega _{6}$ while the weight $\xi _{55}$ of the bottom
state $\left \vert \xi _{55}\right \rangle $ is precisely $-\omega _{6}$,
meaning that we have $\xi _{0}+\xi _{55}=0$.
\newline
Under the Levi decomposition associated to the minuscule $\mu $, the
fundamental representation \textbf{56} decomposes as a reducible sum of $%
so_{2}\oplus e_{6}$ representations as follows
\begin{equation}
\begin{tabular}{lll}
$\mathbf{56}_{0}$ & $=$ & $\mathbf{28}_{+}\oplus \mathbf{28}_{-}$ \\
$\mathbf{28}_{+}\oplus \mathbf{28}_{-}$ & $=$ & $\mathbf{1}_{3/2}\oplus
\mathbf{27}_{+1/2}\oplus \mathbf{27}_{-1/2}\oplus \mathbf{1}_{-3/2}$%
\end{tabular}
\label{28}
\end{equation}
where we have four $e_{6}$ representations, two singlets $\mathbf{1}_{\pm
3/2}$ and two fundamentals $\mathbf{27}_{\pm 1/2}$.\newline
In the diagram of Figure \textbf{\ref{127}}, the 28 weights of $\mathbf{28}%
_{+}$ are labeled by the subset $W_{+}=\left \{ \left \vert \xi
_{i}\right
\rangle \right \} _{0\leq i\leq 27} $ and the 28 weights of the $%
\mathbf{28}_{-}$ by $W_{-}=\left \{ \left
\vert \xi _{i}\right \rangle
\right \} _{28\leq i\leq 55}$. Weights $\xi _{i} $ in the set $W_{+}\cup
W_{-}$ obey some special features that characterize this exceptional algebra
and that will be helpful for the construction of the operator $\mathcal{L}%
_{e_{7}}^{\mathbf{\mu }}$, they are listed below%
\begin{equation}
\begin{tabular}{lllllll}
$\xi _{27}$ & $=$ & $\xi _{0}-\beta _{\max }$ & , & $\xi _{27}+\xi _{28}$ & $%
=$ & $\xi _{0}+\xi _{55}$ \\
$\xi _{28}$ & $=$ & $\xi _{55}+\beta _{\max }$ & , & $\xi _{i}+\xi _{55-i}$
& $=$ & $\xi _{0}+\xi _{55}$ \\
$\xi _{i}$ & $=$ & $\xi _{0}-\gamma _{i}$ & , & $\xi _{55-i}$ & $=$ & $\xi
_{55}+\gamma _{i}$%
\end{tabular}%
\end{equation}%
for a generic root $\gamma _{i}$ in the nilpotent $\mathbf{27}_{+}$ and
where $\beta _{\max }$ is given by%
\begin{equation}
\beta _{\max }=2\alpha _{1}+3\alpha _{2}+4\alpha _{3}+3\alpha _{4}+2\alpha
_{5}+\alpha _{6}+2\alpha _{7}
\end{equation}%
We also have%
\begin{equation}
\begin{tabular}{lllllll}
$\xi _{0}-\xi _{55}$ & $=$ & $2\omega _{6}$ & , & $\xi _{i}-\xi _{55-i}$ & $%
= $ & $2\omega _{6}-2\gamma _{i}$%
\end{tabular}%
\end{equation}%
The list of the ten weights $\xi _{A},A=1,...,10$ represented by blue dots
in the Figure\textbf{\ \ref{56}} is given in the following table in terms of
the seven $\omega _{i}$'s,
\begin{equation}
\begin{tabular}{lllllll}
$\xi _{1}$ & $=$ & $\omega _{5}-\omega _{6}$ & , & $\xi _{6}$ & $=$ & $%
\omega _{2}-\omega _{7}$ \\
$\xi _{2}$ & $=$ & $\omega _{4}-\omega _{5}$ & , & $\xi _{7}$ & $=$ & $%
\omega _{1}+\omega _{3}-\omega _{7}-\omega _{2}$ \\
$\xi _{3}$ & $=$ & $\omega _{3}-\omega _{4}$ & , & $\xi _{8}$ & $=$ & $%
\omega _{1}+\omega _{4}-\omega _{3}$ \\
$\xi _{4}$ & $=$ & $\omega _{7}+\omega _{2}-\omega _{3}$ & , & $\xi _{9}$ & $%
=$ & $\omega _{1}+\omega _{5}-\omega _{4}$ \\
$\xi _{5}$ & $=$ & $\omega _{1}+\omega _{7}-\omega _{2}$ & , & $\xi _{10}$ &
$=$ & $\omega _{1}+\omega _{6}-\omega _{5}$%
\end{tabular}%
\end{equation}%
while the next sixteen states (8+8) represented in the Figure\textbf{\ \ref%
{56}} by yellow and magenta colored dots (from $\xi _{19}$ to $\xi _{26}$)
are listed here
\begin{equation}
\begin{tabular}{lllllll}
$\xi _{11}$ & $=$ & $\omega _{7}-\omega _{1}$ & , & $\xi _{15}$ & $=$ & $%
\omega _{5}+\omega _{2}-\omega _{4}-\omega _{1}$ \\
$\xi _{12}$ & $=$ & $-\omega _{7}-\omega _{1}$ & , & $\xi _{16}$ & $=$ & $%
\omega _{5}+\omega _{3}-\omega _{4}-\omega _{2}$ \\
$\xi _{13}$ & $=$ & $\omega _{2}+\omega _{4}-\omega _{3}-\omega _{1}$ & , & $%
\xi _{17}$ & $=$ & $\omega _{7}+\omega _{5}-\omega _{3}$ \\
$\xi _{14}$ & $=$ & $\omega _{4}-\omega _{2}$ & , & $\xi _{18}$ & $=$ & $%
\omega _{5}-\omega _{7}$%
\end{tabular}%
\end{equation}%
and%
\begin{equation}
\begin{tabular}{lllllll}
$\xi _{19}$ & $=$ & $\omega _{2}+\omega _{6}-\omega _{1}-\omega _{5}$ & , & $%
\xi _{23}$ & $=$ & $\omega _{7}+\omega _{6}-\omega _{4}$ \\
$\xi _{20}$ & $=$ & $\omega _{3}+\omega _{6}-\omega _{5}-\omega _{2}$ & , & $%
\xi _{24}$ & $=$ & $\omega _{3}+\omega _{6}-\omega _{4}-\omega _{7}$ \\
$\xi _{21}$ & $=$ & $\omega _{7}+\omega _{4}+\omega _{6}-\omega _{5}-\omega
_{3}$ & , & $\xi _{25}$ & $=$ & $\omega _{2}+\omega _{6}-\omega _{3}$ \\
$\xi _{22}$ & $=$ & $\omega _{4}+\omega _{6}-\omega _{5}-\omega _{7}$ & , & $%
\xi _{26}$ & $=$ & $\omega _{1}+\omega _{6}-\omega _{2}$%
\end{tabular}%
\end{equation}%
The last 27-th weight is equal to $\xi _{27}=\omega _{6}-\omega _{1}.$%

\subsection{Constructing the $\mathcal{L}_{\mathbf{56}}^{\mathbf{\protect\mu
}}$}

Now, we consider the minuscule 't Hooft line embedded in the E$_{7}$ CS
theory crossing a Wilson line $W_{e_{7}}^{\boldsymbol{R}}$ with electric
weight given by the representation $\mathbf{56}$. To construct the
L-operator $\mathcal{L}_{\mathbf{56}}^{\mathbf{\mu }}$ describing these
topological lines' coupling, we follow the same approach adopted before for
the study A-, D- and E$_{6}$ type theories.

\subsubsection{Realising the generators of the \textbf{n}$_{\pm 27}$
subalgebras}

We begin by recalling that in the L-operator formula for the E$_{7}$
symmetry, namely $\mathcal{L}_{\mathbf{56}}^{\mathbf{\mu }}=e^{X}z^{\mathbf{%
\mu }}e^{Y}\label{e7}$, the $\mathbf{\mu }$ is the minuscule coweight given
in (\ref{E7}) and X and Y are nilpotent matrices expanding as%
\begin{equation}
X=\sum_{\beta =1}^{27}b^{^{\beta }}X_{\beta }\qquad ,\qquad Y=\sum_{\beta
=1}^{27}c_{\beta }Y^{\beta }  \label{lcc}
\end{equation}%
Here, the twenty seven $b^{\beta }$ and twenty seven $c_{\beta }$ are the
Darboux coordinates of the phase space of the E$_{7}$-type 't Hooft line.
The realisation of the nilpotent generators $X_{\beta }$ and $Y^{\beta }$
can be first written using simple representation language like
\begin{equation}
\begin{tabular}{lll}
$X_{\beta }$ & $\equiv $ & $\left \vert \mathbf{1}_{+}\right \rangle \left
\langle \mathbf{27}_{+}\right \vert +\left \vert \mathbf{27}_{+}\right
\rangle \left \langle \mathbf{27}_{-}\right \vert +\left \vert \mathbf{27}%
_{-}\right \rangle \left \langle \mathbf{1}_{-}\right \vert $ \\
$Y^{\beta }$ & $\equiv $ & $\left \vert \mathbf{1}_{-}\right \rangle \left
\langle \mathbf{27}_{-}\right \vert +\left \vert \mathbf{27}_{-}\right
\rangle \left \langle \mathbf{27}_{+}\right \vert +\left \vert \mathbf{27}%
_{+}\right \rangle \left \langle \mathbf{1}_{+}\right \vert $%
\end{tabular}
\label{xby}
\end{equation}%
where we dropped the charges from $\mathbf{1}_{\pm 3/2}$ and $\mathbf{27}%
_{\pm 1/2}$ for simplicity. The explicit form of these generators in terms
of the weight states $\left \vert \xi _{A}\right \rangle $ and their duals $%
\left \langle \xi _{A}\right \vert $ is given by
\begin{equation}
\begin{tabular}{lll}
$X_{\beta }$ & $=$ & $\left \vert \xi _{0_{+}}\right \rangle \left \langle
\xi _{\beta _{+}}\right \vert +\left \vert \xi _{\delta _{+}}\right \rangle
\Gamma _{\beta }^{\delta _{+}\gamma _{-}}\left \langle \xi _{\gamma
_{-}}\right \vert +\left \vert \xi _{\beta _{-}}\right \rangle \left \langle
\xi _{0_{-}}\right \vert $ \\
$Y^{\beta }$ & $=$ & $\left \vert \xi _{0_{-}}\right \rangle \left \langle
\xi ^{\beta _{-}}\right \vert +\left \vert \xi ^{\gamma _{-}}\right \rangle
\Gamma _{\gamma _{-}\delta _{+}}^{\beta }\left \langle \xi ^{\delta
_{+}}\right \vert +\left \vert \xi ^{\beta _{+}}\right \rangle \left \langle
\xi _{0_{+}}\right \vert $%
\end{tabular}
\label{bb}
\end{equation}%
where $\Gamma _{\beta }^{\delta _{+}\gamma _{-}}$ and $\Gamma _{\gamma
_{-}\delta _{+}}^{\beta }$ are couplings between states in the $\mathbf{27}$
representations of E$_{6}$; these tensors are allowed by the tensor product
of E$_{6}$ representations \textrm{\cite{61}}. The adjoint form of the
minuscule coweight used is given by%
\begin{equation}
\mathbf{\mu }=\frac{3}{2}\varrho _{\mathbf{1}_{+}}+\frac{1}{2}\varrho _{%
\mathbf{27}_{+}}-\frac{1}{2}\varrho _{\mathbf{27}_{-}}-\frac{3}{2}\varrho _{%
\mathbf{1}_{-}}  \label{mm}
\end{equation}%
where the four $\varrho _{\boldsymbol{R}_{i}}$'s are projectors on the e$%
_{6} $ representations $\boldsymbol{R}_{i}$ within the $\mathbf{56}$ of $%
e_{7}$, they read as follows%
\begin{equation}
\varrho _{\mathbf{1}_{q}}=\left \vert \xi _{0_{q}}\right \rangle \left
\langle \xi _{0_{q}}\right \vert \qquad ,\qquad \varrho _{\mathbf{27}%
_{q}}=\left \vert \xi _{27_{q}}\right \rangle \left \langle \xi
_{27_{q}}\right \vert
\end{equation}%
with $q=\pm $ and $\left \langle \xi _{0_{q}}|\xi _{0_{q}}\right \rangle
=\left \langle \xi _{27_{q}}|\xi _{27_{q}}\right \rangle =1$. They can also
be written in formal notations as
\begin{equation}
\varrho _{\mathbf{1}_{q}}=\left \vert \mathbf{1}_{q}\right \rangle \left
\langle \mathbf{1}_{q}\right \vert \qquad ,\qquad \varrho _{\mathbf{27}%
_{q}}=\left \vert \mathbf{27}_{q}\right \rangle \left \langle \mathbf{27}%
_{q}\right \vert
\end{equation}%
Now, we need to compute the powers of the generators $X_{\beta }$ and $%
Y^{\beta }$ that will appear in the expansion of the L-operator. We find
using the realisation (\ref{xby}-\ref{bb}) that the non vanishing monomials
are
\begin{equation}
\begin{tabular}{lllllll}
$X_{\alpha }X_{\beta }$ & $\equiv $ & $\left \vert \mathbf{1}_{+}\right
\rangle \left \langle \mathbf{27}_{-}\right \vert +\left \vert \mathbf{27}%
_{+}\right \rangle \left \langle \mathbf{1}_{-}\right \vert $ & , & $%
X_{\alpha }X_{\beta }X_{\gamma }$ & $\equiv $ & $\left \vert \mathbf{1}%
_{+}\right \rangle \left \langle \mathbf{1}_{-}\right \vert $ \\
$Y^{\alpha }Y^{\beta }$ & $\equiv $ & $\left \vert \mathbf{1}_{-}\right
\rangle \left \langle \mathbf{27}_{+}\right \vert +\left \vert \mathbf{27}%
_{-}\right \rangle \left \langle \mathbf{1}_{+}\right \vert $ & , & $%
Y^{\alpha }Y^{\beta }Y^{\gamma }$ & $\equiv $ & $\left \vert \mathbf{1}%
_{-}\right \rangle \left \langle \mathbf{1}_{+}\right \vert $%
\end{tabular}%
\end{equation}%
while the fourth order powers vanish identically. For the powers of the
linear combinations $X=b^{^{\beta }}X_{\beta }$ and $Y=c_{\beta }Y^{\beta }$%
, we find%
\begin{equation}
\begin{tabular}{lll}
$X^{2}$ & $=$ & $2S^{\beta _{-}}\left \vert \xi _{0_{+}}\right \rangle \left
\langle \xi _{\beta _{-}}\right \vert +2S^{\beta _{+}}\left \vert \xi
_{\beta _{+}}\right \rangle \left \langle \xi _{0_{-}}\right \vert $ \\
$Y^{2}$ & $=$ & $2R_{\alpha _{+}}\left \vert \xi _{0_{-}}\right \rangle
\left \langle \xi ^{\alpha _{+}}\right \vert +2R_{\alpha _{-}}\left \vert
\xi ^{\alpha _{-}}\right \rangle \left \langle \xi _{0_{+}}\right \vert $%
\end{tabular}
\label{xx}
\end{equation}%
and%
\begin{equation}
\begin{tabular}{lll}
$X^{3}$ & $=$ & $6\mathcal{E}\left \vert \xi _{0_{+}}\right \rangle \left
\langle \xi _{0_{-}}\right \vert $ \\
$Y^{3}$ & $=$ & $6\mathcal{F}\left \vert \xi _{0_{-}}\right \rangle \left
\langle \xi _{0_{+}}\right \vert $%
\end{tabular}%
\end{equation}%
and of course, $X^{4}=Y^{4}=0$. The realisation (\ref{xby}-\ref{bb}) does
also obey the commutation relations $\left[ \mathbf{\mu ,}X_{\beta }\right]
=X_{\beta }$ and $\left[ \mathbf{\mu ,}Y^{\beta }\right] =-Y^{\beta }$ from
which we deduce that
\begin{equation}
\left[ \mathbf{\mu ,}X\right] =X\qquad ,\qquad \left[ \mathbf{\mu ,}Y\right]
=-Y
\end{equation}%
as required by the Levi decomposition with respect to $\mu $.

\subsubsection{The L-operator $\mathcal{L}_{\mathbf{56}}^{\mathbf{\protect%
\mu }}$}

Finally, to obtain the expression of $\mathcal{L}_{\mathbf{56}}^{\mathbf{\mu
}}$ in terms of the 27+27 Darboux coordinates $b^{\beta }$ and $c_{\beta },$
we use the nilpotency properties mentioned above to write
\begin{equation}
\mathcal{L}_{\mathbf{56}}^{\mathbf{\mu }}=\left( I+X+\frac{1}{2}X^{2}+\frac{1%
}{6}X^{3}\right) z^{\mathbf{\mu }}\left( I+Y+\frac{1}{2}Y^{2}+\frac{1}{6}%
Y^{3}\right)  \label{le}
\end{equation}%
and substitute with
\begin{equation}
z^{\mathbf{\mu }}=z^{\frac{3}{2}}\varrho _{\mathbf{1}_{+}}+z^{\frac{1}{2}%
}\varrho _{\mathbf{27}_{+}}+z^{-\frac{1}{2}}\varrho _{\mathbf{27}_{-}}+z^{-%
\frac{3}{2}}\varrho _{\mathbf{1}_{-}}
\end{equation}%
We moreover need to take into account the special properties of the X and Y
matrices, like for example $X\varrho _{\mathbf{1}_{+}}=0$ and $\varrho _{%
\mathbf{1}_{+}}Y=0$, to reduce the monomials of this L-operator down to 30
as given below%
\begin{equation}
\begin{tabular}{lll}
$\mathcal{L}_{\mathbf{56}}^{\mathbf{\mu }}$ & $=$ & $z^{\frac{3}{2}}\varrho
_{\mathbf{1}_{+}}+z^{\frac{1}{2}}\varrho _{\mathbf{27}_{+}}+z^{-\frac{1}{2}%
}\varrho _{\mathbf{27}_{-}}+z^{-\frac{3}{2}}\varrho _{\mathbf{1}_{-}}$ \\
&  & $z^{\frac{1}{2}}X\varrho _{\mathbf{27}_{+}}+z^{-\frac{1}{2}}X\varrho _{%
\mathbf{27}_{-}}+z^{-\frac{3}{2}}X\varrho _{\mathbf{1}_{-}}+$ \\
&  & $z^{\frac{1}{2}}\varrho _{\mathbf{27}_{+}}Y+z^{-\frac{1}{2}}\varrho _{%
\mathbf{27}_{-}}Y+z^{-\frac{3}{2}}\varrho _{\mathbf{1}_{-}}Y+$ \\
&  & $+\frac{1}{2}X^{2}z^{-\frac{1}{2}}\varrho _{\mathbf{27}_{-}}+\frac{1}{2}%
z^{-\frac{3}{2}}X^{2}\varrho _{\mathbf{1}_{-}}+\frac{1}{6}z^{-\frac{3}{2}%
}X^{3}\varrho _{\mathbf{1}_{-}}+$ \\
&  & $\frac{1}{2}z^{-\frac{1}{2}}\varrho _{\mathbf{27}_{-}}Y^{2}+\frac{1}{2}%
z^{-\frac{3}{2}}\varrho _{\mathbf{1}_{-}}Y^{2}+\frac{1}{6}z^{-\frac{3}{2}%
}\varrho _{\mathbf{1}_{-}}Y^{3}+$ \\
&  & $z^{\frac{1}{2}}X\varrho _{\mathbf{27}_{+}}Y+z^{-\frac{1}{2}}X\varrho _{%
\mathbf{27}_{-}}Y+z^{-\frac{3}{2}}X\varrho _{\mathbf{1}_{-}}Y$ \\
&  & $\frac{1}{2}z^{-\frac{1}{2}}X\varrho _{\mathbf{27}_{-}}Y^{2}+\frac{1}{2}%
z^{-\frac{3}{2}}X\varrho _{\mathbf{1}_{-}}Y^{2}+$ \\
&  & $\frac{1}{2}z^{-\frac{1}{2}}X^{2}\varrho _{\mathbf{27}_{-}}Y+\frac{1}{2}%
z^{-\frac{3}{2}}X^{2}\varrho _{\mathbf{1}_{-}}Y++\frac{1}{6}z^{-\frac{3}{2}%
}X\varrho _{\mathbf{1}_{-}}Y^{3}+$ \\
&  & $\frac{1}{6}z^{-\frac{3}{2}}X^{3}\varrho _{\mathbf{1}_{-}}Y+\frac{1}{12}%
z^{-\frac{3}{2}}X^{2}\varrho _{\mathbf{1}_{-}}Y^{3}+\frac{1}{12}z^{-\frac{3}{%
2}}X^{3}\varrho _{\mathbf{1}_{-}}Y^{2}+$ \\
&  & $\frac{1}{4}z^{-\frac{1}{2}}X^{2}\varrho _{\mathbf{27}_{-}}Y^{2}+\frac{1%
}{4}z^{-\frac{3}{2}}X^{2}\varrho _{\mathbf{1}_{-}}Y^{2}+\frac{1}{36}z^{-%
\frac{3}{2}}X^{3}\varrho _{\mathbf{1}_{-}}Y^{3}$%
\end{tabular}%
\end{equation}%
The explicit form of $\mathcal{L}_{\mathbf{56}}^{\mathbf{\mu }}$ given in
\textrm{\cite{54}} is obtained by replacing $X=b^{\beta }X_{\beta }$, $%
Y=c_{\beta }Y^{\beta }$ and $\mathbf{\mu }$ by their explicit realisations (%
\ref{bb},\ref{mm},\ref{xx}). This is clearly a cumbersome expression, that's
why we use the quiver gauge description to exhibit the interesting
information encoded in $\mathcal{L}_{\mathbf{56}}^{\mathbf{\mu }}$ and help
visualize the key role of the Darboux coordinates.

\subsection{Topological gauge quiver Q$_{\mathbf{56}}^{\mathbf{\protect\mu }%
} $}

\begin{figure}[h]
\begin{center}
\includegraphics[width=10cm]{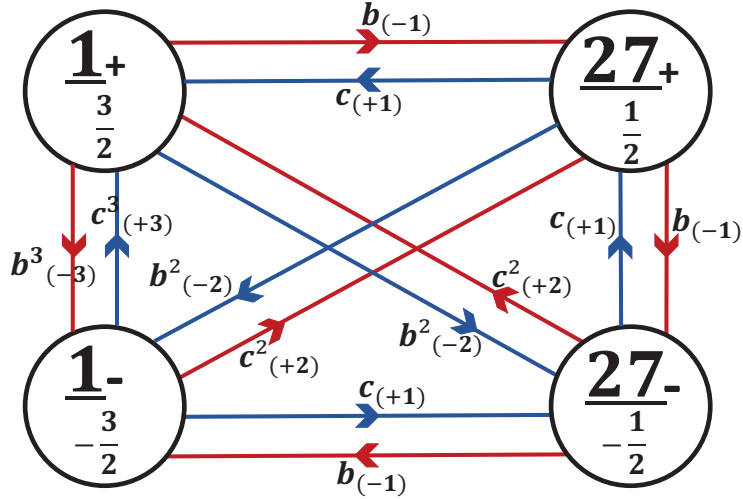}
\end{center}
\par
\vspace{-0.5cm}
\caption{The topological quiver Q$_{\mathbf{56}}^{\mathbf{\protect\mu }}$
representing $\mathcal{L}_{\mathbf{56}}^{\mathbf{\protect\mu }}$. It has 4
nodes and 12 links. The nodes describe self-dual topological matter. The
links describe bi-matter in $\left( R_{i},\bar{R}_{j}\right) $ of E$_{6}$
charged under $SO\left( 2\right) $ with charges $\pm 1,\pm 2,$ $\pm 3.$}
\label{QE7}
\end{figure}

The shape of the gauge quiver Q$_{\mathbf{56}}^{\mathbf{\mu }}$ associated
to the $\mathcal{L}_{\mathbf{56}}^{\mathbf{\mu }}$ operator can be directly
deduced from properties of the $e_{7}$ algebra by comparison with the
previously built quivers for $sl_{N}$, $so_{2N}$ and $e_{6}.$ Firstly, we
can say that the Q$_{\mathbf{56}}^{\mathbf{\mu }}$ has four nodes $\mathcal{N%
}_{i}$ in 1:1 correspondence with the four projectors $\varrho _{\mathbf{1}%
_{\pm }}$ and $\varrho _{\mathbf{27}_{\pm }}$, and 12 links $L_{ij}$
connecting the pairs $\left( \mathcal{N}_{i},\mathcal{N}_{j}\right) $.
Therefore, we can begin by visualizing this Q$_{\mathbf{56}}^{\mathbf{\mu }}$
as given in the Figure \textbf{\ref{QE7} }, and then move on to explicitly
derive it and extract its features.
\newline
We represent the $\mathcal{L}_{\mathbf{56}}^{\mathbf{\mu }}$ in the
projector basis using the $\varrho _{\mathbf{R}_{q}}$ ordered like $\left(
\varrho _{\mathbf{1}_{+}},\varrho _{\mathbf{27}_{+}},\varrho _{\mathbf{27}%
_{-}},\varrho _{\mathbf{1}_{-}}\right) $%
\begin{equation}
\mathcal{L}_{\mathbf{56}}^{\mathbf{\mu }}=\left(
\begin{array}{cccc}
\varrho _{\mathbf{1}_{+}}\mathcal{L}\varrho _{\mathbf{1}_{+}} & \varrho _{%
\mathbf{1}_{+}}\mathcal{L}\varrho _{\mathbf{27}_{+}} & \varrho _{\mathbf{1}%
_{+}}\mathcal{L}\varrho _{\mathbf{27}_{-}} & \varrho _{\mathbf{1}_{+}}%
\mathcal{L}\varrho _{\mathbf{1}_{-}} \\
\varrho _{\mathbf{27}_{+}}\mathcal{L}\varrho _{\mathbf{1}_{+}} & \varrho _{%
\mathbf{27}_{+}}\mathcal{L}\varrho _{\mathbf{27}_{+}} & \varrho _{\mathbf{27}%
_{+}}\mathcal{L}\varrho _{\mathbf{27}_{-}} & \varrho _{\mathbf{27}_{+}}%
\mathcal{L}\varrho _{\mathbf{1}_{-}} \\
\varrho _{\mathbf{27}_{-}}\mathcal{L}\varrho _{\mathbf{1}_{+}} & \varrho _{%
\mathbf{27}_{-}}\mathcal{L}\varrho _{\mathbf{27}_{+}} & \varrho _{\mathbf{27}%
_{-}}\mathcal{L}\varrho _{\mathbf{27}_{-}} & \varrho _{\mathbf{27}_{-}}%
\mathcal{L}\varrho _{\mathbf{1}_{-}} \\
\varrho _{\mathbf{1}_{-}}\mathcal{L}\varrho _{\mathbf{1}_{+}} & \varrho _{%
\mathbf{1}_{-}}\mathcal{L}\varrho _{\mathbf{27}_{+}} & \varrho _{\mathbf{1}%
_{-}}\mathcal{L}\varrho _{\mathbf{27}_{-}} & \varrho _{\mathbf{1}_{-}}%
\mathcal{L}\varrho _{\mathbf{1}_{-}}%
\end{array}%
\right)  \label{mat}
\end{equation}%
The diagonal terms $\varrho _{\boldsymbol{R}_{i}}\mathcal{L}\varrho _{%
\boldsymbol{R}_{i}}$ are depicted by the four nodes $\mathcal{N}_{%
\boldsymbol{R}_{i}}$ of Q$_{\mathbf{56}}^{\mathbf{\mu }},$ while the off
diagonal terms $\varrho _{\boldsymbol{R}_{i}}\mathcal{L}\varrho _{%
\boldsymbol{R}_{j}}$ with $i\neq j$ are associated to the twelve links $%
L_{ij}$.
\begin{equation}
\mathcal{N}_{\boldsymbol{R}_{i}}\equiv \varrho _{\boldsymbol{R}_{i}}\mathcal{%
L}\varrho _{\boldsymbol{R}_{i}}\qquad ,\qquad L_{ij}=\varrho _{\boldsymbol{R}%
_{i}}\mathcal{L}\varrho _{\boldsymbol{R}_{j}}
\end{equation}%
As the explicit calculation of these quantities is cumbersome, we decompose
the matrix $\mathcal{L}_{\mathbf{56}}^{\mathbf{\mu }}$ (\ref{mat}) into four
blocks $A_{\mathbf{56}}^{\mathbf{\mu }},B_{\mathbf{56}}^{\mathbf{\mu }},C_{%
\mathbf{56}}^{\mathbf{\mu }}$ and $D_{\mathbf{56}}^{\mathbf{\mu }}$ as
follows%
\begin{equation}
\mathcal{L}_{\mathbf{56}}^{\mathbf{\mu }}=\left(
\begin{array}{cc}
A_{\mathbf{56}}^{\mathbf{\mu }} & B_{\mathbf{56}}^{\mathbf{\mu }} \\
C_{\mathbf{56}}^{\mathbf{\mu }} & D_{\mathbf{56}}^{\mathbf{\mu }}%
\end{array}%
\right)
\end{equation}

\begin{itemize}
\item \emph{the block }$A$ : concerns the sector $\mathbf{28}_{+}$ of $%
\mathbf{56}$ :%
\begin{equation}
A_{\mathbf{56}}^{\mathbf{\mu }}=\left(
\begin{array}{cc}
\varrho _{\mathbf{1}_{+}}\mathcal{L}\varrho _{\mathbf{1}_{+}} & \varrho _{%
\mathbf{1}_{+}}\mathcal{L}\varrho _{\mathbf{27}_{+}} \\
\varrho _{\mathbf{27}_{+}}\mathcal{L}\varrho _{\mathbf{1}_{+}} & \varrho _{%
\mathbf{27}_{+}}\mathcal{L}\varrho _{\mathbf{27}_{+}}%
\end{array}%
\right) =\left(
\begin{array}{cc}
A_{I}^{I} & A_{I}^{II} \\
A_{II}^{I} & A_{II}^{II}%
\end{array}%
\right)
\end{equation}%
with%
\begin{equation}
\begin{tabular}{lll}
$A_{I}^{I}$ & $=$ & ${\small z}^{\frac{3}{2}}{\small \varrho }_{\mathbf{1}%
_{+}}{\small +z}^{\frac{1}{2}}{\small X\varrho }_{\mathbf{27}_{+}}{\small Y}+%
\frac{1}{4}z^{-\frac{1}{2}}X^{2}\varrho _{\mathbf{27}_{-}}Y^{2}+\frac{1}{36}%
z^{-\frac{3}{2}}X^{3}\varrho _{\mathbf{1}_{-}}Y^{3}$ \\
$A_{I}^{II}$ & $=$ & ${\small z}^{\frac{1}{2}}{\small X\varrho }_{\mathbf{27}%
_{+}}{\small +}\frac{1}{2}{\small z}^{-\frac{1}{2}}{\small X}^{2}{\small %
\varrho }_{\mathbf{27}_{-}}{\small Y}+\frac{1}{12}z^{-\frac{3}{2}%
}X^{3}\varrho _{\mathbf{1}_{-}}Y^{2}$ \\
$A_{II}^{I}$ & $=$ & ${\small z}^{\frac{1}{2}}{\small \varrho }_{\mathbf{27}%
_{+}}{\small Y+}\frac{1}{2}{\small z}^{-\frac{1}{2}}{\small X\varrho }_{%
\mathbf{27}_{-}}{\small Y}^{2}+\frac{1}{12}z^{-\frac{3}{2}}X^{2}\varrho _{%
\mathbf{1}_{-}}Y^{3}$ \\
$A_{II}^{II}$ & $=$ & ${\small z}^{\frac{1}{2}}{\small \varrho }_{\mathbf{27}%
_{+}}{\small +z}^{-\frac{1}{2}}{\small X\varrho }_{\mathbf{27}_{-}}{\small Y}%
+\frac{1}{4}z^{-\frac{3}{2}}X^{2}\varrho _{\mathbf{1}_{-}}Y^{2}$%
\end{tabular}%
\end{equation}%
The $A_{I}^{I}$ and $A_{II}^{II}$ are associated to the nodes $\mathcal{N}_{%
\mathbf{1}_{3/2}}$ and $\mathcal{N}_{\mathbf{27}_{1/2}}$, while the
sub-blocks $A_{I}^{II}$ and $A_{II}^{I}$ describe links between these nodes.

\item \emph{the block }$D$ : concerns the sector $\mathbf{28}_{-}$ of the
representation $\mathbf{56}$ :%
\begin{equation}
D_{\mathbf{56}}^{\mathbf{\mu }}=\left(
\begin{array}{ll}
{\small z}^{-\frac{1}{2}}{\small \varrho }_{\mathbf{27}_{-}}{\small +z}^{-%
\frac{3}{2}}{\small X\varrho }_{\mathbf{1}_{-}}{\small Y} & {\small z}^{-%
\frac{3}{2}}{\small X\varrho }_{\mathbf{1}_{-}} \\
{\small z}^{-\frac{3}{2}}{\small \varrho }_{\mathbf{1}_{-}}{\small Y} &
{\small z}^{-\frac{3}{2}}{\small \varrho }_{\mathbf{1}_{-}}%
\end{array}%
\right)
\end{equation}%
Where $D_{I}^{I}$ and $D_{II}^{II}$ are associated to $\mathcal{N}_{\mathbf{%
27}_{-1/2}}$ and $\mathcal{N}_{\mathbf{1}_{-3/2}}$ and $D_{I}^{II}$ and $%
D_{II}^{I}$ are associated to links between them.

\item \emph{the} \emph{blocks }$B$\emph{\ and }$C$\emph{:} Describe
couplings between sectors $28_{+}$ and $28_{-}$ :%
\begin{equation}
B_{\mathbf{56}}^{\mathbf{\mu }}=\left(
\begin{array}{ll}
\frac{1}{2}{\small X}^{2}{\small z}^{-\frac{1}{2}}{\small \varrho }_{\mathbf{%
27}_{-}}+\frac{1}{6}z^{-\frac{3}{2}}X^{3}\varrho _{\mathbf{1}_{-}}Y & \frac{1%
}{6}{\small z}^{-\frac{3}{2}}{\small X}^{3}{\small \varrho }_{\mathbf{1}_{-}}
\\
{\small z}^{-\frac{1}{2}}{\small X\varrho }_{\mathbf{27}_{-}}{\small +}\frac{%
1}{2}{\small z}^{-\frac{3}{2}}{\small X}^{2}{\small \varrho }_{\mathbf{1}%
_{-}}{\small Y} & \frac{1}{2}{\small z}^{-\frac{3}{2}}{\small X}^{2}{\small %
\varrho }_{\mathbf{1}_{-}}%
\end{array}%
\right)
\end{equation}%
\begin{equation*}
\end{equation*}%
\begin{equation*}
C_{\mathbf{56}}^{\mathbf{\mu }}=\left(
\begin{array}{ll}
\frac{1}{2}{\small z}^{-\frac{1}{2}}{\small \varrho }_{\mathbf{27}_{-}}%
{\small Y}^{2}+\frac{1}{6}z^{-\frac{3}{2}}X\varrho _{\mathbf{1}_{-}}Y^{3} &
{\small z}^{-\frac{1}{2}}{\small \varrho }_{\mathbf{27}_{-}}{\small Y+}\frac{%
1}{2}{\small z}^{-\frac{3}{2}}{\small X\varrho }_{\mathbf{1}_{-}}{\small Y}%
^{2} \\
\frac{1}{6}{\small z}^{-\frac{3}{2}}{\small \varrho }_{\mathbf{1}_{-}}%
{\small Y}^{3} & \frac{1}{2}{\small z}^{-\frac{3}{2}}{\small \varrho }_{%
\mathbf{1}_{-}}{\small Y}^{2}%
\end{array}%
\right)
\end{equation*}%
Entries of these matrices give 4+4 links between the nodes' pairs $\left(
\mathcal{N}_{\mathbf{1}_{3/2}},\mathcal{N}_{\mathbf{27}_{1/2}}\right) $ and $%
\left( \mathcal{N}_{\mathbf{27}_{-1/2}},\mathcal{N}_{\mathbf{1}%
_{-3/2}}\right) $.
\end{itemize}

And so indeed, the topological gauge quiver Q$_{\mathbf{56}}^{\mathbf{\mu }}$
associated with $\mathcal{L}_{\mathbf{56}}^{\mathbf{\mu }}$ has four nodes $%
\mathcal{N}_{i}$ corresponding to the $e_{6}$ representations
\begin{equation}
\begin{tabular}{lllllll}
$\mathcal{N}_{1}$ & : & $\underline{\mathbf{1}}_{+3/2}$ & , & $\mathcal{N}%
_{3}$ & : & $\underline{\mathbf{27}}_{-1/2}$ \\
$\mathcal{N}_{2}$ & : & $\underline{\mathbf{27}}_{+1/2}$ & , & $\mathcal{N}%
_{4}$ & : & $\underline{\mathbf{1}}_{-3/2}$%
\end{tabular}%
\end{equation}%
and describing self-dual topological gauge matter. It also has $12$ links $%
L_{ij}$ describing topological bi-fundamental gauge matter $\left \langle
\boldsymbol{R}_{i},\boldsymbol{R}_{j}\right \rangle $ as collected in the
following tables%
\begin{equation}
\begin{tabular}{|l|l|l||l|l|l|}
\hline
{\small link} & {\small \ \ \ Repres} & {\small bi-matter} & {\small link} &
{\small \ \ \ Repres} & {\small bi-matter} \\ \hline
$L_{1\rightarrow 2}$ & $\left \langle \boldsymbol{1}_{+3/2},\boldsymbol{27}%
_{-1/2}\right \rangle $ & \ $\ \ \ \ b^{\alpha }$ & $L_{2\rightarrow 3}$ & $%
\left \langle \boldsymbol{27}_{+1/2},\boldsymbol{27}_{+1/2}\right \rangle $
& \ $\ \ \ \ b^{\alpha }$ \\ \hline
$L_{1\rightarrow 3}$ & $\left \langle \boldsymbol{1}_{+3/2},\boldsymbol{27}%
_{+1/2}\right \rangle $ & \ $\ \ \ \ \mathcal{B}^{\alpha }$ & $%
L_{2\rightarrow 4}$ & $\left \langle \boldsymbol{27}_{+1/2},\boldsymbol{1}%
_{+3/2}\right \rangle $ & \ $\ \ \ \ \mathcal{B}^{\alpha }$ \\ \hline
$L_{1\rightarrow 4}$ & $\left \langle \boldsymbol{1}_{+3/2},\boldsymbol{1}%
_{+3/2}\right \rangle $ & \ $\ \ \mathcal{B}_{\alpha }b^{\alpha }$ & $%
L_{3\rightarrow 4}$ & $\left \langle \boldsymbol{27}_{-1/2},\boldsymbol{1}%
_{+3/2}\right \rangle $ & \ $\ \ \ \ b_{\alpha }$ \\ \hline
\end{tabular}%
\end{equation}%
and%
\begin{equation}
\begin{tabular}{|l|l|l||l|l|l|}
\hline
{\small link} & {\small \ \ \ Repres} & {\small bi-matter} & {\small link} &
{\small \ \ \ Repres} & {\small bi-matter} \\ \hline
$L_{1\longleftarrow 2}$ & $\left \langle \boldsymbol{1}_{-3/2},\boldsymbol{27%
}_{+1/2}\right \rangle $ & \ $\ \ \ \ c_{\alpha }$ & $L_{2\longleftarrow 3}$
& $\left \langle \boldsymbol{27}_{-1/2},\boldsymbol{27}_{-1/2}\right \rangle
$ & \ $\ \ \ \ c^{\alpha }$ \\ \hline
$L_{1\longleftarrow 3}$ & $\left \langle \boldsymbol{1}_{+3/2},\boldsymbol{27%
}_{+1/2}\right \rangle $ & \ $\ \ \ \ \mathcal{C}_{\alpha }$ & $%
L_{2\longleftarrow 4}$ & $\left \langle \boldsymbol{27}_{-1/2},\boldsymbol{1}%
_{-3/2}\right \rangle $ & \ $\ \ \ \ \mathcal{C}^{\alpha }$ \\ \hline
$L_{1\longleftarrow 4}$ & $\left \langle \boldsymbol{1}_{+3/2},\boldsymbol{1}%
_{+3/2}\right \rangle $ & \ $\ \ \ \ c_{\alpha }^{\alpha }\mathcal{C}$ & $%
L_{3\longleftarrow 4}$ & $\left \langle \boldsymbol{27}_{+1/2},\boldsymbol{1}%
_{-3/2}\right \rangle $ & \ $\ \ \ \ c_{\alpha }$ \\ \hline
\end{tabular}%
\end{equation}%
In these tables, $\mathcal{B}^{\gamma }$ stands for $b^{\alpha }\Gamma
_{\alpha \beta }^{\gamma }b^{\beta }$ having charge $-2$, and $\mathcal{C}%
_{\gamma }$ refers to $c_{\alpha }\bar{\Gamma}_{\gamma }^{\alpha \beta
}c_{\beta }$ having charge $+2.$ The composites $\mathcal{B}_{\alpha
}b^{\alpha }$ and $c_{\alpha }^{\alpha }\mathcal{C}$ have charges $-3$ and $%
+3$ respectively.

\section{Conclusion and comments}

The results presented in this paper are based on the correspondence between
two dimensional integrable models and four dimensional Chern-Simons gauge
theory as formulated in \cite{21}. In the $\boldsymbol{M}_{4}= \mathbb{R}%
^{2}\times \mathbb{CP}^{1}$ of the gauge theory, one can build an integrable
lattice model by implementing a set of line defects looking like curves on $%
\mathbb{R}^{2}$ and points on $\mathbb{CP}^{1}$. In such construction, the
integrability of the corresponding low-dimensional system constrained by the
Yang Baxter or RLL equation is a direct result of the mixed
topological-holomorphic nature of the line defects and the diffeomorphism
invariance in four dimensions. The RLL equation for example, corresponds to
the graphical equivalence of the intersections in different orders of two
electric Wilson lines with one magnetic 't Hooft line, see Figure \textbf{%
\ref{2L}}. In this image, the explicit Feynman diagrams calculation for the
intersection of two Wilson lines in 4D CS yields the first order expansion
of the R-matrix acting on the two quantum spaces carried by the electrically
charged lines \cite{19}-\cite{21}. The L-operator is realised as the
intersection of an electric Wilson line with a magnetic 't Hooft line whose
oscillator phase space acts as an auxiliary space \cite{34A}.\

This Wilson/'t Hooft coupling in the 4D CS theory is the particularly
interesting ingredient of our current investigation, it allows to realise
the Lax matrix as a building block of the transfer matrix generating
conserved commuting quantities of the spin chain. This important quantity is
calculated in the integrability literature using Yangian representations
based techniques that can be cumbersome and inefficient in cases with
complicated symmetries. Surprisingly, it was shown in \cite{34A} that the
oscillator realisation of these L-operators for an XXX spin chain having the
internal symmetry $g$ can be recovered from the analysis of solutions to the
equations of motion of the 4D CS theory with gauge symmetry $G$, in the
presence of interacting Wilson and 't Hooft lines. A general formula
describing the coupling of a Wilson line with electric charge in a
representation $\boldsymbol{R}$ of $G$ and a 't Hooft line with magnetic
charge given by a minuscule coweight $\mu $ of $G$ reads as $\mathcal{L}_{%
\boldsymbol{R}}^{\mu } =e^{X_{\boldsymbol{R}}}z^{\mathbf{\mu }}e^{Y_{%
\boldsymbol{R}}}$. This yields a matrix representation in terms of harmonic
oscillators in $X_{\boldsymbol{R}}$ and $Y_{\boldsymbol{R}}$ with sub-blocks
following from the Levi decomposition of $\boldsymbol{R}$ with respect to $%
\mu $. \

The first part of our contribution concerned the exploitation of this
formula to explicitly calculate this coupling for different types of 't
Hooft and Wilson line defects in 4D Chern-Simons theories with $SL_{N}$, $%
SO_{2N}$, E$_{6}$ and E$_{7}$ gauge symmetries. In particular, we
investigated the splitting of various representations under the action of
minuscule coweights as a first step towards the construction of L-operators
in representations beyond the fundamental for ADE Lie algebras. Therefore, a
better understanding of the effect of the Dirac-like singularity on the
gauge field bundles behavior and the internal quantum states of a spin
chain.\

We remarked that the L-operators have unified intrinsic features that can be
represented by topological quiver diagrams Q$_{\boldsymbol{R}}^{\mathbf{\mu }%
}$ having a formal similarity with the well known graphs Q$_{G}^{susy}$ of
supersymmetric quiver gauge theories embedded in type II strings. This
formal link gives an interesting interpretation of the Darboux coordinates $%
\left( b^{\alpha },c_{\beta }\right) $ of the phase space of the L-operators
in terms of\ topological bi-fundamental matter. In this regard, we gave
several examples to $\left( i\right) $ explain the strong aspects of this
diagrammatic approach, and $\left( ii\right) $ to show how it can be used to
forecast the general form of the matrix representation of L-operators by
indicating the action of its sub-blocks and their charges in terms of
combinations of Darboux coordinates.\

\begin{figure}[ph]
\begin{center}
\includegraphics[width=8cm]{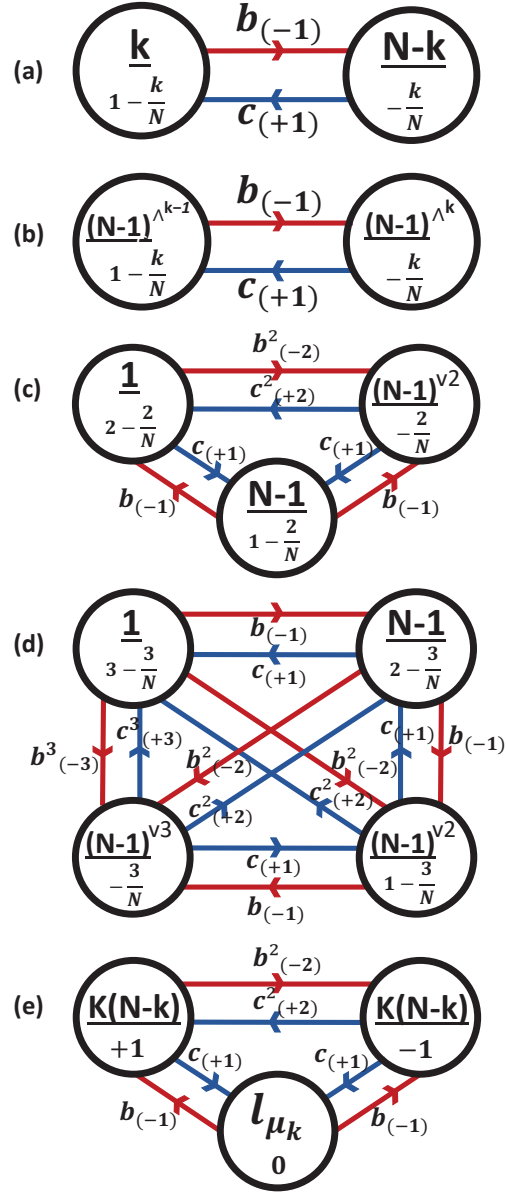}
\end{center}
\par
\vspace{-0.5cm}
\caption{Leading elements of topological quiver diagrams for the L-operators
of A- type. These quivers are classified by the magnetic charge $\protect\mu %
_{k}$ of the 't Hooft line and the representation $\boldsymbol{R}$. (\textbf{%
a}) Wilson line with charge $\boldsymbol{R=N.}$ (\textbf{b}) Wilson line
with $\boldsymbol{R=N}^{\wedge k}.$ (\textbf{c}) Wilson line with $%
\boldsymbol{R=N}^{\vee 2}.$ (\textbf{d}) Wilson line with $\boldsymbol{R=N}%
^{\vee 3}.$ (\textbf{e}) Wilson line with charge $\boldsymbol{R=adj}sl_{N}.$}
\label{TA}
\end{figure}

In particular, For the A-type Chern-Simons theory, all fundamental coweights
are minuscule, and therefore we give in Figure \textbf{\ref{TA}}, for a
generic magnetic charge $\mu _{k}$ of $sl_{N}$, four quiver diagrams
describing L-operators classified by representations $\boldsymbol{R}$ of the
Wilson line.

\begin{figure}[ph]
\begin{center}
\includegraphics[width=10cm]{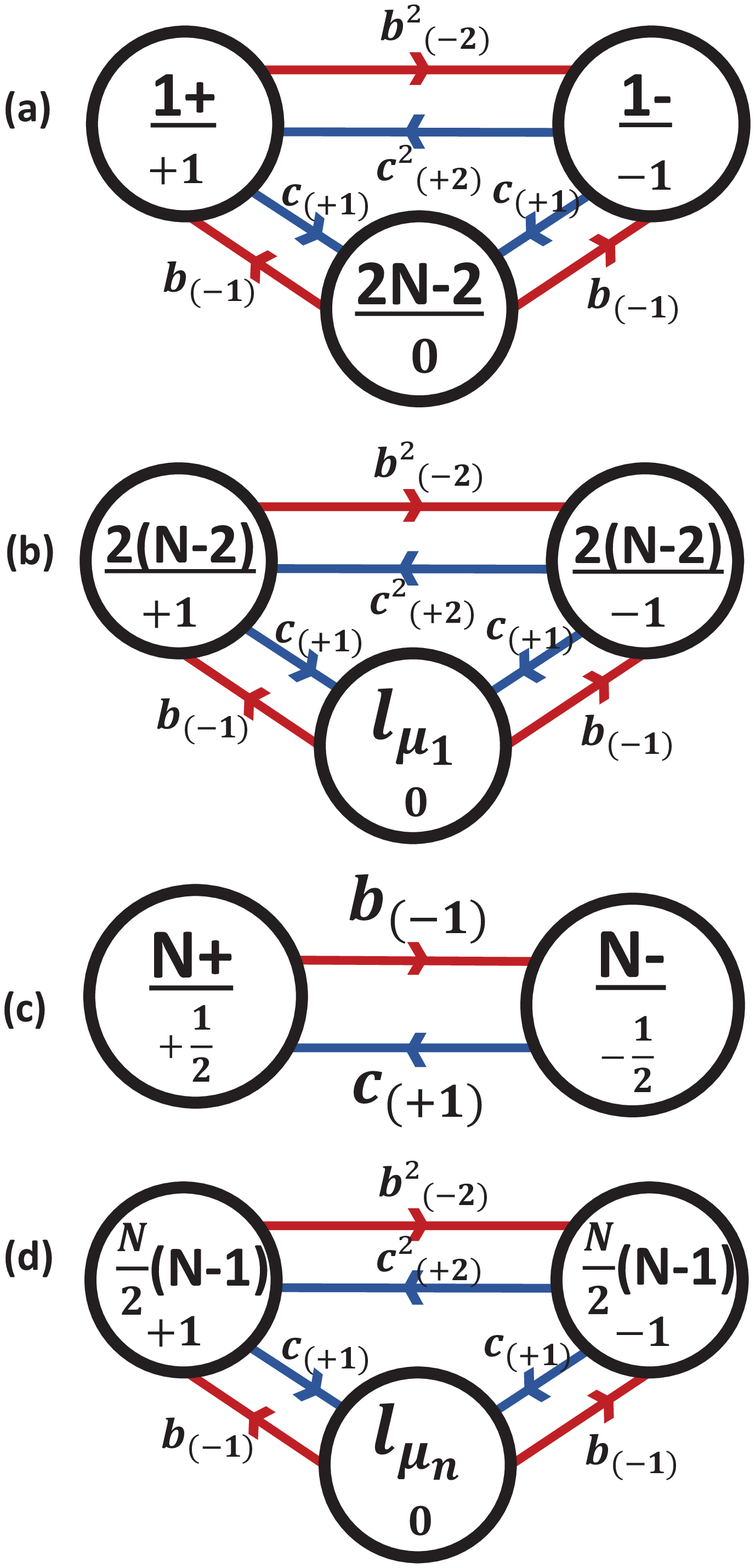}
\end{center}
\par
\vspace{-0.5cm}
\caption{Leading elements of topological quiver diagrams for the L-operators
of D- type. The first two quivers correspond to the Levi decomposition with
respect to the (vectorial) minuscule coweight $\protect\mu _{1}$: (\textbf{a}%
) Wilson line with charge $\boldsymbol{R=2N.}$ (\textbf{b}) Wilson line with
$\boldsymbol{R=adjso}_{\boldsymbol{2N}}.$ The other two quivers correspond
to the Levi decomposition with respect to the (spinorial) minuscule coweight
$\protect\mu _{N}$: (c) Wilson line with $\boldsymbol{R=2}^{\boldsymbol{N-1}%
}.$ (d) Wilson line with $\boldsymbol{R=adjso}_{\boldsymbol{2N}}.$}
\label{TD}
\end{figure}

In the case of D-type symmetry, we have two types of minuscule 't Hooft lines
associated to the vectorial and spinorial coweights of the $SO_{2N}$ gauge
symmetry. In the figure \textbf{\ref{TD}}, we give quiver diagrams
describing four possibilities of Wilson/'t Hooft couplings: a magnetic
charge $\mu _{1}$ with electric $\boldsymbol{R=2N}$ and with $\boldsymbol{%
R=adjso}_{\boldsymbol{2N}}$, and magnetic $\mu _{N}\sim \mu _{N-1}$ with
electric $\boldsymbol{R=2}^{\boldsymbol{N-1}}$ and with $\boldsymbol{R=adjso}%
_{\boldsymbol{2N}}$.
\begin{figure}[ph]
\begin{center}
\includegraphics[width=10cm]{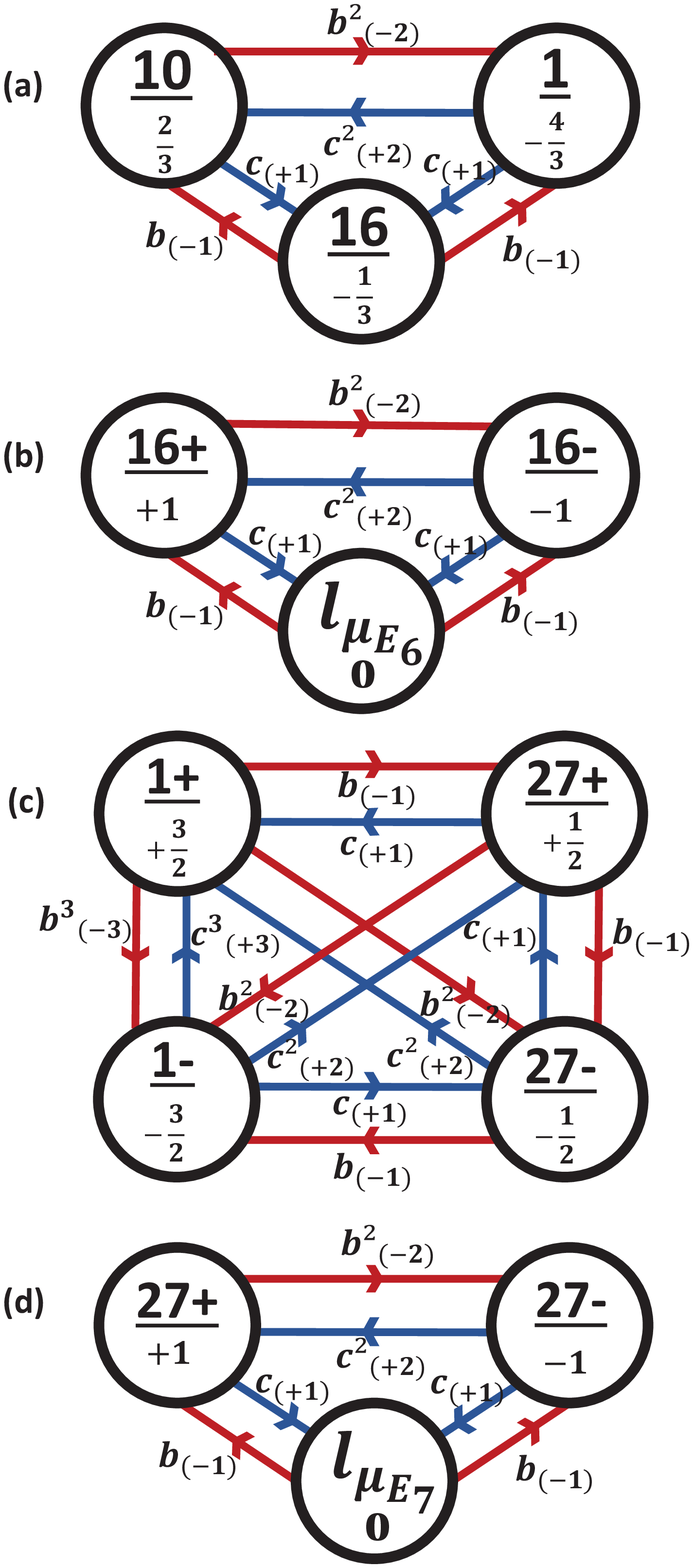}
\end{center}
\par
\vspace{-0.5cm}
\caption{Leading elements of topological quiver diagrams for the L-operators
of E- type. The first two quivers for the E$_{6}$ gauge theory. (\textbf{a})
for the fundamental $\mathbf{27}$ of E$_{6}$; and (\textbf{b}) for the
adjoint representation. The last two quivers regard the E$_{7}$ Chern-Simons
theory. (\textbf{c}) for the fundamental $\mathbf{56}$ of E$_{7}$ and (%
\textbf{d}) for the adjoint representation.}
\label{TE}
\end{figure}

The Figure \textbf{\ref{TE}} represents quiver gauge diagrams of exceptional
type where we gave for each one of the E$_{6}$ and E$_{7}$ 4D CS theories
the graphical descriptions for the coupling of the minuscule 't Hooft line
with Wilson lines in the fundamental and in the adjoint representations.
Notice however, that not all the representations studied here for the three
types of symmetries lift to the Yangian; the corresponding L-operators are
interpreted semi-classically in the integrability language.

Moreover, this construction can be extended for the investigation of other
L-operators that are still missing in the spin chain literature; and the
interpretations associated to the components of the L-operator can also be
used to link the diagrammatic description presented here to quiver diagrams
associated to the realisation of 't Hooft line defects in supersymmetric
quiver theories; in particular the ADE quiver gauge theories describing the
phase spacse of t' Hooft lines as the Coulomb branches as in \cite{34A}.

\begin{figure}[ph]
\begin{center}
\includegraphics[width=17cm]{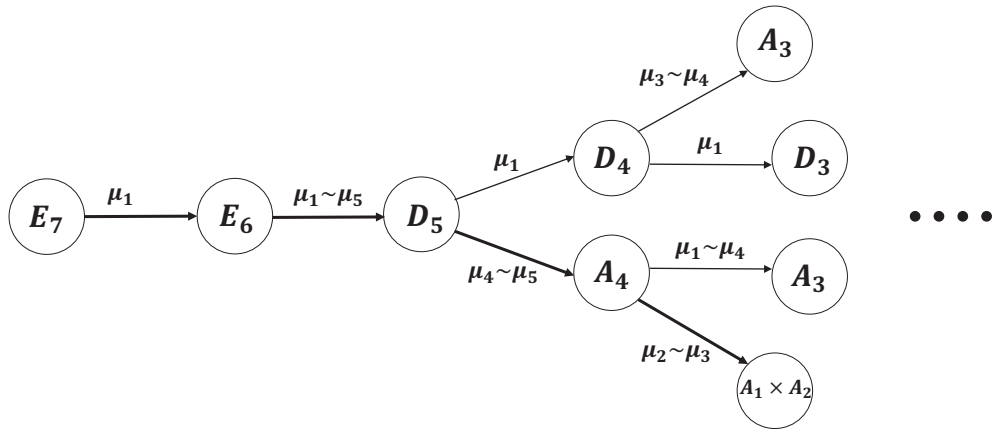}
\end{center}
\par
\vspace{-0.5cm}
\caption{Breaking chains of E$_{7}$ symmetry as given by Levi decompositions
with respect to minuscule coweights. The bold arrows describe the
exceptional sequence leading to the Standard model-like group. The minuscule
coweights $\protect\mu $ correspond to the Lie algebra at which the arrow
starts}
\label{SM}
\end{figure}

Another exquisite property of this graphical quiver description in the 4D
Chern-Simons topological theory is the natural appearance of a unified
theory structure where the minuscule L-operators can be connected and
classified in a larger E$_{7}$ 4D CS theory. In fact, the Lie algebras'
decompositions with respect to minuscule coweights link the E$_{7}$ symmetry
to the E$_{6}$ and then to the family of D$_{N}$ symmetries with $N\leq 5$
and/or the A$_{N}$ with $N\leq 4$. These chains of Levi decompositions lead
to different possible paths for the E$_{7}$ symmetry breaking as described
in Figure \textbf{\ref{SM}} \cite{34L}. To visualize this from the quiver
descriptions of L-operators, we can focus on those corresponding to the
fundamental representations and notice that the Q$_{\boldsymbol{56}}^{\mu
_{e_{7}}}$ has a node corresponding to the $\boldsymbol{27}$ of E$_{6}$;
this node can be therefore imagined as including the Q$_{\boldsymbol{27}%
}^{\mu _{e_{6}}}$ which in turn includes the Q$_{\boldsymbol{10}\left(
so_{10}\right) }^{vect}$ and so on. Finally, notice that the calculation of
minuscule L-operators in 4D CS theories with $SO_{2N+1}$ and $SP_{2N}$
symmetries having each only one minuscule coweight, shows that for $%
\boldsymbol{R}=fundamental$, the $\mathcal{L}_{\boldsymbol{R}\left(
so_{2N+1}\right) }^{{\small vect}}$ matrix is very similar to $\mathcal{L}_{%
\boldsymbol{R}\left( so_{2N}\right) }^{{\small vect}}$ while the $\mathcal{L}%
_{\boldsymbol{R}\left( sp_{2N}\right) }^{{\small spin}}$ is similar to $%
\mathcal{L}_{\boldsymbol{R}\left( so_{2N}\right) }^{{\small spin}}$ \cite%
{54A}. This means that the corresponding quivers look like Q$_{\boldsymbol{2N%
}}^{\mu _{1}}$ and Q$_{\boldsymbol{2N}}^{\mu _{N}}$ which allows to include
the B and C -type symmetries into this unified classification.

\clearpage

\section*{Appendix}

In this appendix, we give complementary tools regarding the construction of
the Lax matrix from the associated graphical quiver description introduced
in section $3$. Recall that a topological quiver diagram in the 4D CS gauge
theory is defined by the data\ $(g,\boldsymbol{R},\mu );$ $g$ is the Lie
algebra of the gauge symmetry $G,$ having a Levi decomposition under a
minuscule coweight $\mu $ reading as $g=n_{-}\oplus l_{\mu }\oplus n_{+}.$
The $\boldsymbol{R}$ is some representation of $g$ decomposing under $\mu $
as follows%
\begin{equation}
\boldsymbol{R}=\sum\limits_{i=1}^{p-1}\boldsymbol{R}_{m_{i}}
\end{equation}%
The $m_{i}$'s are Levi charges appearing in the adjoint action of $\mathbf{%
\mu }$ reading in terms of projectors as $\mathbf{\mu =}\sum\limits_{i}m_{i}%
\Pi _{i}.$ We begin by elaborating the general derivation of $\mathcal{L}$
using a quiver Q; then we illustrate the construction through the particular
example of $\mathcal{L}_{\boldsymbol{adj}}^{\mu _{k}}$ for\emph{\ }$g=sl_{N}$%
.\newline
In fact, given a topological gauge quiver with $p$ nodes $\mathcal{N}_{i}$ $%
(1\leq i\leq p)$ where sit representations $\boldsymbol{R}_{m_{i}}$, and
links $L_{ij}\ \left( i\neq j\right) $ interpreted as the bi-fundamentals $%
\left\langle \boldsymbol{R}_{m_{i}},\boldsymbol{R}_{m_{j}}\right\rangle $;
the corresponding Lax matrix is obtained as follows. The contributions of
the nodes having no Levi charge are given by the polynomials P$_{n}(x)$ with
argument $x=:\mathbf{bc:}$ and order $0\leq n\leq p-1$ as follows%
\begin{equation}
\begin{tabular}{lll}
$\mathcal{L}_{1}^{1}$ & $=$ & $\alpha _{11}z^{m_{1}}+\alpha _{12}z^{m_{2}}%
\mathbf{bc}+...+\alpha _{1p}z^{m_{p}}\mathbf{b}^{p-1}\mathbf{c}^{p-1}$ \\
& $\vdots $ &  \\
$\mathcal{L}_{i}^{i}$ & $=$ & $\alpha _{i1}z^{m_{i}}+\alpha _{i2}z^{m_{i+1}}%
\mathbf{bc}+...+\alpha _{ip}z^{m_{p}}\mathbf{b}^{p-i}\mathbf{c}^{p-i}$ \\
& $\vdots $ &  \\
$\mathcal{L}_{p}^{p}$ & $=$ & $\alpha _{p1}z^{m_{p}}$%
\end{tabular}%
\end{equation}%
where $\alpha _{ij}$ are some real numbers$.$\newline
The contributions of the links carry non-trivial integer Levi charges; they
are given by polynomials in $\mathbf{b}$ and $\mathbf{c}$ such as%
\begin{equation}
\begin{tabular}{llll}
$\mathcal{L}_{i+1}^{i}$ & $=$ & $z^{m_{i+1}}\mathbf{b}+z^{m_{i+2}}\mathbf{b}%
^{2}\mathbf{c+}...+z^{m_{p}}\mathbf{b}^{p-i}\mathbf{c}^{p-\left( i+1\right)
} $ &  \\
$\mathcal{L}_{i}^{i+1}$ & $=$ & $z^{m_{i+1}}\mathbf{c}+z^{m_{i+2}}\mathbf{bc}%
^{2}...+z^{m_{p}}\mathbf{b}^{p-\left( i+1\right) }\mathbf{c}^{p-i}$ &  \\
\multicolumn{4}{l}{} \\
$\mathcal{L}_{j}^{i}$ & $=$ & $z^{m_{j}}\mathbf{b}^{j-i}+z^{m_{j+1}}\mathbf{b%
}^{j-i+1}\mathbf{c}+...+z^{m_{p}}\mathbf{b}^{p-i}\mathbf{c}^{p-j}$ & $;\quad
j>i$ \\
$\mathcal{L}_{i}^{j}$ & $=$ & $z^{m_{j}}\mathbf{c}^{j-i}+z^{m_{j+1}}\mathbf{%
bc}^{j-i+1}+...+z^{m_{p}}\mathbf{b}^{p-j}\mathbf{c}^{p-i}$ & $;\quad j>i$%
\end{tabular}%
\end{equation}%
However, since the phase space coordinates $\mathbf{b}$\ and $\mathbf{c}$
can be given by vectors or tensors depending on the realisation of $X$ and $%
Y $ generating $n_{\pm }$, these terms could be accompanied with metrics to
contract indices, thus homogenizing the tensor structure of each block.%
\newline
The Lax matrices associated to the topological quivers in Figures \textbf{%
\ref{TA}},\textbf{\ref{TD}} and \textbf{\ref{TE}} can be constructed using
these general expressions and by mimicking the example given below.

\emph{Example of }$\mathcal{L}_{\boldsymbol{adj}}^{\mu _{k}}$ \emph{for }$%
sl_{N}$\emph{:}\newline
In Figure \textbf{\ref{TA}-e}, we drawn the gauge quiver Q$_{\boldsymbol{adj}%
}^{\mu _{k}}$ in the 4D Chern-Simons theory with A- type gauge symmetry. It
corresponds to the Levi-decomposition (\ref{app})\ of $\boldsymbol{adj}%
\left( sl_{N}\right) $ with respect to a minuscule coweight $\mu =\mu _{k}$\
with $2\leq k\leq N-2$, it has three nodes $\mathcal{N}_{1},\mathcal{N}_{2},%
\mathcal{N}_{3}$ and six links $L_{ij}$ with $i\neq j.$ The nodes correspond
to the representations%
\begin{equation}
\begin{tabular}{lll}
$\boldsymbol{adj}\left( sl_{N}\right) $ & $=$ & $\boldsymbol{R}%
_{m_{1}}\oplus \boldsymbol{R}_{m_{2}}\oplus \boldsymbol{R}_{m_{3}}$ \\
$\boldsymbol{R}_{m_{1}}$ & $=$ & $\left[ k(N-k)\right] _{-}$ \\
$\boldsymbol{R}_{m_{2}}$ & $=$ & $\left[ N^{2}-2kN+2k^{2}-1\right] _{0}$ \\
$\boldsymbol{R}_{m_{3}}$ & $=$ & $\left[ k(N-k)\right] _{+}$%
\end{tabular}%
\end{equation}%
where Levi charges $m_{i}$ are as given by the sub-labels $0,\pm 1$.\newline
The Lax operator associated to the quiver \textbf{\ref{TA}-e} is represented
by $\left( N^{2}-1\right) \times \left( N^{2}-1\right) $ matrix divided into
three sub-blocks of dimensions $d_{1}=d_{3}=k(N-k)$ and $%
d_{2}=N^{2}-2kN+2k^{2}-1.$ The contributions of the nodes are given by%
\begin{equation}
\begin{tabular}{lllll}
Node &  & \multicolumn{3}{l}{Contribution} \\
$\mathcal{N}_{1}$ &  & $\left( \mathcal{L}\right) _{d_{1}\times d_{1}}$ & $=$
& $\left( z+\mathbf{bc}+z^{-1}\mathbf{b}^{2}\mathbf{c}^{2}\right) \Pi _{1}$
\\
$\mathcal{N}_{2}$ &  & $\left( \mathcal{L}\right) _{d_{2}\times d_{2}}$ & $=$
& $\left( \boldsymbol{1}+z^{-1}\mathbf{bc}\right) \Pi _{2}$ \\
$\mathcal{N}_{3}$ &  & $\left( \mathcal{L}\right) _{d_{3}\times d_{3}}$ & $=$
& $z^{-1}\Pi _{3}$%
\end{tabular}%
\end{equation}%
And the contributions of the links are as follows%
\begin{equation}
\begin{tabular}{lllll}
Link &  & \multicolumn{3}{l}{Contribution} \\
$\mathcal{L}_{12}$ &  & $\left( \mathcal{L}\right) _{d_{1}\times d_{2}}$ & $%
= $ & $\mathbf{b}+z^{-1}\mathbf{b}^{2}\mathbf{c}$ \\
$\mathcal{N}_{23}$ &  & $\left( \mathcal{L}\right) _{d_{2}\times d_{3}}$ & $%
= $ & $z^{-1}\mathbf{b}$ \\
$\mathcal{N}_{13}$ &  & $\left( \mathcal{L}\right) _{d_{1}\times d_{3}}$ & $%
= $ & $z^{-1}\mathbf{b}^{2}$%
\end{tabular}%
\qquad ,\qquad
\begin{tabular}{lllll}
Link &  & \multicolumn{3}{l}{Contribution} \\
$\mathcal{L}_{21}$ &  & $\left( \mathcal{L}\right) _{d_{2}\times d_{1}}$ & $%
= $ & $\mathbf{c}+z^{-1}\mathbf{bc}^{2}$ \\
$\mathcal{N}_{32}$ &  & $\left( \mathcal{L}\right) _{d_{3}\times d_{2}}$ & $%
= $ & $z^{-1}\mathbf{c}$ \\
$\mathcal{N}_{31}$ &  & $\left( \mathcal{L}\right) _{d_{3}\times d_{1}}$ & $%
= $ & $z^{-1}\mathbf{c}^{2}$%
\end{tabular}%
\end{equation}


\begin{thebibliography}{99}
\bibitem{1A} M. Jimbo and T. Miwa, Algebraic Analysis of Solvable Lattice
Models, vol. 85. American Mathematical Soc. (1994).

\bibitem{2} V. V. Bazhanov, S. L. Lukyanov, A. B. Zamolodchikov, Integrable
Structure of Conformal Field Theory II. Q- operator and DDV equation,
Commun. Math. Phys. 190 (1997) 247--278, arXiv:hep-th/9604044.

\bibitem{1G} V. Turaev, The Yang-Baxter equation and invariants of links,
Invent. Math. 92 (1988) 527--553, DOI: 10.1007/BF01393746.

\bibitem{10A} E.H Saidi, M.B Sedra, Hyper-Kaehler Metrics Building and
Integrable Models, Mod.Phys.Lett. A9 (1994) 3163-3174, arXiv:hep-th/0512220.

\bibitem{1C} R. Borsato, O. Ohlsson Sax, A. Sfondrini, B. Stefanski, A.
Torrielli, The all-loop integrable spin-chain for strings on $AdS_{3}\times
S^{3}\times T^{4}$: the massive sector, JHEP 08 (2013) 043, arXiv:1303.5995
[hep-th].

\bibitem{1E} F. Delduc, S. Lacroix, M. Magro, and B. Vicedo, Integrable
Coupled Sigma-Models, Phys. Rev. Lett. 122 no. 4, (2019) 041601,
arXiv:hep-th/1811.12316.

\bibitem{1H} E.H Saidi, MB Sedra, On N= 4 integrable models International
Journal of Modern Physics A 9 (06), 891-913.

\bibitem{6} N. Yu. Reshetikhin, Hamiltonian structures for integrable field
theory models. II. Models with O(n) and Sp(2k) symmetry on a one-dimensional
lattice, Theor. Math. Phys. 63 (1985) 455-462.

\bibitem{7A} E.H Saidi, Matrix representation of higher integer conformal
spin symmetries, Journal of Mathematical Physics 36 (8), 4461-4475.

\bibitem{11} R. J. Baxter, Exactly Solved Models in Statistical Mechanics.
Academic Press, 1982.

\bibitem{11A} C. N. Yang, Some exact results for the many-body problem in
one dimension with repulsive delta-function interaction, Physical Review
Letters 19 (23) (1967) 1312.

\bibitem{12} A. B. Zamolodchikov, A. B. Zamolodchikov, Factorized S-matrices
in two dimensions as the exact solutions of certain relativistic quantum
field theory models, Annals of Physics 120 (2) (1979) 253--291.

\bibitem{5} H. J. de Vega and M. Karowski, Exact Bethe ansatz solution of
O(2n) symmetric theories, Nucl. Phys. B280 (1987) 225-254.

\bibitem{9} L. D. Faddeev, How Algebraic Bethe Ansatz works for integrable
model, Les-Houches lectures, 59 pages, 1996, hep-th/9605187.

\bibitem{13A} E. K. Sklyanin, Some algebraic structures connected with the
Yang-Baxter equation, Functional Analysis and its Applications 16 (4) (1982)
263--270.

\bibitem{3} A. G. Bytsko and J. Teschner, Quantization of models with
non-compact quantum group symmetry: Modular XXZ magnet and lattice
sinh-Gordon model, J. Phys. A39 (2006) 12927--12981, arXiv:hep-th/0602093.

\bibitem{14} V. V. Bazhanov, R. Frassek, T. L ukowski, C. Meneghelli, M.
Staudacher, Baxter Q-Operators and Representations of Yangians,
arXiv:math-ph/1010.3699.

\bibitem{15} T. Lukowski, C. Meneghelli, M. Staudacher, A Shortcut to the
Q-Operator, J. Stat. Mech. 1011 (2010) P11002, arXiv:hep-th/1005.3261.

\bibitem{17} R. Frassek, T. Lukowski, C. Meneghelli, M. Staudacher, Baxter
Operators and Hamiltonians for 'nearly all' Integrable Closed gl(n) Spin
Chains, Nucl. Phys. B874 (2013) 620--646, arXiv:math-ph/1112.3600.

\bibitem{18} R. Frassek, V. Pestun, A. Tsymbaliuk, Lax matrices from
antidominantly shifted Yangians and quantum affine algebras,
arxiv:math-ph/2001.04929.

\bibitem{19} K. Costello, E. Witten, M. Yamazaki, Gauge theory and
integrability, I, ICCM Not. 6 (2018) 46 [1709.09993].

\bibitem{20} K. Costello, E. Witten, M. Yamazaki, Gauge theory and
integrability, II, ICCM Not. 6 (2018) 120 [1802.01579].

\bibitem{21} K. Costello, M. Yamazaki, Gauge Theory And Integrability, III,
1908.02289.

\bibitem{Y1} K. Costello, Integrable lattice models from four-dimensional
field theories, in String-Math 2013, arXiv:hep-th/1308.0370.

\bibitem{20A} K. Costello, Junya Yagi, Unification of integrability in
supersymmetric gauge theories, Adv. Theor. Math. Phys. 24 (2020) 1931-2041,
arXiv:1810.01970.

\bibitem{22} B. Vicedo, Holomorphic Chern-Simons theory and affine Gaudin
models, 1908.07511.

\bibitem{28} N. Dorey, S. Lee, T. J. Hollowood, Quantization of integrable
systems and a 2d/4d duality, JHEP 10 (2011) 077 [1103.5726].

\bibitem{33} E. Witten, Integrable Lattice Models From Gauge Theory,
Advances in Theoretical and Mathematical Physics 21(7):1819-1843,
arXiv:hep-th/1611.00592.

\bibitem{35} R. Bittleston, D. Skinner, Gauge theory and boundary
integrability, JHEP 05 (2019) 195, 52 [1903.03601].

\bibitem{24A} Y. Boujakhrout, E. H Saidi, R. Ahl Laamara, L. B Drissi, Lax
Operator and superspin chains from 4D CS gauge theory, J.Phys.A 55 (2022)
41, 415402, arXiv:hep-th/2209.07117.

\bibitem{24AB} Y. Boujakhrout, E. H Saidi, R. Ahl Laamara, L. B Drissi,
Embedding Integrable Superspin Chain in String Theory, To appear in Nuclear
Physics B (2023).

\bibitem{23} O. Fukushima, J.-i. Sakamoto, K. Yoshida, Yang-Baxter
deformations of the AdS5$\times $ S5 supercoset sigma model from 4D CS
theory, JHEP 09 (2020).

\bibitem{20C} El Hassan Saidi, Computing the Scalar Field Couplings in 6D
Supergravity,Nucl.Phys.B803:323-362,2008, arXiv:0806.3207.

\bibitem{20G} S. Katz, P. Mayr, and C. Vafa, Mirror symmetry and exact
solution of 4-D N=2 gauge theories: 1., Adv.Theor.Math.Phys. 1 (1998)
53-114, arXiv:hep-th/9706110.

\bibitem{13C} Saidi E.H., Sedra M.B., Zerouaoui J., On D = 2 (1/3,1/3)
supersymmetric theories. I, Classical Quantum Gravity 12 (1995), 1567-1580.

\bibitem{20H} EH Saidi, Twisted 3D supersymmetric YM on deformed lattice
Journal of Mathematical Physics 55 (1), 012301.

\bibitem{20F} E.H Saidi, L.B Drissi, 5D N = 1 super QFT: symplectic quivers,
Nucl Phys B 2021.

\bibitem{20D} P. Mattioli, Sanjaye Ramgoolam, Quivers, Words and
Fundamentals, Journal of High Energy Physics; Heidelberg Vol. 2015, N3,
2015, arXiv:hep-th/1412.5991.

\bibitem{36A} E.H Saidi Chiral rings in the N= 4 SU (2) conformal theory,
Physics Letters B 300 (1-2), 84-91. 1993.

\bibitem{20N} E.H Saidi, Mutation Symmetries in BPS Quiver Theories:
Building the BPS Spectra, JHEP, 2012, Volume 2012, Number 8, 18,
arXiv:1204.0395.

\bibitem{20B} K. Costello, B. Stefanski, The Chern-Simons Origin of
Superstring Integrability, Phys. Rev. Lett. 125, 121602 (2020).

\bibitem{24} N. Ishtiaque, S.F Moosavian, S. Raghavendranc, J. Yagid,
Superspin chains from superstring theory, arXiv:hep-th/2110.15112.

\bibitem{26} N. Nekrasov, Open-closed (little) string duality and
Chern-Simons-Bethe/gauge correspondence. Talk at String Math 2017, July
24--28, 2017.

\bibitem{27} M. Ashwinkumar, M.-C. Tan, Q. Zhao, Branes and categorifying
integrable lattice models, Adv. Theor. Math. Phys. 24 (2020) 1 [1806.02821].

\bibitem{29} A. Kapustin, Wilson-'t Hooft operators in four-dimensional
gauge theories and S-duality, Physical Review D, 74(2), 025005 (2006).

\bibitem{33A} E. H Saidi, Gapped gravitinos, isospin 12 particles and N=2
partial breaking, Prog Theor Exp Phys (2019).

\bibitem{33B} A. Kapustin, E. Witten, Electric-Magnetic Duality And The
Geometric Langlands Program, arXiv:hep-th/0604151.

\bibitem{30} J. Yagi, Surface defects and elliptic quantum groups, JHEP 06
(2017) 013 [1701.05562].

\bibitem{36} T. Okuda, Line operators in supersymmetric gauge theories and
the 2d-4d relation. In New dualities of super gauge theories (pp. 195-222).
Springer, Cham. (2016).

\bibitem{34} E.H Saidi, Quantum line operators from Lax pairs, Journal of
Mathematical Physics 61, 063501 (2020), arXiv:hep-th/1812.06701.

\bibitem{31} K. Maruyoshi, T. Ota, J. Yagi, Wilson-'t Hooft lines as
transfer matrices, JHEP 01 (2021) Paper No. 072, 30 [2009.12391].

\bibitem{32} K. Maruyoshi, J. Yagi, Surface defects as transfer matrices,
Prog. Theor. Exp. Phys. (2016) 113B01, 52 [1606.01041].

\bibitem{34A} K. Costello, D. Gaiotto, J. Yagi, Q-operators are 't Hooft
lines, arXiv:hep-th/2103.01835.

\bibitem{34C} H. Hayashia, T. Okuda, Y. Yoshida, ABCD of 't Hooft operators,
10.1007/JHEP04(2021)241, arXiv:hep-th/2012.12275.

\bibitem{34D} T.D Brennan, A. Dey, G.W Moore, 't Hooft Defects and Wall
Crossing in SQM, arXiv:1810.07191.

\bibitem{37} N. Nekrasov, Superspin chains and supersymmetric gauge
theories, JHEP 03 (2019) 102 [1811.04278].

\bibitem{38} N. Nekrasov, S. L. Shatashvili, Quantum integrability and
supersymmetric vacua, Prog. Theor. Phys. Suppl. 177 (2009) 105 [0901.4748].

\bibitem{39} N. Nekrasov, S. L. Shatashvili, Supersymmetric vacua and Bethe
ansatz, Nucl. Phys. B Proc. Suppl. 192/193 (2009) 91 [0901.4744].

\bibitem{40} N. Nekrasov, Samson L. Shatashvili, Bethe/Gauge correspondence
on curved spaces, arXiv:hep-th/1405.6046.

\bibitem{54} Y. Boujakhrout, E.H Saidi, On Exceptional 't Hooft Lines in
4D-Chern-Simons Theory, Nuclear Physics B (2022), arXiv:hep-th/2204.12424.

\bibitem{54A} Y. Boujakhrout, E.H Saidi, Minuscule ABCDE Lax Operators from
4D Chern-Simons Theory, Nucl.Phys.B 981 (2022) 115859,
arXiv:hep-th/2207.14777.

\bibitem{51} B.H. Gross, On minuscule representations and the principal $%
SL_{2}$, Represent. Theory 4 (2000) 225.

\bibitem{55} S. Franco, A. Hanany, K. D. Kennaway, D. Vegh, B. Wecht, Brane
dimers and quiver gauge theories, JHEP 01 (2006) 096, arXiv:hep-th/0504110.

\bibitem{56} A. Hanany, C. P. Herzog, D. Vegh, Brane tilings and exceptional
collections, JHEP 0607 (2006) 001, arXiv:hep-th/0602041.

\bibitem{57} L. B. Drissi, E. H. Saidi, Domain walls in topological
tri-hinge matter, Eur. Phys. J. Plus (2021) 136: 68.

\bibitem{58} L B Drissi, E H Saidi, A signature index for third order
topological insulators, J. Phys.: Condens. Matter (2020) 32 365704.

\bibitem{FrA} R. Frassek, Oscillator realisations associated to the D-type
Yangian: towards the operatorial Q-system of orthogonal spin chains, Nuclear
Phys. B 956 (2020) 115063, 22 [2001.06825].

\bibitem{FrB} G. Ferrando, R. Frassek, V. Kazakov, QQ-system and Weyl-type
transfer matrices in integrable SO(2r) spin chains, 2008.04336.

\bibitem{59} S.L Cacciatori, B.L Cerchiai, A. Marrani, Magic coset
decompositions, Adv. Theor. Math. Phys. 17(5): 1077-1128, 2013,
arXiv:hep-th/1201.6314.

\bibitem{60} M. Esole, S. Pasterski, D4-flops of the E7-model,
arXiv:hep-th/1901.00093.

\bibitem{61} R.Slansky, Group theory for unified model building, Physics
Reports, Volume 79, Issue 1, 1981, Pages 1-128.

\bibitem{34L} B. Nasmith, An Exceptional Combinatorial Sequence and Standard
Model Particles, arXiv:math.CO/2012.03933.
\end{thebibliography}
\end{document}